%% file: main.tex
\documentclass[12pt]{article}

\input{preamble}

\title{Corporate Stances on Societal Issues in the U.S.\thanks{We thank Evelyn Cheng, Michelle Liu, Joie Liu, Joshua Olatubosun, Anna Mikhaylyants and Nicole Fang for excellent research assistance. We also thank the participants of the CBS-Princeton Money in Politics Conference and the American Political Science Association Meeting for helpful comments and suggestions.}}

\author{Jacob Conway\\{University of Chicago} \and
        Andrew Hall\\{Stanford University} \and
        Magdalena Larreboure\\{Harvard University} \and
        Caroline Le Pennec\\{HEC Montr\'eal and CEPR} \and
        Vincent Pons\\{Harvard University, CEPR and NBER; Corresponding author: vpons@hbs.edu} \and
        Anna Sun\\{Stanford University}}
\makeatletter
\renewcommand{\maketitlehookb}{\def\@author{{Jacob Conway}\thanks{University of Chicago} \hspace{0.5em}
        {Andrew Hall}\thanks{Stanford University} \hspace{0.5em}
        {Magdalena Larreboure}\thanks{Harvard University} \\
        {Caroline Le Pennec}\thanks{HEC Montr\'eal and CEPR} \hspace{0.5em}
        {Vincent Pons}\thanks{Harvard University, CEPR and NBER; Corresponding author: vpons@hbs.edu} \hspace{0.5em}
        {Anna Sun}\thanks{Stanford University}}}
\makeatother

\date{November 13, 2026}

\begin{document}

\begin{singlespace}
\setlength{\droptitle}{-2em}
\maketitle

\begin{abstract}
\noindent When do firms take public positions on divisive societal issues and why? We assemble a comprehensive dataset on corporate stance-taking in the United States, using large language models to read 52,000 news articles and press releases about 836 companies that belonged to the S\&P 500 at any point between 2008 and 2025. We identify 9,917 stances on nine partisan issues such as climate change, abortion, and racial discrimination. We show that stance-taking increased fourfold after 2008 and peaked in 2021. It is overwhelmingly liberal and concentrated among the largest firms. A quarter of stances give no explicit business justification. Stance-taking is only weakly related to lobbying and campaign contributions and, unlike them, does not follow the electoral cycle. Instead, firms tend to speak when media coverage and public interest in an issue peak. Together, these findings suggest that stance-taking can be understood as a contribution to broad policy debates rather than a strategy to shape policies that affect firms' core business. Finally, we investigate whose views corporate stances represent. Consumer-facing firms take more stances, and firms whose consumers lean Democratic take more liberal stances on issues other than climate. Companies with larger shares of Democratic employees, executives, and board members also take more liberal stances. Exploiting within-firm variation, we find that having more Democrats among a company's leadership increases liberal stance-taking, while having more women increases stance-taking on gender-related issues and abortion. 

\vspace{1em}

\vspace{0.5em}
\end{abstract}

\end{singlespace}

\newpage

\section{Introduction} \label{sec:intro}

Companies routinely use lobbying and campaign contributions to influence elections and regulations that may affect their operations and profitability. These non-market strategies advance firms' private interests while also conveying private information about certain costs and benefits of policy choices. In recent years, companies have increasingly engaged in a new form of political communication: taking public stances on societal issues that are at best loosely related to their core business. High-profile examples include Disney's opposition to Florida's "don't say gay" laws and corporate pledges to cover travel costs for employees seeking out-of-state abortions after the Supreme Court overturned the federal right to abortion. Existing research estimates the effects of some types of corporate stances on consumer and employee behavior \citep{adrjan2026we, conway2024consuming}, but comprehensive evidence on the prevalence of stance-taking and the motivations behind it remains limited.

In this paper, we construct the first comprehensive dataset on corporate sociopolitical activism in the United States from 2008 to 2025 and present three sets of results. First, we provide stylized facts on how the frequency of corporate stance-taking has changed over time, the societal issues these stances address, and their liberal orientation. Second, we show that stance-taking is better understood as a contribution to broader policy debates than as a strategy for shaping policies that directly affect firms’ core business. Third, we assess whose views these stances represent by examining their relationship with the preferences and sociodemographic characteristics of firms' executives, board members, employees, and consumers.

Studying corporate stance-taking is challenging. Unlike lobbying reports and campaign finance records, firms' stances on societal issues are scattered across press releases, CEO statements, petitions, and media coverage. 
We study all firms that were included in the S\&P 500 index at any point between January 2008 and June 2025. For each firm, we collect articles and press releases mentioning both the firm as well as keywords related to nine prominent issues on which the Democratic and Republican Party platforms present opposite views: climate change, LGBTQ rights, gender, race, minimum wage, immigration, abortion, guns, and voting rights. We then use text clustering and large language models (LLMs) to identify stories in which the firm or its CEO publicly supports or opposes one side of one of these issues. The resulting dataset contains 9,917 stances taken by 836 firms over 17 years. For each stance, we observe the date at which it first appeared in the news, its topic, ideological leaning, whether it is related to its business practices and justified by its interests, whether it was taken by the company or the CEO, whether it was communicated through a press release or a news article, and whether it was a response to criticism or an attack.

We combine this new dataset with a rich firm-level monthly panel. We collect lobbying expenditures from the Senate Lobbying Disclosure data and political contributions from the DIME database \citep{bonica2024dime}. We obtain workforce characteristics from several sources: a proprietary database from Revelio Labs, which provides information from employees' LinkedIn profiles; the Execucomp database; and the ISS Board Members database. We measure the political affiliation of employees, executives, and board members by matching them to voter records, and complement these measures with data on executives' and directors' political contributions. We also classify firms as business-to-consumer or business-to-business and use Twitter/X follower data to proxy for the partisan composition of firms' consumer bases \citep{schoenmueller2023frontiers}. Finally, we match our dataset with data on NGO campaigns against firms as well as peaks in news articles and Google searches on our nine topics of interest. 

Our first set of analyses provides broad stylized facts about corporate stance-taking. The yearly number of stances per firm was relatively stable from 2008 to 2018, at about 0.5 stances per year, then increased sharply and peaked in 2021, at about two stances per year, before falling back to 0.7 by 2024. The rise and fall of stance-taking is driven mostly by climate change, which accounts for 63\% of all stances, and by diversity, equity, and inclusion (DEI) issues such as gender, race, and LGBTQ rights, which account for another 30\%. Stance-taking on minimum wage, immigration, abortion, guns, and voting rights is less frequent and more episodic, with peaks around specific events such as the 2018 gun-control debate and the 2022 abortion debate. Stance-taking is also overwhelmingly liberal: depending on the topic, the share of stances coded as liberal ranges from 86 percent for minimum wage to more than 98 percent for LGBTQ rights, voting rights, racial issues, and climate change.

Our second set of analyses examines whether stance-taking is best understood as a strategy for shaping policies that affect firms’ core business, as with lobbying and campaign contributions, or as a contribution to broader policy debates. We present several pieces of evidence supporting the latter interpretation. Firms that spend more on lobbying also tend to give more in campaign contributions, consistent with the idea that these strategies are complements used by firms seeking to influence policymaking. By contrast, stance-taking is much more weakly related to either lobbying or donations. 
The timing of stance-taking also differs from the timing of traditional non-market strategies. Corporate political donations rise before midterm and presidential elections, and lobbying expenditures tend to increase when policymaking is most active during the presidential cycle. By contrast, stance-taking does not follow the electoral cycle. It is not concentrated before elections, when firms might try to influence voters or support candidates, nor after elections, when firms might try to affect the agenda of newly elected officials. 

In addition, close to a quarter of stances express support for policies that are unrelated to the firm's activity. The vast majority of other stances are public statements about changes firms made to their operations that do not aim to directly shape the policies affecting their business either. Only 3\% of stances express support for a certain policy by citing the company’s interests as a justification. Some stances mention a criticism that the firm received, but we do not find any evidence that firms take more stances after NGO campaigns that directly target them or their sector of activity. Overall, our evidence suggests that most stances are best characterized as expressions of support for societal goals and contributions to policy debates that are not directly related to firms’ core business.

We bring additional support for this interpretation by showing that stance-taking follows the news cycle. Using an event-study design, we find that peaks in media coverage or Google searches generate contemporaneous increases in the number of stances taken by firms on the same topic. The increase is visible in the month of the peak and fades afterward. Rather than lagging behind public debate, firms take stances when an issue becomes salient to journalists and the public. 

Our third set of analyses investigates whose views corporate stances represent. There is a long-standing debate between the shareholder and stakeholder models of the firm’s mission: do and should companies focus solely on maximizing profits \citep{friedman1970business} or should they also serve a broader set of interests, including those of suppliers, employees, consumers, and society at large \citep{freeman1984strategic}? We examine how closely corporate stances on societal issues align with the views of executives and board members, which may reflect either their personal preferences or shareholders’ interests, and with those of other stakeholders. 

Only 6\% of stances are coming from the firm's CEO personally, suggesting that CEOs are not the only relevant voice. The identity of the most active firms points to the importance of public-facing stakeholders: top stance-takers include well-known consumer brands and firms that provide essential services to households. Consistent with this interpretation, business-to-consumer firms take 0.24 more stances per year than business-to-business firms, controlling for year effects and firm size, a difference equal to about 40\% of the average yearly number of stances. The gap is especially large for DEI-related issues. 

We then relate liberal stance-taking to the partisan composition of firms' employees, executives, board members, and consumers. Firms with more Democratic employees, executives, and board members take more liberal stances, and, for issues other than climate, firms with more Democratic consumers do so as well. While most of these relationships become smaller when we include firm fixed effects, the relationship with board partisanship remains positive and significant. We also find that, within firm, a larger share of female executives and board members increases the number of stances on gender-related issues. A larger share of female executives also increases the number of stances on abortion. These patterns suggest that stance-taking reflects the preferences of the people in the firm, especially those who are in power.

This paper first contributes to the growing literature on corporate socio-political activism. Conceptual work has emphasized that this form of engagement differs from corporate social responsibility because it is partisan and potentially divisive \citep{wettstein2016why, nalick2016corporate, hydock2019consumer, bhagwat2020corporate}. Empirical work has focused on a non-representative selection of salient stances and investigated their effects on consumers \citep{klostermann2022effect,conway2024consuming}, investors \citep{bhagwat2020corporate}, and workers \citep{wowak2022employees,adrjan2026we}.
We complement this literature by providing systematic evidence on corporate stance-taking and by studying the factors motivating this form of communication rather than its effects. Collecting data on a wide range of stances rather than focusing on one specific topic or a few salient events allows us to ask which firms choose to take stances, on which issues, when, and for what reason. While a few recent papers have focused on specific communication channels to study firms' political speech, such as earnings calls, annual reports, securities disclosures or social media \citep{ottonello2024anatomy, cassidy2025partisan, dottling2026talking, mccarty2026dei}, our database records \textit{all} instances in which a firm's position on one of our nine societal issues appeared in the media, whether it was communicated through a press release, a CEO announcement during an interview, a post on social media, or anything else.

Second, the paper contributes to a broader literature on firms' non-market strategies and political influence. A large body of work studies lobbying, campaign contributions, political connections, and other channels through which firms seek to influence policy in their favor \citep{hillman2004corporate, lux2011mixing, kalla2016campaign, fowler2020quid, bertrand2021hall, teso2025influence}. We show that public stance-taking on societal issues is empirically distinct from more standard forms of corporate political activity. Firms' public engagement with societal issues is not simply a new channel for pursuing firm-specific policy benefits. Instead, it is closer to a form of public participation in broad policy debates. 

Third, the paper speaks to research on the role of politics inside the firm. A first set of papers shows that firms led or governed by more Democratic-leaning executives and directors engage more with social and environmental issues \citep{di2014red, steel2026political}. A second set of papers shows that firms and their leaders can mobilize workers politically, encourage them to support particular candidates or causes, and shape their campaign contributions \citep{hertel2017american, babenko2020ceos, stuckatz2022workplace}. A third set of papers shows that workers care about firms' values, culture, diversity, and social commitments \citep{hedblom2019toward, huang2024communicating, choi2023jobseekers, colonnelli2023polarizing,adrjan2026we}. Employees and executives sort across firms along partisan lines, and ideological fit within the workplace predicts retention and career choices \citep{bermiss2018ideological, fos2022political, frake2026political, chinoy2024political, colonnelli2025politics}. We contribute to this literature by studying a highly visible political action from the firm and linking it to the partisan composition of several stakeholder groups. The results suggest that corporate stances are not merely expressions of CEOs' personal ideology: they are more often taken in the name of the company, and they are systematically related to the preferences of board members, executives, employees, and consumers. More broadly, our findings suggest that corporate stance-taking may both reflect and reinforce political sorting across firms.

The rest of the paper proceeds as follows. Section \ref{sec:data} describes the construction of our dataset and the other firm-level data sources. Section \ref{sec:facts} documents stylized facts about the timing, topics, ideological leaning, and firm-level distribution of corporate stance-taking. Section \ref{sec:goal} compares stance-taking to traditional non-market strategies and studies its relationship with the electoral cycle, the news cycle, and NGO campaigns. Section \ref{sec:audience} investigates which voices and stakeholders corporate stances appear to represent. The final section concludes.

\section{Data} \label{sec:data}

The contribution of this paper is to assemble a 17-year-long monthly panel of publicly-traded firms measuring their stances on societal issues. We merge this panel with information on these firms' sector and revenues, their lobbying and campaign donations, and the characteristics of their workforce, leadership, and consumers.

\subsection{Corporate Stances on Societal Issues} \label{subsec:data_stances}

\paragraph{Set of firms and topics} We record all the instances in which a company included in the S\&P 500 stock market index took a stance on a societal issue between January 1st, 2008 and June 1st, 2025. We focus on the nine following topics: climate change, LGBTQ, gender, race, minimum wage, immigration, abortion, gun control, and voting. We selected these topics based on the Democratic and Republican Party platforms from the 2008,
2012, 2016, and 2020 presidential elections.\footnote{These platforms are publicly accessible on the official websites of the Democratic Party (democrats.org) and the Republican Party (gop.com) prior to each election. Archived versions are available through Ballotpedia and the American Presidency Project at UCSB.} These documents outline the parties' positions on domestic and foreign affairs and the issues they intend to focus on during the presidential election. We read the platforms to extract the most prominent topics on which the two parties disagree and a set of keywords for each of them. Our topic selection method is described in more detail in Appendix \ref{ap:topic_selection}, and the list of keywords for each topic is shown in Appendix Table \ref{tab:topics}.

Our sample includes all firms that were part of the S\&P 500 index at any point between January 2008 and June 2025. For each firm, we collect data from either the beginning of our sample period or the firm’s founding date (if it was founded later than January 2008), and until either the end of the sample period or the firm's bankruptcy date (if it went bankrupt before June 2025). Appendix \ref{ap:sample_firms} describes how we handle firms that changed names, declared bankruptcy, merged, were acquired, or split during the sample period.
Our final sample comprises 836 publicly traded companies, of which 83.5\% existed throughout the entire sample period and 29.2\% were included in the S\&P 500 index throughout the entire period.

\paragraph{Stance extraction}

We record public stances taken by these companies or their CEOs based on news articles from the three main global news agencies (Associated Press, Reuters News, and Agence France Presse), as well as press releases distributed via four services (GlobeNewswire, PR Newswire, Dow Jones Institutional News, and Business Wire). The three news agencies are the wholesale producers of U.S.\ news that other outlets republish. Excluding other news outlets avoids multiplying near-duplicate articles about the same event and limits the ideological and geographic bias present in partisan or local newspapers. The four press-release services are the standard channels through which firms disseminate their own statements. Using the research tool \textit{Factiva}, we find 74,222 articles and press releases referring to one of the companies over the sample period and mentioning at least one of the nine topics' keywords.\footnote{See Appendix \ref{ap:factiva} for more details.}

Since it is very common for several sources to publish articles about the same story, we use a clustering algorithm to identify duplicate articles or repeated reporting of the same event and record the date at which the first article relating that story was published. Specifically, we group news articles and press releases that refer to the same story based on the semantic similarity of their text. We represent each article by a sentence embedding computed with Sentence-BERT (SBERT), a modification of the pretrained BERT network designed to produce sentence embeddings that can be compared to one another using cosine similarity. We then group articles whose embeddings are highly similar. More details are provided in Appendix \ref{ap:factiva}. This leaves us with a total of 51,978 articles, corresponding to the first (or only) occurrence of a story that mentions both a company of our sample and a keyword related to one of our topics. 

The vast majority of these stories do not correspond to instances in which a company took a stance on a topic. For instance, many articles mention the company's name in one paragraph and one of the keywords several paragraphs later to report a piece of news that is unrelated to the company. Therefore, we use the assistance of OpenAI's large language model GPT version 4o-mini to perform the following tasks. First, we ask the model to determine, for each article, whether it refers to an instance of corporate socio-political activism (CSA), following the definition proposed by \citet{bhagwat2020corporate}: ``A firm's public demonstration (statements or actions) of support or opposition to one side of a partisan sociopolitical issue.'' We then run a second prompt to confirm that the company of interest (or its CEO) is one of the main actors mentioned in the article, and that the article is primarily about the topic of interest. That second prompt also asks the LLM to extract the following information about the stance: whether it was taken by the company or by the CEO, whether it was leaning liberal or conservative, and whether it was taken in response to a criticism or an attack (e.g., from employees or from an NGO). 
In a third prompt, we ask the LLM to put each stance in one of three buckets: stances in which the company publicizes current or future business practices that align with its position on a societal issue, stances that express the company’s position on a topic based on the company's interests, and statements that frame the company’s position on a topic without reference to a business rationale.

Our final sample contains 9,917 stances. Appendices \ref{ap:gpt_prompts} and \ref{ap:validation} show our prompts, describe how we fine-tuned them, and outline the tests we performed to quantify measurement errors. The comparison between our sample of stances and smaller sets of stances from alternative sources suggests that we miss less than 10\% of the stances taken by an S\&P 500 company on one of our nine issues, while a human reading of a random subset of 200 stances reveals that 3.5\% of them are not actual stances.

\paragraph{Facebook posts} To assess the completeness of our sample, we use an alternative source of information on corporate stances: social media posts. Using the Meta Content Library, we identify Facebook pages for 36.6\% of the companies in the sample. We collect all the posts related to the nine topics of interest and perform the same LLM classification to determine whether each post corresponds to the company taking a stance on the topic. The resulting dataset contains 2,088 posts.

\subsection{Political Activity}

We complete our data on stance-taking, our main outcome of interest, with data on two other non-market strategies: lobbying and campaign donations.

We collect lobbying data using the Senate’s Lobbying Disclosure API.\footnote{\citet{lda2026}: https://lda.gov/api/} We track both the total lobbying expenditure as well as the number of lobbyists for the 651 firms in the sample (78\%) that filed at least one lobbying report between 2008 and 2025. The remaining 185 companies never appear in the Senate data, which means they did not register lobbying activity.\footnote{Registration and quarterly reporting are mandatory above a threshold (currently \$16,000 per quarter of in-house lobbying expenses, \$3,500 for an outside lobbying firm), so companies absent from the Senate data spent no more than these amounts.} 

In addition, we measure the campaign contributions made by each company to each party using the Database on Ideology, Money in Politics, and Elections \citep{bonica2024dime}.\footnote{As firms in the DIME database often donate under various names (e.g., Verizon donates as ``Verizon Communications,'' ``Verizon Communications Inc PAC,'' and ``Verizon Wireless,'' among others), we use fuzzy string matching to match firms to DIME donors. Specifically, we adopt a two-step matching procedure. First, we perform an exact name match between DIME donors and our firms. Next, we relax the matching requirement between the donor and firm name, keeping fuzzy matches with a Jaccard similarity score above 0.72. We manually check the matched results to correct any obvious mismatches.} Our measure includes both donations made directly to political parties and donations to individual candidates and organizations affiliated with these parties. According to our matching procedure, 74.0\% of firms in the sample made at least one political contribution in election cycles ranging from 2008 to 2024, the last election cycle year for which DIME data are available as of September 2026.

\subsection{Workforce, Leadership, and Consumers}

We combine multiple existing databases to collect information on the sociodemographic characteristics and partisanship of three types of stakeholders: the workforce, leadership, and consumers of the firms in our sample. 

We obtain firms' monthly headcount and their employees' gender and ethnicity composition from Revelio Labs, a company that collects and aggregates public LinkedIn profiles.\footnote{Appendix Section \ref{ap:revelio} explains how we construct the corresponding variables.}  
To measure employees' political affiliation, we merge each LinkedIn profile from Revelio to the corresponding voter file record from the data vendor L2. We adopt 50 combinations of individual name, age, and work location to perform our matching procedure, following \citet{chinoy2024political}. We provide more details in Appendix \ref{ap:matching_l2}. The Revelio data show 21,549,040 employees working for firms in our sample. We exclude non-US employees, employees with missing first or last names, and employees whose LinkedIn job entry contains missing start date, end date, or location, and restrict the sample to the 33 states recording voters' party registration, resulting in 12,247,449 unique LinkedIn profiles. We successfully match 41.5\% of them to a voter record, which is similar to the 40.4\% match rate obtained by \citet{chinoy2024political}. As an individual can switch parties over time, we allow each individual's partisanship to vary between jobs, resulting in 9,274,952 employee-firm observations. We classify employees' partisanship as Democrat, Republican, non-partisan, or other.

We collect similar data on the top executives and board members of each company. We obtain the age, gender, and race of board members from the ISS Board Members database \citep{iss_directors2026}, and the age and gender of the most highly paid executives of each company from the Execucomp database \citep{execucomp2026}. We accessed both datasets via Wharton Research Data Services (WRDS). We complement them with data from Fortune and Investopedia listing top executives who are Black and with data from Fortune Magazine and Out Leadership listing both executives and board members who publicly identify as LGBTQ+. 
We use the same matching with voter records as above to collect the party affiliation of 4,641 executives and 6,670 board members, and leverage the BoardEx-DIME crosswalk developed by \citet{steel2026political} to match 3,216 top executives and 5,350 board members with data on their political contributions to each party up to 2024 \citep{bonica2024dime}. The DIME data provide the political donations of each donor by election cycle and party as well as their weighted CFScore, a summary measure of their ideology.
Finally, we gather two types of information on consumers. We ask  OpenAI's GPT model to classify each company as either business-to-consumer (B2C) or business-to-business (B2B), depending on whether its primary activity serves individual consumers or other businesses. In addition, we use the Twitter-based brand political affiliation scores from \citet{schoenmueller2023frontiers} as a proxy for the political leaning of a company's consumer base. This score is defined as the share of a company's Twitter/X followers who also follow the Democratic Party account, out of those who follow the company as well as the Democratic party but not the Republican party or vice versa. This measure is available for about 137 companies and does not vary over time.

\subsection{Other Data} \label{subsec:other_data}

\paragraph{Company Financials and Industry Classification}

We use Compustat, accessed through Wharton Research Data Services (WRDS), to collect each company's annual revenue, which we use as a measure of company size, as well as their industry classification under the Global Industry Classification Standard (GICS), which we use to define industry fixed effects.

\paragraph{Sigwatch}

We identify campaigns run by NGOs against the companies in the sample using a publicly available database from Sigwatch,\footnote{\url{https://sigwatch.com/}.} a European consultancy firm which records cases in which an NGO applies pressure to a ``target" (e.g., issuing publicly a criticism or allegation, organizing a significant public protest, etc). The database provides the date at which a firm is targeted by one or several NGOs along with some information about the general topic of the campaign (e.g., the environment). 

\section{Stylized Facts} \label{sec:facts}

\begin{figure}[H]
    \centering
    \caption{Corporate Stance-Taking over Time}
    \includegraphics[width=0.9\linewidth]{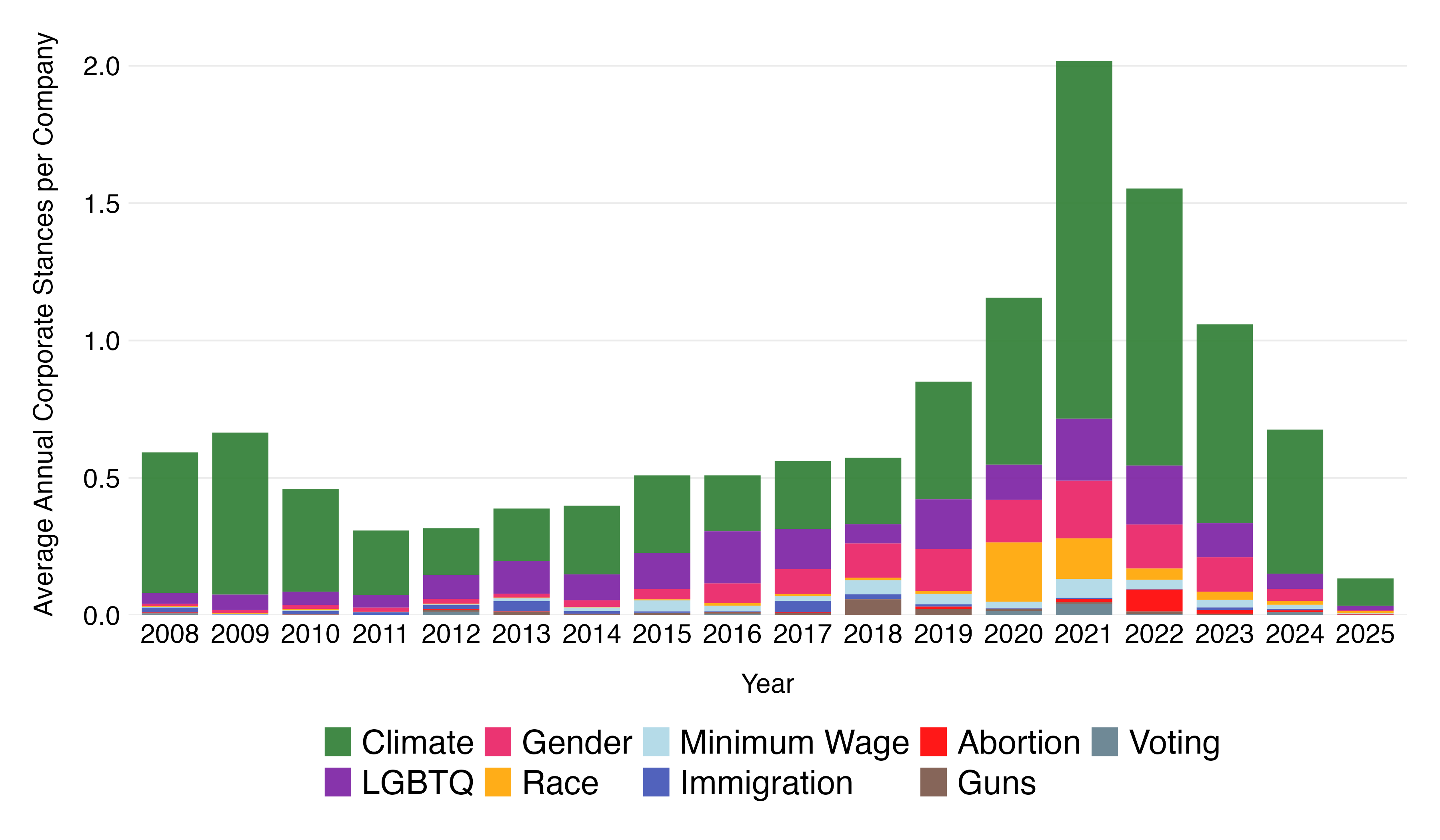}
    \label{fig:timeline}
    \begin{minipage}{0.9\linewidth}
    \footnotesize
        \textit{Notes:} Average yearly number of corporate stances by S\&P 500 companies, 2008-2025, by topic. 2025 includes data only up to June 1st.
    \end{minipage}
\end{figure}

We start by establishing several stylized facts about corporate stance-taking. First, as shown in Figure \ref{fig:timeline}, the average yearly number of stances taken by a firm remained stable from 2008 to 2018, around 0.5. It then rapidly increased and peaked in 2021, with an average of 2.0 stances per year, before going back to 0.7 stances per year in 2024. This pattern is predominantly driven by an increase followed by a decrease in the number of stances on climate-related issues, which represent 63\% of all stances in the sample. The increase in the overall number of stances taken between 2008 and the peak of 2021 was also fueled by an increase in the number of DEI-related stances on gender issues (from 0.01 to 0.21 yearly stances, on average), racial issues (from 0.005 to 0.21 yearly stances), and LGBTQ issues (from 0.05 to 0.22 yearly stances). Overall, DEI issues represent 30\% of all stances. The number of stances on the five remaining topics (i.e., minimum wage, immigration, abortion, guns, and voting) has remained comparatively small and stable over time, with a few sporadic peaks (e.g., a peak in gun-related stances in 2018, and a peak in stances on abortion in 2022).

Identifying corporate stances from companies' social media activity rather than news articles yields remarkably similar patterns. Restricting our main sample to the 223 companies for which we observe Facebook posts, we find that the trend in the number of stances taken on Facebook closely follows the trend in stances observed in Figure \ref{fig:timeline}, with a correlation of 0.88 between the two quarterly series (Appendix Figure \ref{fig:meta_val_timeline}). In addition, the companies that take stances on Facebook tend to be the same as those that take stances according to our news data: for the 223 companies for which we have both news data and Facebook posts, 94\% of the companies that take a stance on Facebook also take a stance in the news, and 95\% of those that take a stance in our news data also take a stance on Facebook. Among companies taking stances in both sources, controlling for companies' headcount and revenue, a 10\% increase in the number of stances in our news data is associated with a 4.7\% increase in the number of stances observed on Facebook (Appendix Figure \ref{fig:meta_val_loglog}). 
In other words, the corporate stances that we collect from news articles are representative of stance-taking in general, including on social media. However, our main dataset is much more comprehensive: it contains 4.6 times as many stances, taken by almost three times as many companies. While stances on climate, LGBTQ, gender, and racial issues are well-represented in our Facebook sample, there are only three posts that represent stances on the minimum wage, four on abortion, seven on immigration, and eight on gun control. Focusing on social media would therefore miss many stances that are reported in the news, including stances on topics that are less common but on which companies still take positions publicly.

\begin{figure}[H]
    \centering
    \caption{Characteristics of Corporate Stance-Taking}
    \includegraphics[width=0.9\linewidth]{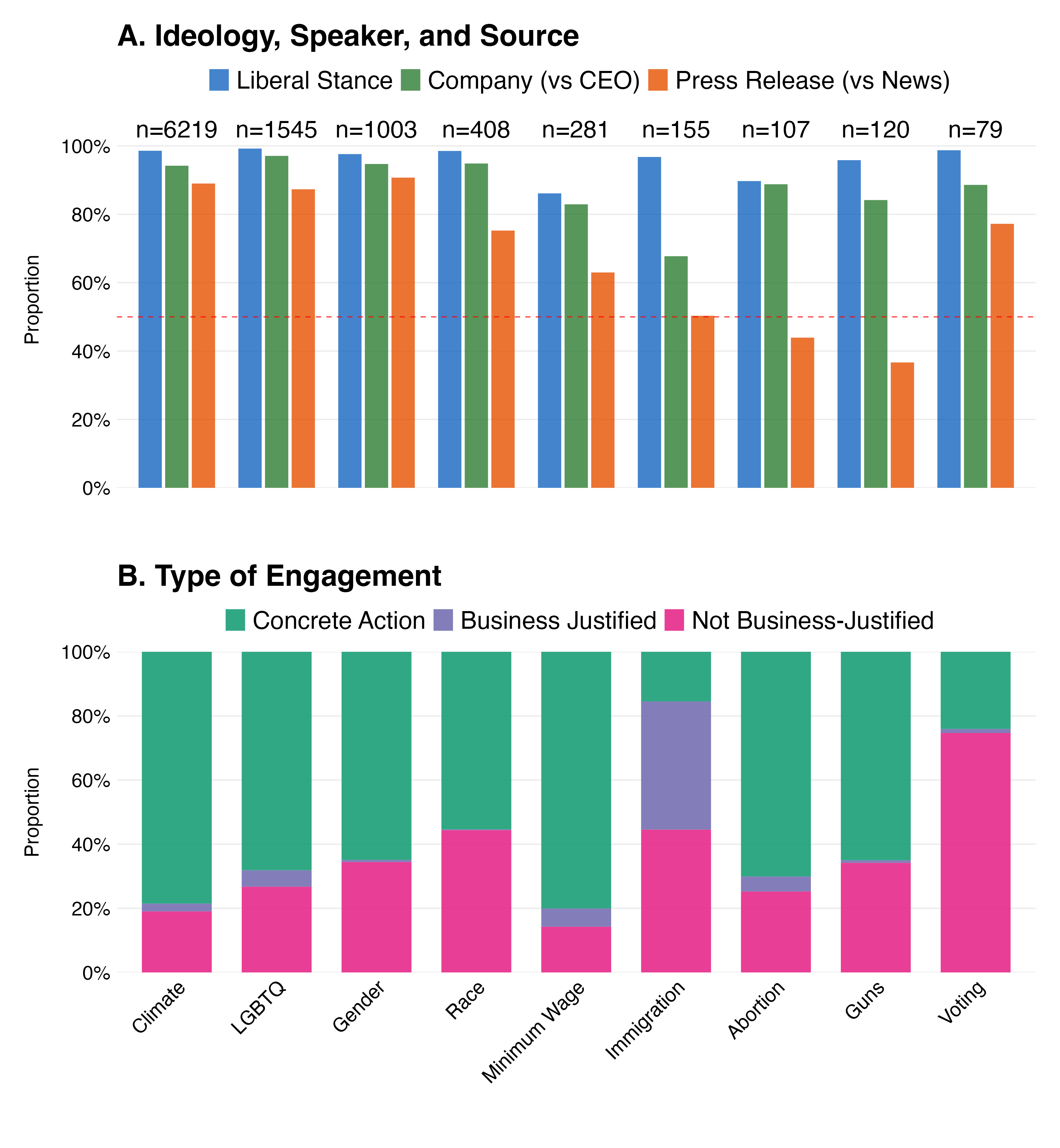}
    \label{fig:characteristics}
    \begin{minipage}{0.9\linewidth}
        \footnotesize
        \textit{Notes:} Panel A shows the proportion of corporate stances by S\&P 500 companies that are liberal vs.\ conservative (blue), from companies vs.\ CEOs (green), and communicated via press releases vs.\ news coverage (orange). Sample sizes shown above each topic. The dashed line indicates the 50\% threshold. Panel B shows the proportion of stances in which the company publicizes current or future business practices that align with its position on a societal issue (teal green) vs.\ stances that express the company's position on a topic based on the company's interests  (purple) vs.\ stances that frame the company's position on a topic without reference to a business rationale (pink).

    \end{minipage}
\end{figure}

Second, for each of our nine topics of interest, the vast majority of stances are liberal-leaning (Figure \ref{fig:characteristics} and Appendix Figure \ref{fig:timeline_chars_prop}, Panel A). This ranges from 86\% for stances on minimum wage to over 98\% for stances on LGBTQ issues, voting (which are typically related to voting ID laws and the disenfranchisement of some racial groups), racial issues, and climate. Despite this overwhelming representation of liberal stances, we still find several instances of conservative stance-taking on our different topics. For example, in February 2015, eBay Inc. publicly opposed a shareholder proposal seeking to address the gender pay gap within the company, despite growing public demands for gender wage equality. Similarly, in January 2025, Wells Fargo \& Co. withdrew from the Net Zero Banking Alliance amidst a shift in the U.S.\ political landscape regarding climate change.

Third, stances tend to be taken in the name of the company, rather than by the CEO personally, and to be communicated through press releases controlled by the company, rather than found in news articles written by a third party (Figure \ref{fig:characteristics}, Panel A). However, this varies across topics. For topics that are the most represented in our sample and on which companies are very likely to take a liberal stance, like climate and DEI issues, press releases issued by the company are predominant (about 88\%) and stances taken by the CEO are relatively rare (about 6\%). For other topics, such as minimum wage, immigration, abortion and guns, press releases are less common (48\% of stances) and CEOs are more likely to step in (19\% of stances). Conservative stances are also relatively more common on those topics (e.g., 14\% of stances on minimum wage and 10\% of stances on abortion lean conservative, respectively), suggesting that companies take fewer stances (Figure \ref{fig:timeline}) on topics on which there is not an overwhelming liberal consensus in the corporate world, and these stances are more likely to be personally taken by the company's leader rather than coming from the PR department (Figure \ref{fig:characteristics}). The share of stances taken by companies rather than CEOs and the share of stances communicated through press releases rather than third-party news articles have remained stable over our study period (Appendix Figure \ref{fig:timeline_chars_prop}, Panels B and C).

\begin{figure}[H]
    \centering
    \caption{Corporate Stance-Taking by Company Size}
    \includegraphics[width=0.9\linewidth]{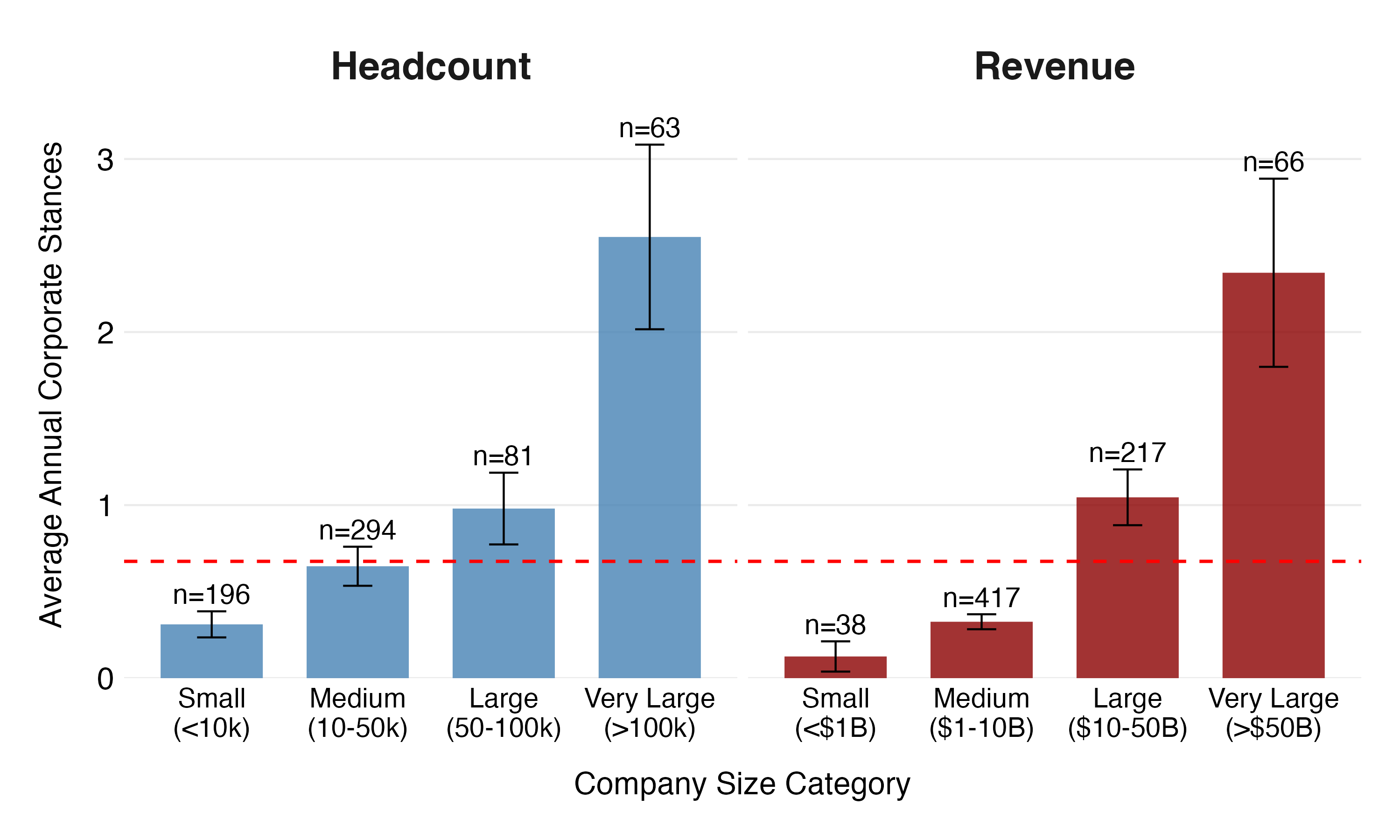}
    \label{fig:companysize}
    \begin{minipage}{0.9\linewidth}
    \footnotesize
        \textit{Notes:} Average annual corporate stances per S\&P 500 company, 2008--2025, categorized by headcount (left panel) and revenue (right panel). Error bars show 95\% confidence intervals. The red dashed line shows the overall average across the sample. Sample sizes differ between panels due to variation in data availability across size measures.
    \end{minipage}
\end{figure}

Fourth, larger companies take a lot more stances. As shown in Figure \ref{fig:companysize}, the 63 companies in our sample that have more than 100,000 employees take an average of 2.6 stances per year, which is almost 10 times more than the 196 firms that employ fewer than 10,000 individuals (0.3 stances per year). Similarly, companies making more than \$50B in average annual revenue take 2.4 stances per year, on average, against 0.1 for companies making less than \$1B. There are also significant differences across industries (Appendix Figure \ref{fig:sector}), with utilities companies, the top stance-taking industry, taking an average of 2.2 stances per year, against 0.3 for health care companies, the lowest stance-taking industry. This heterogeneity in the average number of stances is largely driven by disparities in the likelihood of taking any stance: out of the 836 companies in our sample, 211 have never taken any stance between 2008 and 2025, while the top ten stance-taking companies (see Appendix Figure \ref{fig:top10companies}) have taken more than 1,200 stances over the same period. Among them, we find well-known brands from various sectors, such as Alphabet (Google's parent company), Bank of America, Amazon, Walmart, and Microsoft. The other companies on that podium are either public utilities companies providing electricity and gas (Exelon, PG\&E, Sempra, and Duke Energy) or oil and gas producers (Exxon Mobil). Several of these companies' CEOs are also found among the top ten stance-taking CEOs, along with the CEOs of Apple, Meta (formerly Facebook), Tesla, and several companies from the financial sector (BlackRock, Lehman Brothers, JPMorgan Chase).

\section{Stance-Taking as a Contribution to the Broader Policy Debate} \label{sec:goal}

In this section, we ask whether firms' stances on societal issues are best understood as efforts to shape regulations affecting them and their sector or as contributions to the broader policy debate. While we do not observe firms' intentions directly, several pieces of evidence support the latter interpretation.

\subsection{Stance-Taking and Traditional Non-Market Strategies}

We start by showing that stance-taking substantially differs from campaign donations and lobbying, the two main non-market strategies used by firms to influence their regulatory environment. 

First, we estimate the correlation between a firm's propensity to take stances on our nine topics of interest on one hand and their political donations or the amount they spend on lobbying on the other hand, using the following equation:

\begin{equation} \label{eq:pol_activity}
\ln(S_{it})=\gamma_t+\lambda^I_i+\boldsymbol{\delta}'\!\mathit{Size}_{it}+\beta\,\ln(X_{it})+\epsilon_{it},
\end{equation}

where $S_{it}$ is the number of stances taken by firm $i$ in year $t$; $X_{it}$ is either the dollar amount of political donations made by firm $i$ to the Democratic and Republican parties in year $t$ or its lobbying expenditures; $\gamma_t$ is a year fixed effect; $\lambda^I_i$ is an industry fixed effect; and $\mathit{Size}_{it}$ is a vector of controls including company headcount and revenue.\footnote{These controls are mean-imputed when missing, with corresponding indicator variables flagging the imputed observations.} Standard errors are clustered at the company level. We restrict the sample to observations for which both $S_{it}$ and $X_{it}$ are strictly positive so the coefficient $\beta$ represents the stance elasticity to changes in political donations or lobbying.\footnote{This restriction enables us to take logs and measure elasticities without using transformations such as $\log(\cdot+1)$, which are unreliable \citep{chen_logs_2024}. Appendix Figure~\ref{fig:cpa-loglog-appendix} reports the same estimates on the full sample (including observations for which $S_{it}$ or $X_{it}$ are null), using the $\log(\cdot+1)$ transformation. Results are qualitatively similar.}  
 
\begin{figure}[H]
    \centering
    \caption{Stances and Traditional Non-Market Strategies}
    \includegraphics[width=\linewidth]{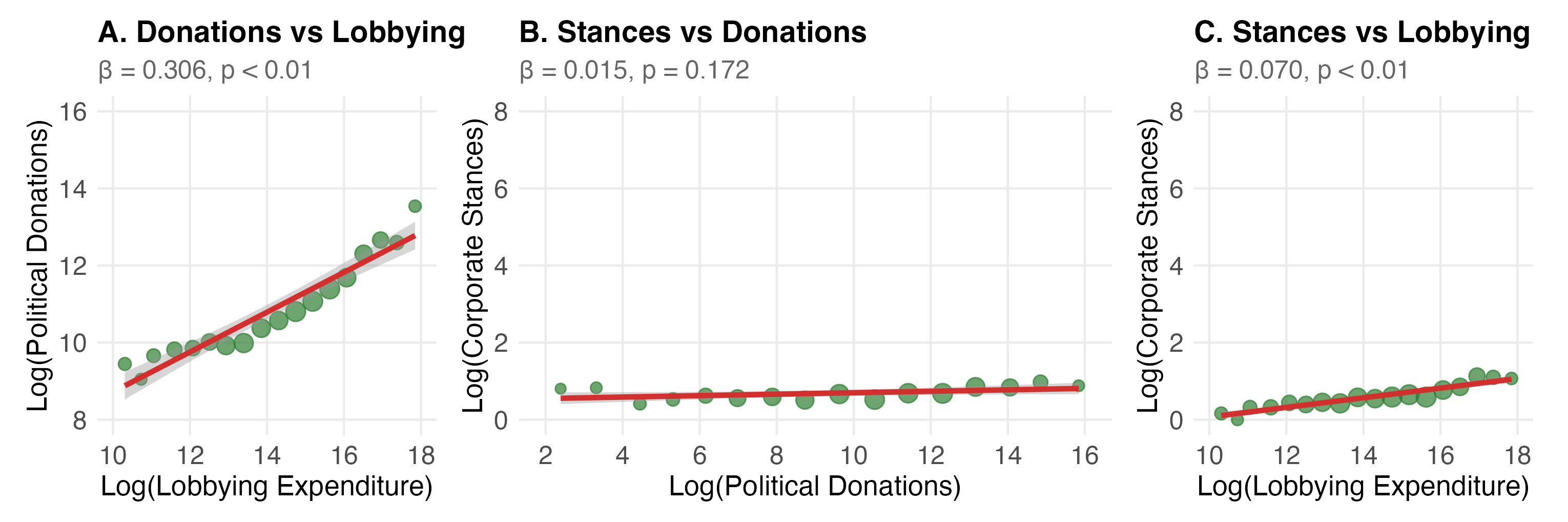}
    \label{fig:political_act}
     \begin{minipage}{\linewidth}
        \footnotesize
        \textit{Notes:} Binscatter plots showing the relationship between different types of corporate political activity, controlling for industry and year fixed effects as well as company headcount and revenue. Each regression uses only observations with strictly positive values of the two variables. Each point represents the mean of 20 equally-sized bins, with point size reflecting the number of observations per bin. The line is a linear fit with its 95\% confidence band. Panel A shows the relationship between the yearly amount of political donations made by a firm and its yearly lobbying expenditure ($\beta$ = 0.306, p < 0.001, partial $R^2$ = 0.032, N = 4,351 company-year observations). Panel B shows the relationship between the yearly number of stances taken by a firm and its yearly political donations ($\beta$ = 0.015, p = 0.172, partial $R^2$ = 0.002, N = 2,338  company-year observations). Panel C shows the relationship between the yearly number of stances taken by a firm and its yearly lobbying expenditure ($\beta$ = 0.070, p < 0.001, partial $R^2$ = 0.016, N = 2,646  company-year observations). Standard errors are clustered at the company level.
    \end{minipage}
\end{figure}

Panel A of Figure \ref{fig:political_act} first shows a strong relationship between donations and lobbying: controlling for industry fixed effects and company size, a 10\% increase in lobbying expenditures is associated with a 3.1\% increase in political donations, which is significant at the 1\% level.\footnote{In this panel, we estimate equation \ref{eq:pol_activity} using donations as the outcome and lobbying as the explanatory variable.} This elasticity, which corroborates existing findings, suggests that firms which seek to influence policymaking use both strategies as complements. By contrast, the relationships between either of these two strategies and stance-taking shown in Panels B and C are much weaker. The elasticity of stances with respect to political donations is statistically indistinguishable from zero. The elasticity of stances with respect to lobbying is positive but small: a 10\% increase in lobbying is associated with a 0.7\% increase in stance-taking. 

The same patterns hold at the extensive margin. After partialling out controls and fixed effects, the partial Pearson correlation between an indicator for making any donation and an indicator for making any lobbying expenditure is $r = 0.25$. The corresponding correlations between an indicator for taking any stance and an indicator for either making any donation or making any lobbying expenditure are less than half as large ($r = 0.11$ and $r = 0.12$, respectively): while the firms that spend on lobbying also tend to make political donations, firms that take stances on societal issues are often different.

\begin{figure}[H]
    \centering
    \caption{Corporate Political Activities Over the Electoral Cycle}
    \includegraphics[width=0.8\linewidth]{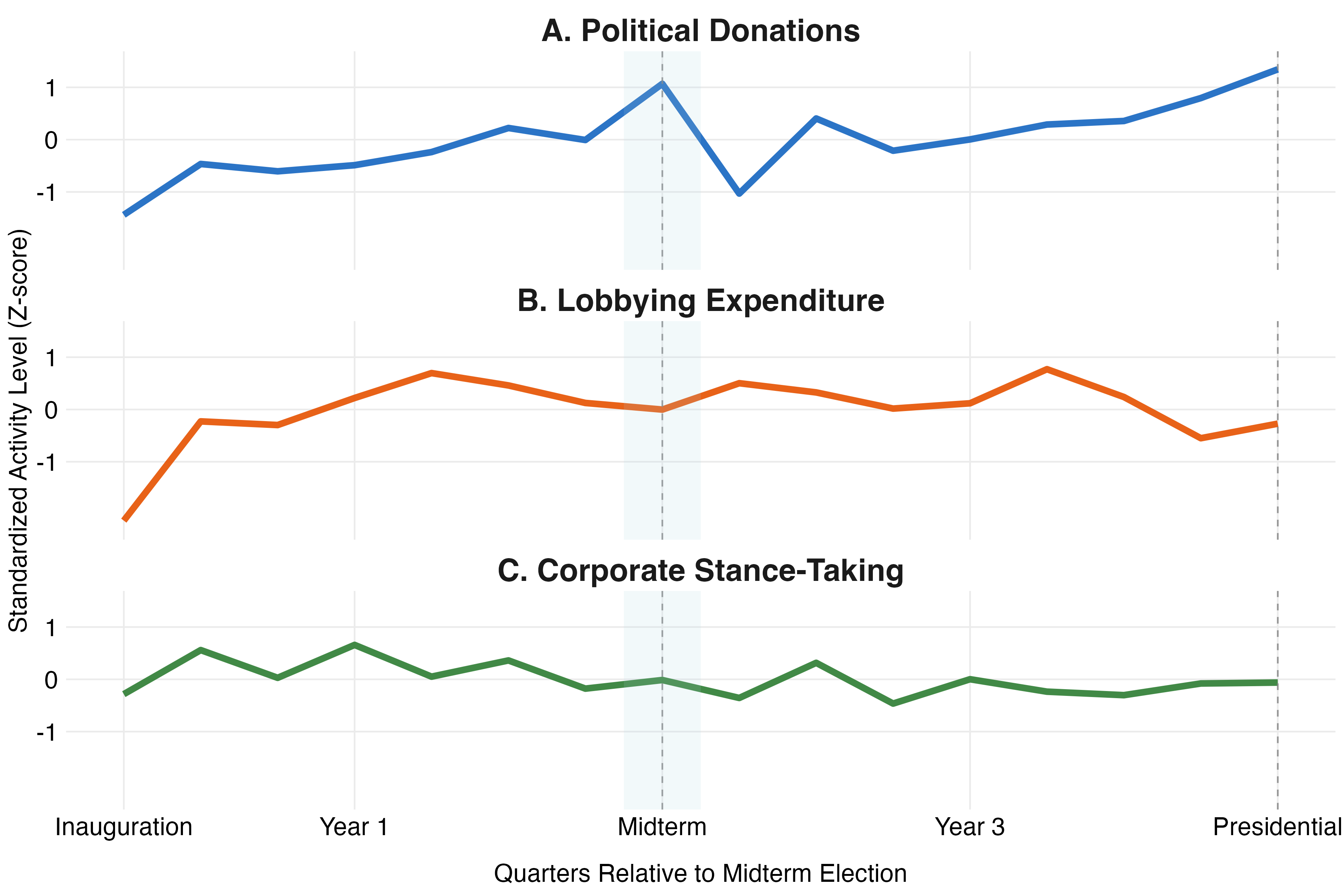}
    \label{fig:electoral_cycle}
     \begin{minipage}{\linewidth}
        \footnotesize
        \textit{Notes:} Average corporate political activities by S\&P 500 companies over the four-year presidential terms between 2009 and 2024. All measures are aggregated at the quarter level, with the first quarter containing September-October-November of each election year, and standardized to enable cross-activity comparison. Shaded areas mark midterm election quarters. N = 48,768 company-quarter observations over four electoral cycles (2009-2024). 
    \end{minipage}
\end{figure}

Next, we contrast the evolution of stance-taking over the electoral cycle to that of political donations and lobbying. Panel A of Figure \ref{fig:electoral_cycle} shows a clear pattern in the (standardized) amount of political donations made by firms in each quarter between the presidential inauguration in January and the next presidential election in November, four years later: donations increase steadily before the midterm elections, drop back down to their pre-midterm level after the election, and peak again right before the presidential election. As shown in Panel B, spending on lobbying increases in the first year after inauguration, then plateaus until the midterm elections, and slowly decreases until the next presidential election: while political donations increase during electoral campaigns, firms spend the most on lobbying in the heart of the presidential cycle, when policy-making is most active, and reduce their lobbying as the current administration nears its end. By contrast, Panel C shows that the number of stances on societal issues does not follow any clear pattern over the electoral cycle, bringing additional suggestive evidence that stance-taking follows a different logic and is not motivated by firms' desire to influence the outcome of the elections or the policy work of elected legislators. 

\subsection{Stance-Taking and Business Activity}
\label{subsec:business_activity}

We now use the LLM prompt described in Section \ref{subsec:data_stances} to distinguish stances in which companies publicize current or future business practices that align with their position on a societal issue from those in which they simply state a position on a topic, e.g., by expressing support for a particular type of policy. For the latter, we further assess whether that position is justified by the company’s interests or expressed without reference to a business rationale, yielding three categories. As shown in Panel B of Figure \ref{fig:characteristics}, 24\% of stances overall, and up to half of the stances on immigration and racial issues and 75\% of the stances on voting rights, fall in the third category. These include publicly supporting or opposing a specific policy based on a certain social goal, advertising donations or pledges to social causes, or signing a petition.\footnote{Specific examples are provided in Appendix \ref{ap:gpt_prompts}.} Another 73\% of stances refer to concrete actions that the company took or is planning to take with respect to its operations (e.g., diversity hiring initiatives, wage increases, investment in infrastructure) and that align with a certain position on a societal issue. These two types of stances do not aim to directly shape policies affecting the firm's business, differently from lobbying efforts pushing for a certain regulation or campaign donations supporting a candidate or party perceived as friendlier to the firm. Only 3\% of stances cite the company's interests to express support for a certain policy  (e.g., supporting immigration reform because of workforce needs). These fractions have remained relatively stable over time (Appendix Figure \ref{fig:timeline_chars_prop}, Panel D).

Stances referring to broad societal goals could still be taken in order to protect the firm's interests when it is under scrutiny. The text analysis of the news articles and press releases citing the stances shows that some of them appear to be taken in response to a criticism or an attack, but that fraction is modest: 11\%. We obtain consistent evidence from a test estimating firms' response to NGO campaigns. Advocacy groups frequently target individual companies with public campaigns, and a common claim is that such pressure pushes firms to take public positions on contested issues. We run an event study around NGO campaigns against the companies in our sample, using data from Sigwatch.

We use one monthly observation per company per campaign $c$, so the same company-month observation can enter the sample multiple times for companies targeted by several NGO campaigns over the period. For companies targeted only once or never targeted, each company-month observation is used once. We estimate the following equation:
\begin{equation}
\label{eq:es_ngo}
Y^c_{it} \;=\; \sum_{\tau \in [-6,-2]} \beta_\tau D^\tau_{ict} +  \sum_{\tau \in [0,6]}\beta_\tau D^\tau_{ict}+ \beta_{7-} D^{7-}_{ict} + \beta_{7+} D^{7+}_{ict} + \gamma_t + \alpha_{ic} + \boldsymbol{\delta}'\!\mathit{Size}_{it} + \epsilon_{ict},
\end{equation}
where $Y^c_{it}$ is the number of stances taken by firm $i$ in month $t$; $D^\tau_{ict}$ is a dummy equal to one if company $i$ was targeted by campaign $c$ $\tau$ months before (for positive values of $\tau$) or after (for negative values) month $t$; $D^{7+}_{ict}$ and $D^{7-}_{ict}$ are dummies equal to one if the company was targeted by the campaign seven months earlier or more or seven months later or more; $\alpha_{ic}$ is a company$\times$campaign fixed effect, $\gamma_t$ is a month fixed effect, and the vector of controls $Size_{it}$ includes company headcount and revenue. The parameters of interest, $\beta_{-6}$ through $\beta_6$, capture the effect of being targeted by an NGO campaign over the months that precede or follow the beginning of the campaign, relative to the reference period of $\tau=-1$. Standard errors are clustered at the company level.

We exclude any campaign that occurs within six months of another NGO campaign against the same firm, to be able to estimate effects up to six months after the campaign, resulting in a sample of 899 NGO campaigns that were run against 357 companies between 2011 and 2023. We also run a specification focusing on campaigns related to the environment and including 915 campaigns against 341 companies.\footnote{Although environmental campaigns are a subset of all NGO campaigns, the requirement that campaigns be at least six months apart retains slightly more of them: environmental campaigns against a given firm are less densely clustered in time than NGO campaigns pooled across all topics, so a larger share survive the sample restriction.}

\begin{figure}[H]
    \centering
    \caption{Effect of NGO Firm-Targeted Campaigns on Corporate Stance-Taking}
    \begin{subfigure}[t]{0.49\linewidth}\centering
        \includegraphics[width=\linewidth]{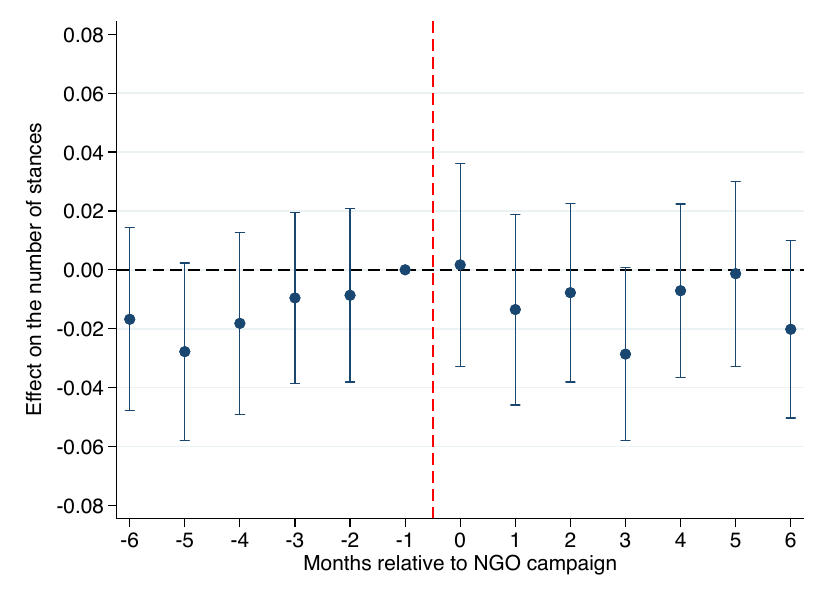}
        \caption{Any NGO campaign}
        \label{fig:sw_any}
    \end{subfigure}\hfill
    \begin{subfigure}[t]{0.49\linewidth}\centering
        \includegraphics[width=\linewidth]{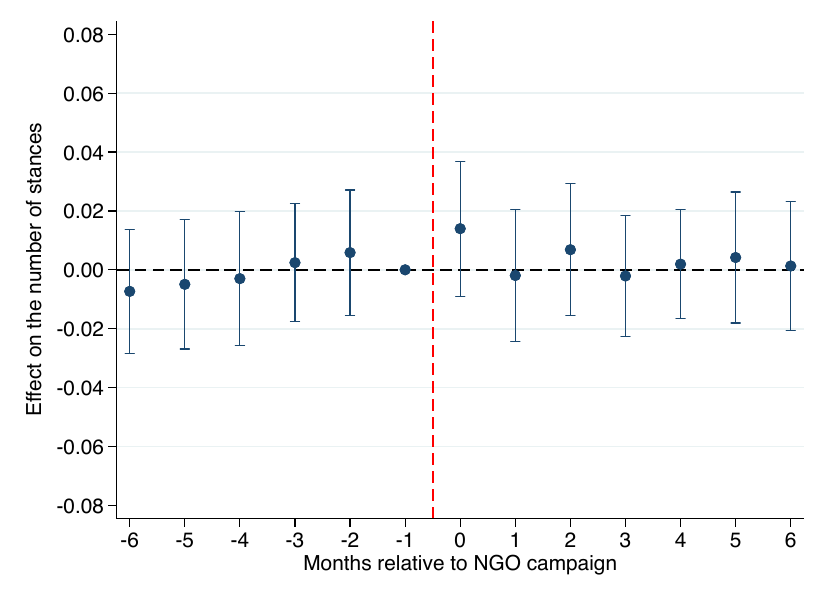}
        \caption{Environmental campaign}
        \label{fig:sw_envclimate}
    \end{subfigure}
   \begin{minipage}{\linewidth}
        \footnotesize
        \textit{Notes:} We show point estimates and 95\% confidence intervals from specifications in the form of equation~\eqref{eq:es_ngo}. We regress the monthly number of stances on all topics (Panel a) or the monthly number of liberal stances on climate change (Panel b) taken by a company on dummy variables for being targeted by an NGO campaign in the same month or for having been targeted one to six months earlier, as well as dummies for being targeted two to six months later. We also include dummies for being targeted seven months earlier or more, and for having been targeted seven months later or more. We omit the dummy for being targeted one month later. We control for company$\times$campaign and month fixed effects as well as for company headcount and revenue. Standard errors are clustered at the company level. $N = 119{,}318$ (Panel a) and $N = 125{,}743$ (Panel b) company-month-campaign observations.
    \end{minipage}
    \label{fig:ngo_eventstudy}
\end{figure}

As shown in Figure~\ref{fig:ngo_eventstudy}, none of the coefficients for the months before a campaign are significant, indicating that firms that are about to be targeted are on a similar trend as those that are not, and facilitating the interpretation of the effects observed after the campaign. Coefficients in the month of the campaign and over the six subsequent months are also not significant. This is true whether we estimate the effects of any NGO campaign on the total number of stances taken by the company (left panel) or focus on the effect of environmental campaigns on pro-climate action stances (right panel).\footnote{We find similar results using the \citet{callaway_difference--differences_2021} heterogeneity-robust estimator (Appendix Figure \ref{fig:ngo_eventstudy_cs}). Equation (\ref{eq:es_ngo}) may suffer from the potential issues of staggered-adoption two-way fixed effect models (e.g., bias arising from heterogeneous effects across companies targeted at different times). The estimator proposed by \citet{callaway_difference--differences_2021} addresses these issues.}

In sum, the vast majority of stances can be characterized neither as an effort to support a business-related policy nor as a response to business-specific attacks. We next show that their timing largely follows the broad policy debate, suggesting that they are best characterized as a contribution to that debate.

\subsection{Stance-Taking and the News Cycle} \label{subsec:news_cycle}

We test whether spikes of public and media attention on one of our nine societal issues lead companies to increase their public stance-taking on that issue. We identify such spikes using two complementary sources: the monthly number of news articles mentioning our topic keywords on the Factiva platform and the monthly number of (relative) Google searches from the Google Trends API. We propose a statistical method to identify peaks in each topic-specific time series and ask LLMs to verify that these peaks correspond to actual events related to these topics. Details are provided in Appendix \ref{ap:peak_detection}. The peak-detection algorithm identifies 48 peaks in news articles and 40 peaks in Google searches across the nine topics over our sample period. We restrict the sample to peaks whose nearest peak on the same topic is more than six months away, leaving us with 40 peaks in news articles and 38 peaks in Google searches. The final lists of peaks and the corresponding events identified by LLMs are shown in Appendix Tables \ref{tab:peaks_factiva} and \ref{tab:peaks_google}.

We estimate the impact of a peak in news articles or Google searches for a given topic on the number of stances taken by companies on that topic using a stacked event study design. 
For each peak, we construct a peak-specific sample comprising all nine topics, all companies, and the months in the $[-36,+36]$ window around the peak. The topic of the peak is the sample's focal topic while the other eight topics serve as controls. We stack these samples and estimate the following equation, where $s$ indexes stacks:

\begin{equation} \label{eq:event_study_news}
    S^s_{ikt} = \sum_{\tau \in [-6, -2]} \beta_\tau D^\tau_{kst} + \sum_{\tau \in [0, 6]} \beta_\tau D^\tau_{kst} + \beta_{7-} D^{7-}_{kst} + \beta_{7+} D^{7+}_{kst} + \alpha_{iks} + \gamma_{st} + \boldsymbol{\delta}'\!\mathit{Size}_{it} + \varepsilon_{ikst},
\end{equation}

where $S^s_{ikt}$ is the number of stances on topic $k$ taken by firm $i$ in month $t$; $D^\tau_{kst}$ is a dummy equal to one if topic $k$ is the focal topic of stack $s$ and the peak on topic $k$ occurred $\tau$ months before (for $\tau\geq0$) or after (for $\tau<0$) month $t$; $\alpha_{iks}$ is a company$\times$topic$\times$stack fixed effect which controls for each company's baseline level of stance activity on each topic; and $\gamma_{st}$ is a stack$\times$month fixed effect, which absorbs any shock common to all topics in a given month, such as an election.\footnote{In Appendix Figure \ref{fig:peaks-iso6-specB}, we replace the company$\times$topic $\times$stack and stack$\times$month fixed effects with the combination of company$\times$stack, topic$\times$stack, and month$\times$stack fixed effects. Results are unchanged.} The vector of controls $Size_{it}$ includes company headcount and revenue. As in equation (\ref{eq:es_ngo}), the parameters of interest are $\beta_{\tau}$ ($-6 \leq \tau \leq 6$) and the reference period is $\tau = -1$. Standard errors are two-way clustered, at the peak level and at the company level.\footnote{Our clustering strategy implies that all the company-month-topic observations neighboring a given peak are included in the same cluster.} 

The identification of the $\beta_\tau$ coefficients relies on comparing the focal topic, which experienced a peak in media or public attention, to the other eight topics. For this comparison to be valid, the control topics must not themselves be experiencing a peak around the same time. To ensure that the topics serving as controls are not simultaneously salient, we exclude from the control group, at the stack level, any topic that has a peak within six months of the focal peak.\footnote{The stacked design of equation (\ref{eq:event_study_news}) compares, within a stack, observations corresponding to a treated topic to unaffected control topics over the same months. This approach avoids the standard staggered-adoption two-way fixed effects bias that could arise in equation (\ref{eq:es_ngo}) \citep{dube2025local}.}

\begin{figure}[H]
\caption{Effect of Topic-Specific Peaks in News Articles and Google Searches on Corporate Stance-taking}
    \label{fig:peaks-iso6-specA}
    \centering
    \begin{subfigure}[t]{0.49\linewidth}
        \centering
        \includegraphics[width=\linewidth]{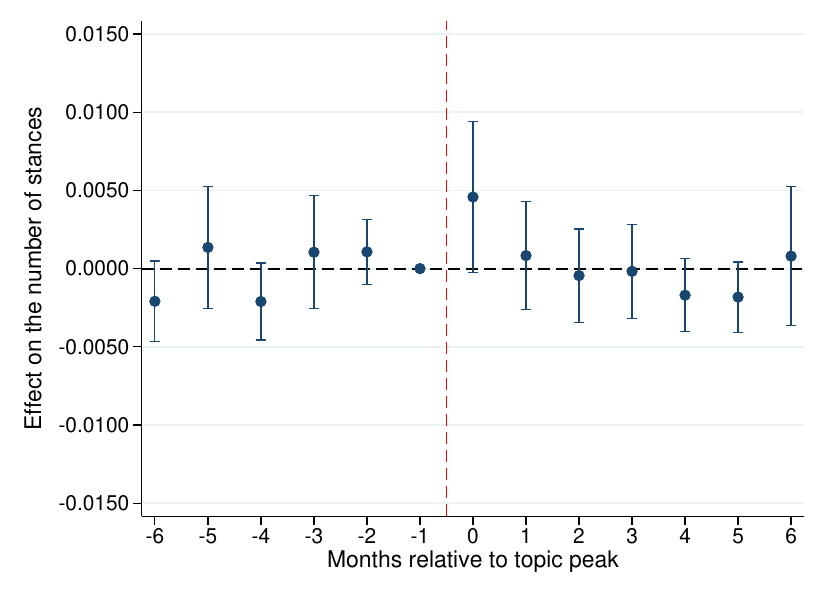}
        \caption{Factiva peaks}
    \end{subfigure}
    \hfill
    \begin{subfigure}[t]{0.49\linewidth}
        \centering
        \includegraphics[width=\linewidth]{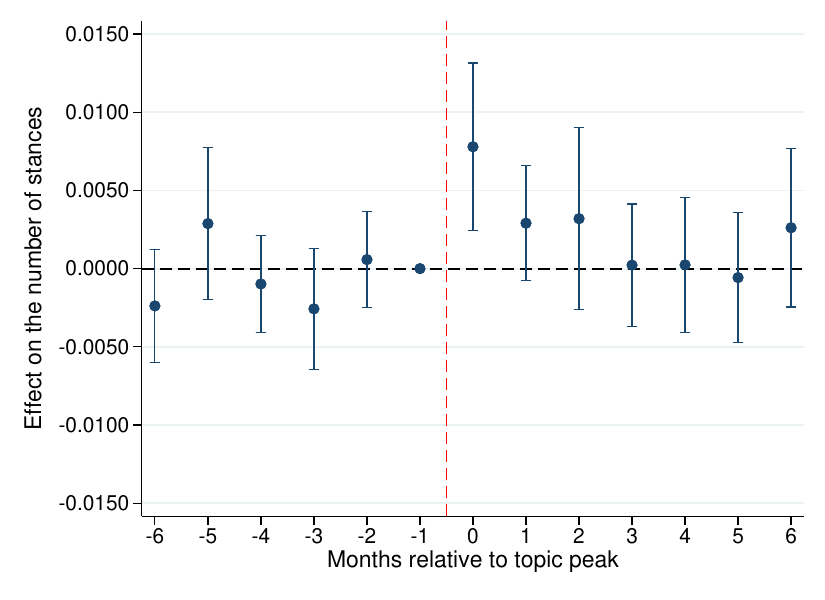}
        \caption{Google Trends peaks}
    \end{subfigure}
    \begin{minipage}{0.97\linewidth}
    \footnotesize
        \textit{Notes:} We show point estimates and 95\% confidence intervals from specifications in the form of equation (\ref{eq:event_study_news}). We regress the monthly number of stances taken by a company on a given topic on dummy variables for the observation being recorded two to six months before a peak in news articles (Panel a) or in Google searches (Panel b), as well as dummies for being recorded in the same month as a peak and one to six months after a peak. We also include dummies for being recorded seven months or more before or after a peak, respectively, and omit the dummy for being recorded one month before a peak. We control for company$\times$topic$\times$stack and stack$\times$month fixed effects, where a stack includes the focal topic on which there was a peak as well as the control topics that did not experience any peak within six months of the focal peak. We also control for company headcount and revenue. Standard errors are clustered two-way by peak and company. N = $13{,}006{,}382$ (Panel a) and $11{,}830{,}420$ (Panel b) company-month-topic-stack observations.
    \end{minipage}
\end{figure}

The results are shown in Figure~\ref{fig:peaks-iso6-specA}. We do not observe any significant pre-trend in stance-taking before either a Factiva peak (Panel a) or a Google Trends peak (Panel b). Instead, the number of corporate stances on a topic jumps in the month corresponding to the peak in news articles or Google searches on a topic, and gradually decreases afterward, indicating that companies' communication gets triggered by the same types of salient events that generate news coverage and public interest. Furthermore, rather than lagging behind, firms tend to take stances on the hot topic of the day at the exact moment in which that topic is attracting attention from the media and the general public.

\section{Stances' Voices and Audience} \label{sec:audience}

Companies' contributions to the policy debate through their stances on societal issues may reflect their leaders' preferences or strategic efforts to appeal to other stakeholders, including their employees and consumers. 

\begin{figure}[H]
    \centering
    \caption{Corporate Stance-Taking by Consumer vs.\ Business Facing Companies}
    \includegraphics[width=1\linewidth]{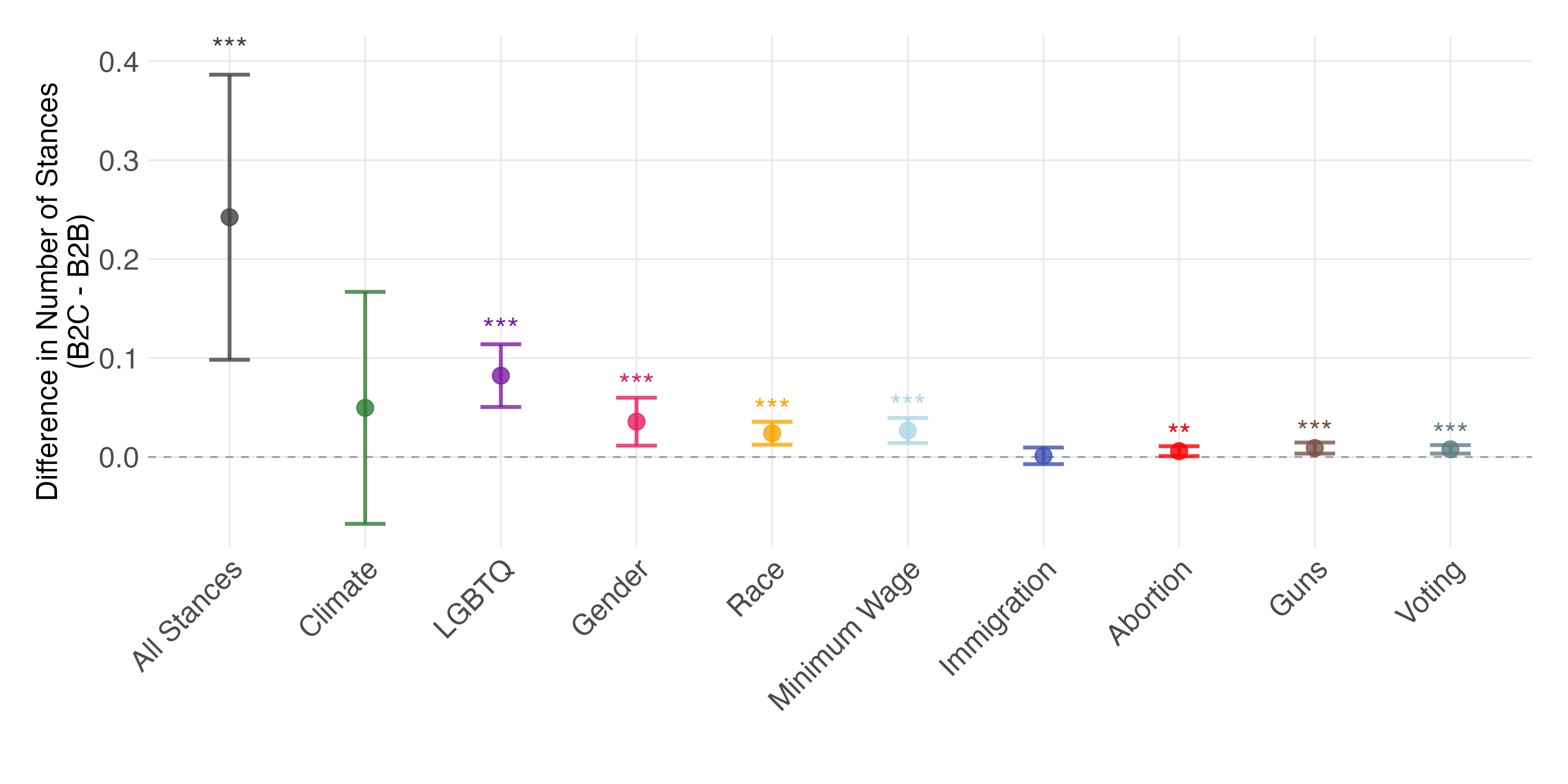}
    \label{fig:b2c}
    \begin{minipage}{\linewidth}
        \footnotesize
        \textit{Notes:} We show point estimates and 95\% confidence intervals from a regression of the yearly number of stances on a dummy variable indicating whether the company is B2C vs.\ B2B. We control for  year fixed effects as well as company headcount and revenue. We do not include industry fixed effects as they are highly correlated with the B2C vs.\ B2B classification. Standard errors are clustered at the company level. N $=$ 13,712 company-year observations. Significance: *** p < 0.01, ** p < 0.05, * p < 0.10.
    \end{minipage}
\end{figure}

As shown in Figure \ref{fig:characteristics}, only a small fraction (6\%) of all stances are taken by CEOs themselves, suggesting that the top executives of the companies are not the only voices represented in these stances. Companies may also use their stances to speak to their employees or their consumers. Indeed, we observe that the companies that take the most stances are firms that are well-known to the public (e.g., Google, Amazon, and Walmart) or that provide essential utilities to a large number of consumers (e.g., Exelon and PG\&E). Figure \ref{fig:b2c} provides further evidence on the possible role of consumers: controlling for company size and year fixed effects, B2C companies take 0.24 more stances per year on average than B2B companies, which is significant at the 1\% level and corresponds to 40\% of the overall average number of yearly stances. The difference between the number of stances taken by B2C and B2B companies is positive for all topics except climate and immigration (for which the difference is not statistically significant), and it is particularly pronounced for minimum wage and DEI issues (gender, race, and LGBTQ).

To determine more systematically which stakeholders companies represent or target when they take a stance on societal issues, we study the relationship between stance-taking and the political leaning of a firm's consumers, employees, executives, and board members. We estimate the following equation:
\begin{equation} \label{eq:multivariate} S_{it}=\gamma_t+\lambda^I_i
       +\beta_1 DemEmp_{it}+\beta_2 DemEx_{it}+\beta_3 DemDir_{it}+\beta_4 DemCons_{i}
       +\boldsymbol{\delta}'\!\mathit{Size}_{it}+\epsilon_{it},
\end{equation}
where $S_{it}$ is the number of liberal stances taken by firm $i$ in year $t$; $DemEmp_{it}$ is the share of employees registered as Democrats on the voter list (out of the total number of employees found in the voter list and registered as either Democrat or Republican); $DemEx_{it}$ and $DemDir_{it}$ are the shares of executives and directors registered as Democrats, respectively; and $DemCons_{i}$ is a proxy for the share of Democratic consumers, defined as the ratio between individuals following both the company and the Democratic Party account on the platform Twitter/X, and individuals following the company and exclusively following either the Democratic or the Republican party, as in \citet{schoenmueller2023frontiers}. The vector of controls $Size_{it}$ includes company headcount and revenue.\footnote{Each of the four explanatory variables and each control is mean-imputed when missing; we include a corresponding set of indicator variables flagging the imputed observations, so identification comes from non-imputed variation. As in equation (\ref{eq:pol_activity}), $\gamma_t$ is a year fixed effect and $\lambda^I_i$ an industry fixed effect.} Standard errors are clustered at the company level.

\begin{figure}[H]
    \centering
    \caption{Stakeholders' Political Preferences and Liberal Corporate Stance-Taking}
    \includegraphics[width=\linewidth]{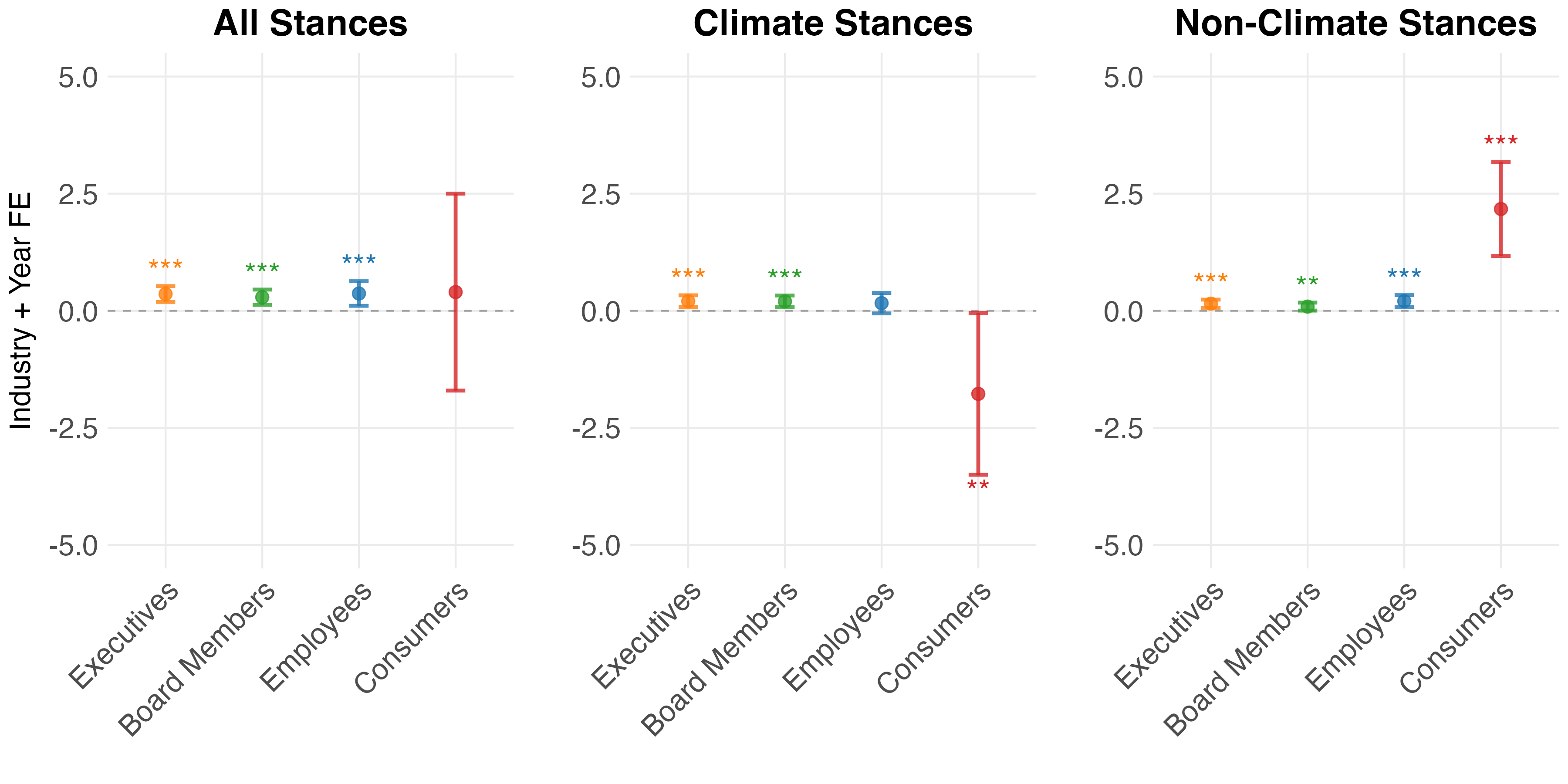}
    \label{fig:mv_stakeholders_politics_main}
    \begin{minipage}{\linewidth}
        \footnotesize
        \textit{Notes:} We show point estimates and 95\% confidence intervals from specifications in the form of equation \ref{eq:multivariate}. We regress the yearly number of liberal stances on the yearly shares of Democratic employees, executives, and board members, as well as the share of Democratic consumers, which does not vary over time. We control for industry and year fixed effects as well as company headcount and revenue. Standard errors are clustered at the company level. N= 13,748 company-year observations.  Significance: $^{***}p<0.01$, $^{**}p<0.05$, $^{*}p<0.10$.

    \end{minipage}
\end{figure}

The left panel of Figure \ref{fig:mv_stakeholders_politics_main} shows a positive relationship between the number of liberal stances taken by a company and the shares of Democrats among its different stakeholders, which is statistically significant for all groups except for consumers. When focusing on liberal stances on climate (middle panel), the relationship remains positive and significant for executives and board members, but it is not significant for employees and even turns negative (and significant at the 5\% level) for consumers. When excluding stances on climate, increases in the shares of Democrats among all types of stakeholders are associated with significantly larger numbers of stances (right panel). The effect is particularly large for consumers: a 10 percentage point increase in the estimated share of Democratic consumers raises the number of liberal stances on topics other than climate by 0.22 (or 86\% of the mean). The estimates for the three other groups are smaller, ranging from a 0.009 increase in the number of liberal stances for a 10 percentage point increase in the share of Democratic board members to a 0.021 increase for a 10 percentage point increase in the share of Democratic employees. All estimates are significant at the 5\% or 1\% level. Measuring the partisanship of both top executives and board members using their political donations as opposed to their party of registration yields similar estimates (Appendix Figure \ref{fig:mv_stakeholders_donations}). These results suggest that the political preferences of both the individuals who are part of the company (from employees to top decision-makers) and its consumers influence stance-taking.

We now exploit within-company changes in the shares of Democratic employees, top executives, and board members to move beyond the correlations shown in Figure \ref{fig:mv_stakeholders_politics_main} toward a causal interpretation. We cannot do this exercise for consumers since our measure of partisanship among consumers does not vary over time. The top panel of Appendix Figure \ref{fig:mv_stakeholders_politics} first verifies that the previous results are robust to excluding the share of Democratic consumers in equation (\ref{eq:multivariate}). The bottom panel estimates a version of equation (\ref{eq:multivariate}) that replaces industry fixed effects with company fixed effects such that all coefficients are exclusively estimated out of variation in Democratic shares over time. Most estimates drop in size and lose their significance, suggesting that the companies that take more liberal stances attract liberal-leaning employees and executives, without the latter necessarily causing the former. However, the coefficients remain positive and significant for board members: having more Democrats among them leads the company to claim liberal positions on societal issues (particularly climate) more often. 

We conduct similar exercises to test whether the shares of women and Black individuals among employees, top executives, and board members affect the company's propensity to take liberal stances on issues related to gender and abortion, which are likely to matter most to women, and to racial discrimination and voting laws, which may matter most to Black individuals. We do not find any significant relationship between a workforce's racial composition and the company's stances on race or voting laws, once we control for company fixed effects. Instead, the shares of female executives and board members both significantly increase the number of stances on gender-related issues, and the share of female executives significantly raises a company's propensity to take stances on abortion (bottom panel of Appendix Figure \ref{fig:mv_stakeholders}).

\paragraph{Event study around changes in leaders' partisanship}
\label{sec:mtf_event_study}

We finally use event studies to estimate the impact of having an additional Democrat sitting on the board or serving among the top executives. A company experiences an event in year $t$ if a new individual joins the board (or becomes a top executive), if that individual is registered as a Democrat, and if their arrival increases the total number of Democrats among board members (or top executives) relative to year $t-1$. Since party affiliation is observed only for individuals matched to the voter records (23\% of new directors and 36\% of new executives), we only count as events those instances in which the party of the newcomer and the party of possible leavers are observed, so that we can be certain that the number of Democrats increased. This leaves us with a sample of 50 events for board members and 65 events for top executives.

For each year between 2012 and 2021 (i.e., each year of our sample excluding the first and last three years), we construct a cohort $c$ comprising the treated companies that experienced an event in that year, except for those that experienced another event within the three preceding or following years.\footnote{The event study designs of equations (\ref{eq:es_ngo}) and (\ref{eq:event_study_news}) excluded treated observations that experienced another event within the six preceding or following months. Since our data on executives and board members varies by year, not by month, we limit the exclusion window to $-3$ and $+3$ years to avoid dropping too many observations.} The cohort also includes control companies that experienced an event neither in year $c$ nor in any of the three preceding or three following years.\footnote{For executives, we keep as controls only companies for which, in every year of the window, the voter records of new and departing executives rule out a Democratic arrival. Because board newcomers are rarely matched to voter records, this restriction would leave too few control companies if applied to board members. In that case, controls are companies with no observed Democratic arrival in the window, which may include unobserved Democratic arrivals and bias the estimates toward zero.} We require that both treated and control companies have their board (or executive team) observed in every year of the $[-3,+3]$ window. We then estimate the following equation:

\begin{equation} \label{eq:event_study_boardex}
    S^c_{it} = \sum_{\tau \in [-3, -2]} \beta_\tau D^\tau_{ict} + \sum_{\tau \in [0, 3]} \beta_\tau D^\tau_{ict} + \beta_{4-} D^{4-}_{ict} + \beta_{4+} D^{4+}_{ict} + \alpha_{ic} + \gamma_{ct} + \boldsymbol{\delta}'\!\mathit{Size}_{it} + \varepsilon_{ict},
\end{equation}

where $S^c_{it}$ is the number of liberal stances taken by company $i$ in year $t$, $D^\tau_{ict}$ is a dummy equal to one if company $i$ is treated in cohort $c$ and year $t$ is $\tau$ years before or after the event, $\alpha_{ic}$ is a company$\times$cohort fixed effect, and $\gamma_{ct}$ is a cohort$\times$year fixed effect. The vector of controls $Size_{it}$ includes company headcount and revenue. The parameters of interest are $\beta_{-3}$ through $\beta_3$ and the reference period is $\tau=-1$. We also include dummies equal to one for treated companies four or more years before the event, or four or more years after the event. Standard errors are clustered at the company level.

\begin{figure}[H]
    \centering
    \caption{Partisan Changes in Leadership and Liberal Stances}
    \label{fig:es_part}
    \begin{subfigure}[t]{0.49\linewidth}\centering
        \includegraphics[width=\linewidth]{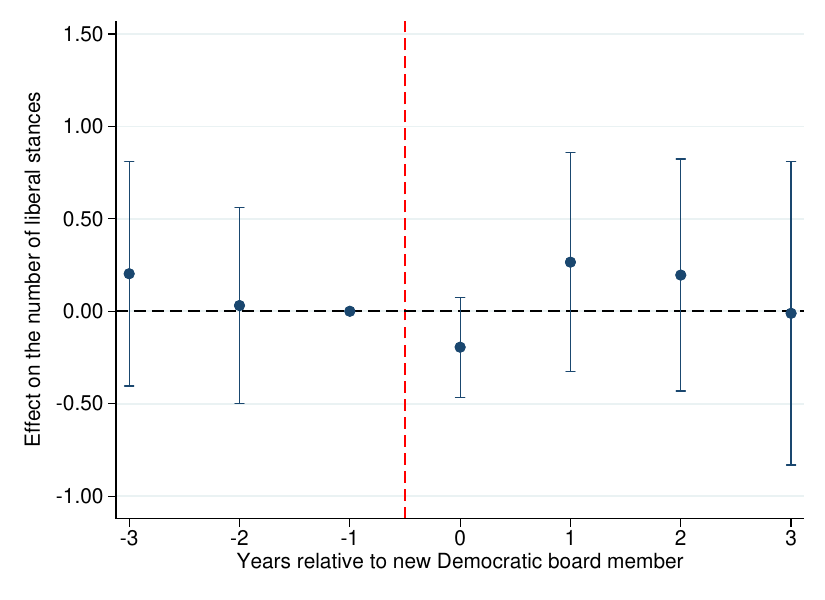}
        \caption{Board members: all liberal stances}
    \end{subfigure}\hfill
    \begin{subfigure}[t]{0.49\linewidth}\centering
        \includegraphics[width=\linewidth]{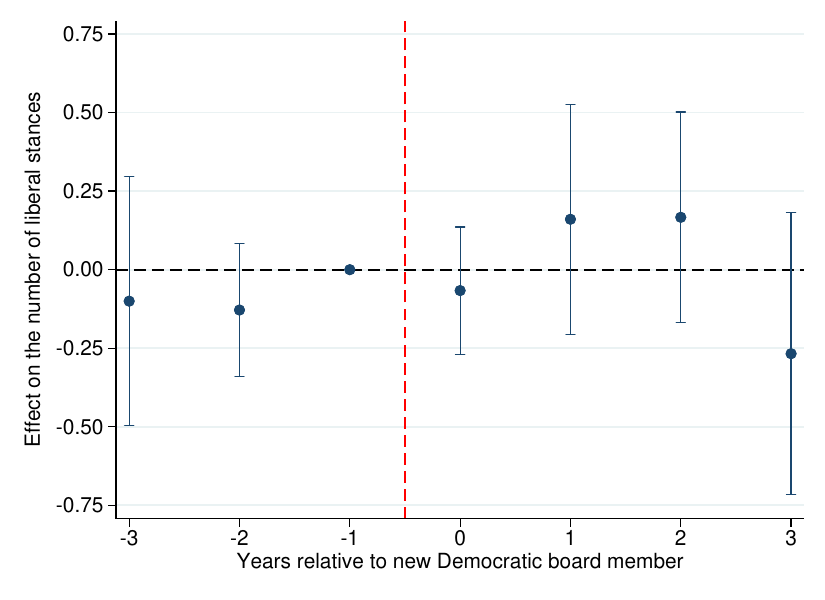}
        \caption{Board members: liberal stances excluding climate}
    \end{subfigure}
    \par\medskip
    \begin{subfigure}[t]{0.49\linewidth}\centering
        \includegraphics[width=\linewidth]{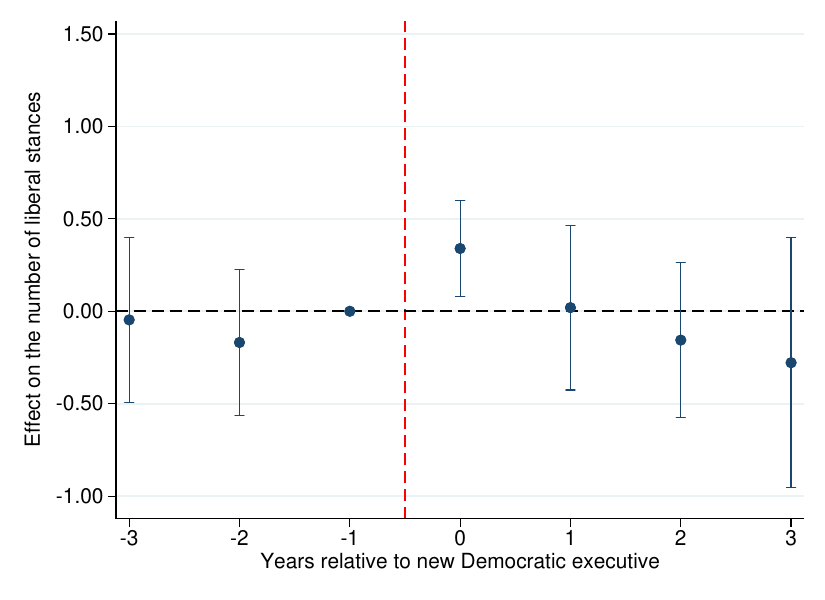}
        \caption{Executives: all liberal stances}
    \end{subfigure}\hfill
    \begin{subfigure}[t]{0.49\linewidth}\centering
        \includegraphics[width=\linewidth]{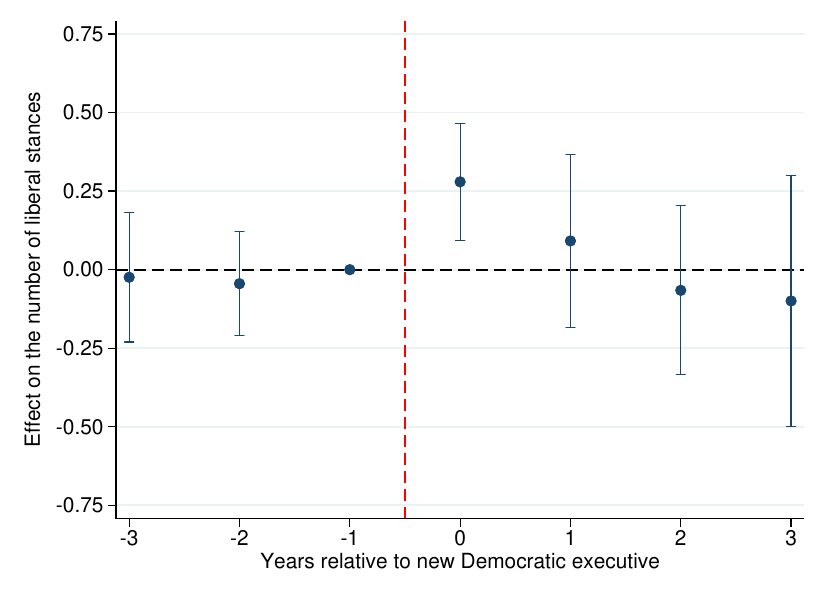}
        \caption{Executives: liberal stances excluding climate}
    \end{subfigure}
    \begin{minipage}{\linewidth}
        \footnotesize
        \textit{Notes:} We show point estimates and 95\% confidence intervals from specifications in the form of equation~\eqref{eq:event_study_boardex}. The outcome is the yearly number of liberal stances (Panels a and c) and the yearly number of liberal stances on topics other than climate (Panels b and d). An event is the arrival of a new board member (Panels a and b) or of a new executive (Panels c and d) registered as a Democrat that increases the number of Democrats on the board or in the executive team. We include dummies for experiencing an event in the same year and for having experienced an event one to three years earlier, as well as dummies for experiencing an event two or three years later. We also include dummies for experiencing an event four years later or more, and for having experienced an event four years earlier or more. We omit the dummy for experiencing an event one year later. We control for company$\times$cohort and cohort$\times$year fixed effects, where a cohort includes all companies experiencing an event in a given year and no other event in the three preceding and following years as well as companies that did not experience any event over that same window. We also control for company headcount and revenue. Standard errors are clustered at the company level. $N = 54{,}194$ (board members) and $5{,}114$ (executives) company-year-cohort observations.
    \end{minipage}
\end{figure}

The coefficients corresponding to two and three years before the addition of a Democrat to the board (Figure \ref{fig:es_part}, Panels a and b) or to the set of top executives (Panels c and d) are not significant, indicating that the trend in stance-taking among firms close to experiencing such events is similar to the trend among firms that are not. The coefficients for the years following the addition of a Democrat to the board are close to zero and not significant either, but they are very imprecisely estimated, preventing us from drawing any definitive conclusions.\footnote{For instance, considering the upper bound of the 95\% confidence interval, we cannot reject that the addition of a Democrat board member increases the number of liberal stances taken by the company in the following year by up to 0.86, or 119\% of the mean.} However, the effects on stances of the addition of a Democrat executive in the year of that addition is positive and significant at the 5\% level. Having one more Democrat leader pushes the company to take 0.34 more liberal stances (Panel c), an increase of 47\% relative to the average yearly number of liberal stances. The effect is short lived: the coefficients one to three years after the event are closer to zero and non-significant. We obtain similar results when we focus on liberal stances unrelated to climate change (Panel d).

Using the same event-study design, we find no significant effect of adding a woman to the board or to the set of top executives on the number of stances taken on either gender-related or abortion-related issues (Appendix Figure~\ref{fig:es_arr_gender}), although the first effect is positive. We also estimate the impact of adding a Black board member on the number of stances on race and voting laws (Appendix Figure~\ref{fig:es_arr_race}).\footnote{ExecuComp does not report the ethnicity of top executives, and the number of Black executives identified from other sources is too small for an event study.} Once again, these coefficients are imprecisely estimated, so we are unable to draw any definitive conclusion.

Overall, we conclude that corporate stances correlate with several stakeholders’ political preferences. The association is strongest for consumers on issues other than climate, whereas climate stances align more closely with executives and board members. Evidence from within-company changes and leadership transitions indicates that individual leaders influence stance-taking, while the alignment between corporate stances and employees’ partisanship may partly reflect political sorting, with employees selecting into companies whose stances align with their views, rather than companies adjusting their communication to employees’ preferences.

\section{Conclusion}

We construct a dataset of 9,917 corporate stances on nine divisive societal issues among 836 firms that belonged to the S\&P 500 between 2008 and 2025. Our approach uses large language models to identify and characterize stances in approximately 52,000 news articles and press releases. This methodology provides systematic coverage across firms and issues, thus making a dispersed and unstructured form of corporate political activity amenable to large-scale empirical analysis. We find that stance-taking is overwhelmingly liberal, increased sharply before peaking in 2021, and subsequently declined. It is only weakly related to lobbying and campaign contributions, and follows the news cycle rather than the electoral cycle. These patterns suggest that corporate stance-taking constitutes a distinct form of participation in public debate rather than a non-market strategy aimed at shaping policies and regulations favorable to the company's business activity. A particularly striking finding is that nearly a quarter of stances express positions or support social causes without referring to the firm's business practices or a business rationale.

This raises questions about the scope of the firm’s mission. The absence of a stated business justification does not establish that a stance provides no benefit to the company's shareholders. Yet when firms intervene in political debates without an evident connection to their activities, the justification for doing so is less clear. Invoking the interests of society at large leaves open whose interests should guide corporate speech, whose preferences firms represent, and to whom they are accountable. 

Our findings indicate that corporate stances align with the political preferences of different stakeholders: executives, board members, employees, and consumers. Further research will be needed to assess the mechanisms behind this alignment and determine which stakeholders matter the most. While leaders may be able to shape stance-taking according to their own views, employees and consumers may be more likely to sort toward firms whose stances already match their views. Firms may strategically choose positions that appeal to these groups, either to increase consumer demand or to recruit and retain employees. We hope our data will help advance the growing literature on these questions.

\pagebreak \newpage
\bibliographystyle{apsr}
\bibliography{references}
\clearpage 

\appendix

\section*{\LARGE{Online Appendix}}
\section*{Table of Contents}
\begin{itemize}
\item Appendix \ref{ap:data}: Additional Details on the Data
\item Appendix \ref{ap:figures}: Additional Figures and Tables
\end{itemize}

\setcounter{figure}{0}
\renewcommand{\thefigure}{A\arabic{figure}}
\setcounter{table}{0}
\renewcommand{\thetable}{A\arabic{table}}
\numberwithin{figure}{section}
\numberwithin{table}{section}

\section{Additional Details on the Data} \label{ap:data}

\subsection{Topic Selection} \label{ap:topic_selection}

The party platforms outline each party's principles, goals, and positions on domestic and foreign affairs, typically announced during the national convention when the presidential nominee is confirmed. These documents articulate the issues the parties prioritize and intend to focus on during the presidential election. We manually reviewed and compared the platforms, identifying the most prominent topics. We identified various terms used over time to refer to the same topic and developed a set of keywords likely to appear in news articles that discuss each of them. The selected topics and associated concepts are detailed in Table \ref{tab:topics}.

\begin{table}[H]
\caption{Topics and Keywords}\label{tab:topics}
\begin{center}
\scalebox{.9}{
\begin{tabular}{ccl}
 & \textbf{Topic} & \textbf{Keywords} \\\hline

\multirow{4}{*}{\textbf{1}} & \multirow{4}{*}{\textbf{Abortion}} 
 & Abortion \\
 &  & Planned Parenthood \\
 &  & Reproductive Health \\
 &  & Roe v. Wade \\\hline

\textbf{2} & \textbf{Climate Change} & Climate Change \\\hline

\multirow{6}{*}{\textbf{3}} & \multirow{6}{*}{\textbf{Gender}} 
 & Equal Pay \\
 &  & Gender Discrimination \\
 &  & Gender Equality \\
 &  & Gender Inequality \\
 &  & Paycheck Fairness Act \\
 &  & Women's Rights \\\hline

\multirow{6}{*}{\textbf{4}} & \multirow{6}{*}{\textbf{Guns}} 
 & Firearms \\
 &  & Gun Ban \\
 &  & Gun Control \\
 &  & Gun Ownership \\
 &  & Gun Prevention \\
 &  & Gun Violence \\\hline

\multirow{8}{*}{\textbf{5}} & \multirow{8}{*}{\textbf{Immigration}} 
 & Border Security \\
 &  & DACA \\
 &  & Deferred Action for Childhood Arrivals \\
 &  & Illegal Immigration \\
 &  & Illegal Immigrants \\
 &  & Immigration Reform \\
 &  & Immigration System \\
 &  & Undocumented Immigrants \\\hline

\multirow{10}{*}{\textbf{6}} & \multirow{10}{*}{\textbf{LGBTQ}} 
 & Defense of Marriage Act \\
 &  & Equality Act \\
 &  & LGBT* \\
 &  & Marriage Equality \\
 &  & Religious Freedom \\
 &  & Religious Liberty \\
 &  & Respect for Marriage Act \\
 &  & Same-Sex Marriage \\
 &  & Sexual Orientation \\\hline

\textbf{7} & \textbf{Minimum Wage} & Minimum Wage \\\hline

\multirow{3}{*}{\textbf{8}} & \multirow{3}{*}{\textbf{Race}} 
 & Black Lives Matter \\
 &  & Racial Discrimination \\
 &  & Racial Justice \\\hline

\multirow{5}{*}{\textbf{9}} & \multirow{5}{*}{\textbf{Voting}} 
 & Right to Vote \\
 &  & Voter ID Laws \\
 &  & Voter Identification Laws \\
 &  & Voting Laws \\
 &  & Voting Rights \\\hline

\end{tabular}}
\end{center}
\end{table}

\subsection{Sample of Firms} \label{ap:sample_firms}

Our target sample consists of every company that was included in the S\&P 500 index at any point between January 2008 and June 2025. Because index
membership turns over, this universe is considerably larger than the 500 companies included in the index on any single date. Our initial list contains 867
distinct stock tickers that belonged to the index at some point during the sample period. Four companies appear under two share classes---Fox
(FOX/FOXA), Alphabet (GOOG/GOOGL), News Corp (NWS/NWSA), and Under Armour
(UA/UAA)---and we keep a single class per company to avoid double-counting.

We collect Factiva articles by searching for each corporate entity using the platform's “company search” feature. We provide a company ticker (e.g., GOOGL) and Factiva returns all articles in which the company Alphabet (or one of its subsidiaries like Google) is automatically recognized.\footnote{We provided further details on this search feature in Appendix
\ref{ap:factiva}.} Corporate structural changes complicate this step because, at the time of data collection, a company may no longer have existed as a searchable entity on Factiva. We first search for each company under its original ticker and, when that ticker is not listed, we identify the appropriate successor entity using the following
rules:
\begin{itemize}
    \item \textbf{Merger:} we search for the entity created by the merger.
    (E.g., Harris Corporation is not listed; in 2019 it merged with L3
    Technologies to form L3Harris Technologies, whose articles we record on behalf
    of Harris Corporation.)
    \item \textbf{Acquisition:} we search for the acquiring (parent) company.
    (E.g., UST Inc.\ is not listed; it was acquired by Altria Group in 2009.)
    \item \textbf{Split:} we combine the articles of all resulting companies.
    (E.g., L Brands is not listed; it split in 2021 into Bath \& Body Works
    and Victoria's Secret \& Co., whose articles we combine.)
    \item \textbf{Renaming:} we search for the new name. (E.g., Dayton Hudson
    Corporation is not listed; it was renamed Target Corporation in 2000.)
    \item \textbf{Bankruptcy:} we search for the reorganized or successor
    entity. (E.g., Alpha Natural Resources is not listed; after its 2015
    bankruptcy its assets passed to Alpha Natural Resources Holdings.)
\end{itemize}
When a company underwent multiple changes, we use the earliest change that yields a searchable entity on Factiva. In total, 83 of the 867 companies
(9.6\%) could not be found under their original ticker and required one of these substitutions, most commonly a renaming or an acquisition.

When constructing the company--month panel, we retain each company for the full study period over which it existed, regardless of whether it was a member of the S\&P 500 index in any given month. A company's
series runs from January 2008 (or its founding date, if later) until the end of the sample period, or until the company ceased to exist as an
independent entity. For mergers, acquisitions, splits, and bankruptcies we retain observations up to the event date and discontinue the
series thereafter. If the successor or acquiring entity is itself an S\&P 500 constituent, it enters the sample as a separate company. 

For example, we collect United Technologies (UTX) articles only through its April 2020 merger with Raytheon, Bear Stearns (BSC) articles only through its acquisition by JPMorgan Chase in May 2008, and Lehman Brothers (LEHMQ) articles only through its September 2008 bankruptcy filing. For renamings, we treat the company as a single continuous entity and drop the obsolete ticker to avoid duplicates---for instance, Facebook (FB) and Meta Platforms are a single company throughout the period. This procedure yields the final sample of 836 companies described in Section \ref{subsec:data_stances}.

\subsection{Factiva Search and Article Processing} \label{ap:factiva}

We used the Factiva platform to manually download news articles referring both to the companies in our sample and to one of our nine topics of interest. We report below the protocol followed to perform this task:

\begin{itemize}
    \item Enter date range: 01/01/2008 to 06/01/2025.
    \item Filter source, selecting: "Associated Press Newswires" Or "Reuters News" Or "Agence France Presse" Or "GlobeNewswire (U.S.)" Or "PR Newswire (U.S.)" Or "Dow Jones Institutional News" Or "Business Wire".
    \item Deselect subject Sport.
    \item Select region North America, click $+$, select United States.
    \item Select language English.
    \item Type in the company's ticker symbol (the unique letter code used by stock exchanges to identify a company) and select the company name from the drop-down list.
    \item Enter the topic concepts in free text search between quotes, separated by 'or' (e.g., "reproductive health" \textbf{or} "planned parenthood" \textbf{or} "abortion" \textbf{or} "roe v wade"; "gun ownership" \textbf{or} "gun control" \textbf{or} "gun ban" \textbf{or} "gun prevention" \textbf{or} "gun violence" \textbf{or} "firearms").
    \item Stick to the duplicates option default "Similar".
    \item Search
\end{itemize}

For each search that returns results:
\begin{itemize}
    \item Select all headlines.
    \item Click RTF and download the searches selecting the option "Article Format".
    \item Transform the .rtf file into a .txt and save it into the "factiva\_output" folder naming the file "Topic\_Ticker.txt", where the Topic standard name should match the ones under the Topic column in Table \ref{tab:topics}, and Ticker is the company code.
    \item If there are more than 100 results, download 100 at a time and add a number for each search to the file name ("Topic\_Ticker1.txt", "Topic\_Ticker2.txt").
\end{itemize}  

If the search does not return any results, go to the next one.

\paragraph{Company Search} This pipeline uses Factiva’s “company search” feature to select articles mentioning a given company, which means that we rely on Factiva's own entity recognition algorithm to identify companies in the text of the articles. The main advantage of this approach is that it ensures all selected articles mention the company we are looking for. It also guarantees that the companies mentioned are linked to the right parent company (e.g., an article mentioning Google will be correctly identified as referring to the ticker GOOGL of the parent company Alphabet). A possible downside is that, if Factiva's algorithm misses some mentions of a company, we may miss articles that would have been relevant to our search. We verified this by running an alternative search, using the company name as a keyword instead of using the automatic company search functionality. This approach yields all articles in which the keyword "company name" appears in the text. An important challenge is that, for companies whose name refers to other objects (e.g., Apple is both the name of a company and a fruit), the keyword search is likely to yield a lot of articles that are not referring to the company at all. Alternatively, for parent companies whose name is less likely to appear in a news article (e.g., an article is more likely to refer to Google than Alphabet), the keyword search is likely to miss articles. Therefore, we performed the keyword search for three companies chosen because their name does not refer to anything else and because they are not the parent company of a more famous subsidiary: Airbnb, Lululemon, and Netflix.

We then compared the articles obtained, for these three companies and for each of our nine topics, using the automated company search vs.\ the company name as a keyword. The latter yields many more articles: 5,405 articles in total (across the nine topics) against 375 with the automated company search. Many of these articles are not relevant to us, since they may mention the company name and our topic keywords without referring to the company taking a stance on that topic. To determine if we are missing truly important articles when using the automated search, we downloaded all these articles and ran them through our GPT pipeline (see Section \ref{ap:gpt_prompts}). We keep only the articles identified as instances of corporate sociopolitical activism. Among those, there were 37 articles found by both searches, 18 articles that were found with the automated company search but not when using the company name as a keyword, and 50 articles that were found when using the company name as a keyword but not when using the automated keyword search. Those represent potential articles of interest that we miss with our preferred automated approach. We read each of these 50 articles to determine whether they should have been included in our dataset or not. The key takeaway from this analysis is that these missed articles tend to refer to the company of interest as one of many. For example, an article might discuss how multiple tech companies, including Netflix, voiced opposition to the Trump administration’s decision to end DACA. 

In conclusion, a limitation of Factiva's automated company search feature is that, when searching for articles referring to a given company, the algorithm may miss articles that refer to that one company among many others. We quantify the magnitude of this issue in Section \ref{ap:validation}. Searching articles with the company name as a keyword may enable us to capture these multi-company articles. However, as explained above, this alternative approach also misses some of the relevant articles that we found with the automated company search. It yields a much larger initial number of articles, which dramatically increases the downloading and processing costs of our pipeline. For three companies whose names are relatively unique, using the company name as keyword yields about 14 times more articles than the automated company search. For a company whose name refers to other things, like Target, using the company name as keyword yields 134 times more articles than the automated company search (65,145 against 486). Therefore, we chose to keep using the automated company search in our protocol.

\paragraph{Clustering} Using the full text of each article and the pre-trained \texttt{all-MiniLM-L6-v2} Sentence-BERT (SBERT) model (used off-the-shelf,
without fine-tuning),\footnote{\url{https://sbert.net/examples/applications/clustering/README.html\#fast-clustering}} we identify clusters of articles that exhibit semantic similarity, applying a cosine-similarity threshold of 0.7. Initially, there are 22,147 clusters of articles. After forming these clusters with the transformer model, we refine them to ensure consistency in terms of topic, company, and time frame. Specifically, we ensure that all articles within a cluster were found using the same topic keywords and the same company ticker, and that they were published within a one-year period. With these additional restrictions, we end up with 51,978 clusters of articles. The number of clusters increases at this step because the embedding-based
clustering groups articles by textual similarity alone. About a third of the
initial clusters contain articles corresponding to different company searches and
roughly one in seven contain articles corresponding to different sets of topic keywords. Requiring consistency in
company, topic, and time frame splits these into separate clusters.

To implement the one-year period temporal restriction, we use the publication date of the earliest article in each cluster as a reference point. We include all subsequent articles that were published within one year of this reference date. If an article's date falls outside this one-year window, we initiate a new cluster, repeating the process with this article as the new reference point. Finally, we keep one article per cluster, corresponding to the oldest one.

\subsection{GPT Prompts} \label{ap:gpt_prompts}

\paragraph{First classification} We used OpenAI's model GPT to assess whether the article refers to Corporate Socio-Political Activism (CSA), following the definition proposed by \citet{bhagwat2020corporate}: "A firm's public demonstration (statements or actions) of support or opposition to one side of a partisan sociopolitical issue."

For each article in our Factiva database, we ran the following prompt:

    \textbf{I will input a news article. Quietly determine whether the article describes an instance in which the company \texttt{company} or its CEO publicly demonstrated, through statements or actions, support or opposition to the issue of \texttt{topic}. Respond 'Yes' or 'No', without further comments.}

To fine-tune the prompt and assess its validity, we compared GPT's answers to human answers. More precisely, two of us separately labeled the 545 articles resulting from the Factiva search for the companies Disney (DIS), Verizon (VZ) and Johnson\&Johnson (JNJ), and for all topics. After deliberation, we reached a consensus on 45 articles identified as CSA. We then tried four different versions of the prompt:

\begin{itemize}
        \item \textbf{v1: short prompt with reasoning} \\
        I will input a news article. Determine whether the article describes an instance in which the company \textit{Disney} or its CEO publicly demonstrated, through statements or actions, support or opposition to the issue of \textit{Abortion}. Respond 'Yes' or 'No', and provide your reasoning in one or two sentences.
        \item \textbf{v2: short prompt without reasoning}
        \item \textbf{v3: 3-part-prompt with reasoning} \\
        We define Corporate Socio-Political Activism (CSA) as: "A firm's public demonstration (statements or actions) of support or opposition to one side of a partisan sociopolitical issue."\\
        I will input a news article. Determine whether:
        (1) It relates to CSA as defined above.
        (2) The company \textit{Disney} or its CEO are one of the main actors in the article.
        (3) The partisan sociopolitical issue discussed is \textit{Abortion}.
        Provide three ‘Yes’ or ‘No’ answers separated by pipes, and one or two sentences explaining your reasoning. For example: No|Yes|Yes|Reasoning.

        \item \textbf{v4: 3-part-prompt without reasoning}
    \end{itemize}

The confusion matrices and the performance statistics of each prompt against our manual labeling are displayed in Tables \ref{tab:confusion} and \ref{tab:performance}.

\begin{table}[H]
\caption{Confusion matrices for 4 prompts using GPT 3.5-0613}\label{tab:confusion}
\begin{tabular}{llcccc}
\textbf{} & \textbf{} & \multicolumn{2}{c}{With Reasoning} & \multicolumn{2}{c}{Without Reasoning} \\ \cline{3-6} 
\textbf{} &  & \multicolumn{1}{l}{Predicted No} & \multicolumn{1}{l}{Predicted Yes} & \multicolumn{1}{l}{Predicted No} & \multicolumn{1}{l}{Predicted Yes} \\ \cline{3-6} 
\multirow{3}{*}{Short-prompt} &  & \multicolumn{2}{c}{v1} & \multicolumn{2}{c}{v2} \\ \cline{3-6} 
 & Actual No & 448 & 52 & 446 & 54 \\
 & Actual Yes & 6 & 39 & 4 & 41 \\ \cline{3-6} 
\multirow{3}{*}{3-part-prompt} &  & \multicolumn{2}{c}{v3} & \multicolumn{2}{c}{v4} \\ \cline{3-6} 
 & Actual No & 484 & 16 & 481 & 19 \\
 & Actual Yes & 10 & 35 & 10 & 35
\end{tabular}
\end{table}

 \begin{table}[H]
 \caption{Performance metrics} \label{tab:performance}
 \begin{center}
    \begin{tabular}{lcccc}
         & \multicolumn{1}{c}{v1} & \multicolumn{1}{c}{v2} & \multicolumn{1}{c}{v3} & \multicolumn{1}{c}{v4} \\
        Accuracy & 0.893 & 0.893 & 0.952 & 0.946 \\
        Precision & 0.428 & 0.431 & 0.686 & 0.648 \\
        Recall & 0.866 & 0.911 & 0.777 & 0.777 \\
        Specificity & 0.896 & 0.892 & 0.968 & 0.962 \\
        F1 Score & 0.573 & 0.585 & 0.729 & 0.707
    \end{tabular}
    \end{center}
    \footnotesize
        \textit{Notes:} Accuracy measures the overall correctness of the model, calculated as the sum of true positives and true negatives divided by the total number of cases. Precision (Positive Predictive Value) measures the correctness of positive predictions, calculated as the number of true positives divided by the sum of true positives and false positives. Recall (Sensitivity or True Positive Rate) measures the ability of the model to detect positive instances, calculated as the number of true positives divided by the sum of true positives and false negatives. Specificity (True Negative Rate) measures the ability of the model to detect negative instances, calculated as the number of true negatives divided by the sum of true negatives and false positives. F1 Score is the harmonic mean of precision and recall, providing a balance between them.
    \end{table}

We selected the prompt that minimizes the number of false negatives (i.e., articles truly referring to a firm taking a stance that the model misses). In this first step of the analysis, we want to maximize our chances of detecting corporate stances on societal issues in news articles (i.e., maximize recall), at the cost of including some false positives (i.e., articles that the model includes as stances although it should not, which we filter out later). Therefore, although the three-part prompt is more holistically accurate, we picked prompt v2 (short prompt without reasoning). We executed this code in August 2025, using GPT's version gpt-3.5-turbo-0125. Out of the 51,979 news articles fed to the model, 14,756 were identified as CSA.

The other prompts enabled us to analyze the reasoning behind GPT's failure to select some news articles that the coders thought were CSA. We found that this was mostly due to a misunderstanding of our topics of interest. Therefore, we clarified the definition of each \texttt{topic} given in the prompt:
\begin{itemize}
    \item Abortion: Abortion (including reproductive health, Planned Parenthood, Roe v Wade)
    \item Climate: Climate Change
    \item Gender: Gender Issues (including Gender Discrimination, Equality, and Women's rights)
    \item Guns: Gun Control (including gun ownership, gun prevention and gun violence)
    \item Immigration: Immigration (including Border Security)
    \item LGBTQ: LGBTQ+ (including the Equality Act, Religious Freedom and Liberty, and Marriage Equality)
    \item MinWage: Minimum Wage
    \item Race: Racial Justice (including the Black Lives Matter movement and Race Discrimination in general)
   \item Voting: Voting Rights, related to U.S.\ elections
\end{itemize}

As an additional test, we tried increasing the temperature level to allow more creativity in GPT's answers but this did not significantly increase the performance of the classification. Therefore, we keep the temperature level at 0 to maximize replicability. We also tried using versions GPT3.5 and GPT4 of the model and the results were virtually the same. Finally, we evaluated two approaches of our full pipeline against our benchmark of manually labeled results. In the first approach, we clustered the news articles (see Section \ref{ap:factiva}), and then classified the resulting articles (i.e., the oldest article of each cluster) using GPT. In the second approach, we first classified all news articles using GPT, and then clustered the articles identified as related to CSA. The first approach minimized the false negatives.

\paragraph{Summary}

After the first classification of news articles, we asked GPT to summarize each
of the articles identified as instances of CSA. We used OpenAI's \texttt{gpt-4o-mini} model,\footnote{Version
\texttt{gpt-4o-mini-2024-07-18}.} which was released while we were developing
the pipeline and outperformed \texttt{gpt-3.5-turbo} on this task, at lower cost. We used the following prompt, where \texttt{topic} is
replaced by the expanded topic definition listed above:

    \textbf{I will input a news article that mentions the company
    \texttt{company} and the topic \texttt{topic}. Provide a short one-sentence
    summary of the news article. Here is the article: \texttt{article}}

Despite clustering events into stories at the start of the pipeline, we encountered numerous duplicates within our CSA events dataset. The summaries help us fix that issue. For each pair combination of news articles that fall within five days of each other, and belong to the same topic and company, we asked the model whether they describe the same event. If duplicates were identified, they were flagged, allowing us to restrict our analysis to only the first article in each batch.

This process resulted in 1,614 duplicates identified, reducing our total number of CSA events to 13,143.

\paragraph{Second classification}

Our initial prompt was designed to minimize false negatives (i.e., cases where GPT classifies an article as not being CSA, but human coders classify it as CSA). However, this approach resulted in a significant number of false positives, which are challenging to identify manually. We use a second GPT prompt to help us detect false positives and to provide more detail on the nature of each stance:

\textbf{I will provide a news article about the company \texttt{company} or its CEO taking a stance around the topic of \texttt{topic\_explanation}.\footnote{The \texttt{topic explanations} are defined as: Abortion (including reproductive health, Planned Parenthood, Roe v Wade), Climate Change, Gender Issues (including Gender Discrimination, Equality, and Women's rights), Gun Control (including gun ownership, gun prevention and gun violence), Immigration (including Border Security), LGBTQ+ (including the Equality Act, Religious Freedom and Liberty, and Marriage Equality), Minimum Wage, Racial Justice (including the Black Lives Matter movement and Race Discrimination in general), Voting Rights, related to U.S.\ elections.} You will need to answer the following questions:}

\textbf{\begin{enumerate}
    \item Is the company \texttt{company} or its CEO a main actor in the article? (Yes/No)
    \item Is the article primarily about the topic of \texttt{topic}? (Yes/No)
    \item Does the company support or oppose the topic of \texttt{topic} \texttt{current\_issue\_side}? (Support/Oppose)
    \item Is it the company or the CEO taking the stance? (Company/CEO)
    \item Was the stance a response to criticism or an attack (e.g., from the public, employees, or a public official)? (Yes/No)
    \item Did multiple actors react to the company’s stance (e.g., employees, competitors, politicians)? (Yes/No)
\end{enumerate}}

\textbf{For each answer, assign a confidence score between 0 and 1, where 0 denotes complete uncertainty and 1 denotes complete certainty.}

\textbf{Please respond in the following format, with each answer separated by ``|'': \\
\textbf{Example format}:\\
\texttt{Yes/No|0–1|Yes/No|0–1|Support/Oppose|0–1|Company/CEO|0–1|\\Yes/No|0–1|Yes/No|0–1}}

\textbf{Here’s the article: \texttt{article}}

\bigskip
The \texttt{issue sides} capture the ideological leaning of the stance. We provide GPT with the following definitions: 

\textbf{(Pro-Abortion would be support), (Pro-Climate Change Action would be support), (Pro-Gender Equality would be support), (Pro-Gun Control would be support, pro-Gun Ownership would be against), (Pro-Immigration would be support), (Pro-LGBTQ+ would be support), (Pro-Minimum Wage would be support), (Pro-Black Lives Matter would be support), (Pro-Voting Rights would be support, pro-Voter ID laws would be against).}

\medskip
We use 119 events that the first GPT prompt identified as CSA for the companies \texttt{DIS}, \texttt{JNJ} and \texttt{VZ}, to assess the reliability of GPT when compared to human coders.
\footnote{The number of CSA events reported in Tables \ref{tab:confusion} and
\ref{tab:performance} use \texttt{gpt-3.5-turbo-0613}, which is the version that we used to fine-tune our first classification prompt. This version was later deprecated and changed to \texttt{gpt-3.5-turbo-0125}, which classifies more articles as CSA. Running it on our initial sample of articles for \texttt{DIS}, \texttt{JNJ} and \texttt{VZ} yields 119 CSA events as opposed to 95. Since we could not reproduce the sample of 95 events obtained with the deprecated version of the model, we fine tuned the second classification prompt and assessed its validity based on the new sample of 119 CSA events.} Two of us manually coded responses for all six questions in the prompt above.

\begin{itemize}
    \item \textbf{Q1:} Is the company \texttt{company} or its CEO a main actor in the article? (Yes/No) \\
    Intercoder reliability: 79\% (24/112 disagree) \\
    GPT and coder 1: 78\% (25/112 disagree) \\
    GPT and coder 2: 72\% (31/112 disagree)
    
    \item \textbf{Q2:} Is the article primarily about the topic of \texttt{topic}? (Yes/No) \\
    Intercoder reliability: 82\% (20/112 disagree) \\
    GPT and coder 1: 81\% (21/112 disagree) \\
    GPT and coder 2: 79\% (23/112 disagree) \\

    We use questions 1 and 2 to identify false positives obtained with the first prompt and exclude them from the analysis.
    
    In what follows, the comparison between human annotations and GPT's answers is restricted to 45 articles that both coders considered to be related to the company and topic.

    \item \textbf{Q3:} Does the company support or oppose the topic of \texttt{topic} \texttt{current\_issue\_side}? (Support/Oppose) \\
    GPT response: 69/69 support \\
    Intercoder reliability: 95\% (2/42 disagree) \\
    GPT and coder 1: 96\% (2/45 disagree) \\
    GPT and coder 2: 95\% (2/42 disagree)
    
    \item \textbf{Q4:} Is it the company or the CEO taking the stance? (Company/CEO) \\
    GPT response: 58 company; 11 CEO \\
    Intercoder reliability: 95\% (2/42 disagree) \\
    GPT and coder 1: 93\% (3/45 disagree) \\
    GPT and coder 2: 93\% (3/43 disagree)
    
    \item \textbf{Q5:} Was the stance a response to a criticism or an attack directed at the company or CEO (e.g., from the public, employees, or a public official)? (Yes/No) \\
    GPT response: 23 yes; 49 no \\
    Intercoder reliability: 95\% (2/44 disagree) \\
    GPT and coder 1: 89\% (5/44 disagree) \\
    GPT and coder 2: 89\% (5/44 disagree)
    
    \item \textbf{Q6:} Did multiple actors react to the company’s stance (e.g., employees, competitors, politicians)? (Yes/No) \\
    GPT response: 26 yes; 46 no \\
    Intercoder reliability: 70\% (13/44 disagree) \\
    GPT and coder 1: 84\% (7/44 disagree) \\
    GPT and coder 2: 68\% (14/44 disagree)
\end{itemize}

For all these questions, the agreement rate between GPT (\texttt{gpt-4o-mini} model) and each human coder is similar to the agreement rate between the two coders, suggesting that GPT is a reliable substitute for human labeling. 

We excluded a few other questions that we initially included in the prompt but for which the agreement rate between GPT and humans was substantially lower than the agreement rate between two humans: Whether the stance refers to a statement, an action, or other. Whether the action or statement is something the company is doing now or has done in the past. Whether the stance was communicated through official channels (e.g., press release, statement from a spokesperson), internal communications (e.g., memo to employees, declaration to shareholders), social media (e.g., Twitter), or another medium. Whether the company's stance is likely to have a large impact (e.g., among employees, among consumers, in the media, etc), yes or no. Whether the events in the article are reported by the company itself or by a third party, like a journalist. 

Instead of asking GPT, we chose to rely on the news source of the article to determine whether it is a press release or an article likely written by a journalist. More precisely, we code as press releases all articles from the sources ``Dow Jones Newswires'', ``GlobeNewswire'', ``PR Newswire'', plus those from ``Dow Jones Institutional News'' that explicitly say ``Press Release'' in the bodytext.

In addition, since simply asking the LLM whether the article describes a statement or an action produced poor results, we designed a third prompt specifically to answer that question.

\paragraph{Third classification}

We thoroughly read a subsample of 95 articles among our final list of stances and identified three situations that an article could describe: the company undertakes a concrete action that is related to one of the nine societal issues and that directly changes how it conducts business; the company does not take an action but engages with one of the nine issues and explicitly links that position to business interests; the company engages with one of the nine issues without explicitly linking the activity to its business interests.

Two annotators manually classified the subsample of 95 articles between these three categories and we used their consensus answers as a benchmark to fine-tune the classification prompt. We used the \texttt{gpt-4.1} model, which is more costly than \texttt{gpt-4o-mini} but performs substantially better for this task.

The final prompt that we used is the following:
\\
\textbf{You are coding corporate socio-political activism (CSA). A stance event is a company, or its CEO, publicly demonstrating, through statements or actions, support for or opposition to one side of a partisan societal issue.
\\
You will receive an article, the company being coded, and the topic. Assign exactly one category using the codebook below. Apply the Quick Rule tests in order and stop at the first category that matches.
\\
Important: do not use the human code or any outside knowledge. Use only the article, company, topic, and the codebook.
\\
Quick Rule: The article mentioned an action or the announcement of an action that affects the way a firm conducts business or its operations inside the firm? = Category 1, otherwise → the stance is explicitly stated and justified in the article for business or strategy purposes? = Category 2, otherwise → Category 3 = stance taken by social, political, or charitable engagement with no stated business justification.
\\
Category 1: Operational Action
\\
Definition: The company undertakes a concrete action that directly changes how it conducts business.
\\
Decision Rule:
\\
Code as Category 1 when the company changes its business practices or operations, regardless of whether the action is accompanied by language surrounding its mission, business or social justification. Code as Category 1 if the operational content is an action already taken or announcement of a future action. Receiving an award for a past action sits under this category, as it is recognition for something a firm has done. If it is a new action, it must change the way a firm conducts business and its operations. Merely funding, endorsing, or lending its name to a coalition, without a commitment or concrete action is not Category 1.
\\
Practical Examples: Committing to or implementing employee benefits, diversity hiring initiatives, wage increases, changes to production, investment in its own operations or infrastructure, changes to products or content offerings.
\\
Specific Examples: Company updates public-facing product or experience in response to racial criticism/public pressure (topic: race). A company provides open or low-cost access to its climate-related patents, tools, or expertise to help other organizations reduce emissions or develop sustainable technologies (topic: climate).
\\
Category 2: Business-Justified Position Taking
\\
Definition: The company expresses support for, opposition to, or engagement with one of the nine societal issues and explicitly links that position to business interests.
\\
Decision Rule:
\\
Code as Category 2 when the company explicitly justifies its position using business-related considerations. Code Category 2 only when the operative content is a position, not an action. If the company is also taking a concrete operational action, Category 1 takes precedence. The business rationale must apply to the specific stance being coded, not to a separate item mentioned elsewhere in the article. Even if an action is justified for business interests, it is coded as Category 1, based on an action being taken.
\\
Practical Examples: Supporting immigration reform because of workforce needs; advocating climate policies because they align with corporate strategy; opposing legislation because it may affect recruiting, retention, production, or customer demand.
\\
Specific Examples: A coalition of CEOs publicly backing immigration reform, citing the need for skilled-worker talent pipelines (topic: immigration). A coalition of corporations publicly opposing anti-LGBT state legislation because they could harm talent attraction and state economic competitiveness (topic: LGBTQ rights).
\\
Category 3: Non-Business Social or Political Expression
\\
Definition: The company engages in advocacy, philanthropy, public statements, charitable or advocacy partnerships, donations, or other forms of social, political, or environmental engagement without explicitly linking the activity to its business interests. The action is justified primarily in terms of social goals rather than explicit business objectives.
\\
Decision Rule:
\\
Code as Category 3 when the company neither changes how it operates nor explicitly connects the position or stance to business interests. When a stance falls between Category 1 and 3, note that donations, grants, and sponsorships do not change the way a firm conducts business, meaning it must fall to Category 3 as non-business justified stance-taking. When a stance falls between Categories 2 and 3, apply the explicit-statement test: code Category 2 only if the business justification is stated in the article. If the business connection is something the coder must infer rather than read, code Category 3. The justification must be on the page, not supplied by the coder.
\\
Practical Examples: Charitable donations, grants, sponsorships, giving out or sponsoring awards, fellowships, climate or social initiatives framed primarily as efforts to benefit society, and public statements on social issues that do not reference business impacts.
\\
Specific Examples: A firm pauses political donations in a state in response to anti-LGBTQ legislation (topic: LGBTQ). A firm joins a campaign, pledging forest protection when consumers commit to buying pre-owned products (topic: climate change).
\\
Respond with ONLY a JSON object, no markdown, in this exact form:
\\
{"reasoning": "<one or two sentences naming the coded stance and the test that decided it>", "category": 1}
\\
where “category” is the integer 1, 2, or 3.}
\\

\subsection{Validation} \label{ap:validation}

We conducted a series of exercises to assess the validity of our final dataset. First, we list examples of stances on different topics and for different ideological leanings, reporting the article summary provided by GPT (see section \ref{ap:gpt_prompts}).

\paragraph{Examples of Company Stances}

We provide an example of a stance for each of the nine societal topics, along with its ideological leaning (liberal or conservative) and the type of engagement it represents (a concrete action that affects its business, a statement justified by business interests, or a statement/action that does not affect the way the company conducts its business and that is not justified by business interests).

\begin{itemize}
    \item \textbf{Abortion (liberal, business-justified):} Biogen, Inc., along with over 300 other pharmaceutical companies, signed an open letter urging the reversal of a federal court ruling that suspended the approval of the abortion pill mifepristone, arguing it undermines the FDA's authority and poses a significant risk to drug development and safety. \textit{[April 2023]}
    \item \textbf{Climate (liberal, not business-justified):} Citigroup, Inc. is among a group of major banks that co-created the Green Bond Principles to promote transparency and integrity in the Green Bond market, thereby encouraging investments in environmentally sustainable projects to combat climate change. \textit{[January 2014]}
    \item \textbf{Gender (conservative, not business justified):} eBay Inc. publicly opposed a shareholder proposal by Arjuna Capital seeking to address the gender pay gap within the company, despite growing public demands for gender wage equality and the recognized benefits of diversity in the workplace. \textit{[February 2015]}
    \item \textbf{Guns (liberal, concrete action):} Kroger has joined Walmart in advocating for gun safety and reform by asking customers not to bring guns into their stores and supporting measures like background checks following a series of mass shootings. \textit{[September 2019]}
    \item \textbf{Immigration (liberal, business-justified):} Caterpillar's Chief Technology Officer, Gwenne Henricks, urged Congress to pass immigration reform to enhance the recruitment of skilled workers in STEM fields essential for the company's growth and competitiveness. \textit{[May 2013]}
    \item \textbf{LGBTQ (conservative, concrete action):} The article discusses conservative activist Robby Starbuck's successful campaign to pressure companies like Deere \& Co. and Tractor Supply to abandon their diversity and inclusion programs, amidst a growing backlash against corporate support for LGBTQ+ initiatives. \textit{[August 2024]}
    \item \textbf{Minimum wage (conservative, concrete action):} Uber is appealing a tribunal ruling in the UK that determined its drivers should be classified as workers entitled to minimum wage and paid holidays, arguing instead that they are self-employed and operate similarly to traditional taxi drivers. \textit{[September 2017]}
    \item \textbf{Race (liberal, not business-justified):} Comcast Corporation has pledged \$2.5 million to Echoing Green's Racial Equity Philanthropic Fund to support over 500 social enterprises aiming to combat racial inequity and promote social innovation. \textit{[February 2021]}
    \item \textbf{Voting (liberal, concrete action):} Vote.org is partnering with companies like Twitter, Blue Apron, and Group Nine Media to promote a campaign encouraging 1,000 businesses to provide paid time off for employees to vote, aiming to enhance voter participation and address obstacles created by COVID-19. \textit{[July 2020]}
\end{itemize}

\paragraph{False negatives and false positives}

Since our database of firm stances on societal issues is novel, there is no clear benchmark to compare it to in order to determine whether we are missing important stances taken by the firms in our sample (i.e., false negatives) or whether we include stories that are not about a firm taking a stance on one of our issues of interest (i.e., false positives). We conduct several exercises to evaluate the share of false negatives and false positives in our final data. 

First, we asked an LLM (ChatGPT) to provide a list of 35 prominent stories where companies from the S\&P 500 took public stances on our nine topics of interest over our sample period. We then manually looked for these events in our own database. Out of the 35 stories provided by ChatGPT, 32 of them (91\%) were found in our dataset while 3 (9\%) were missing. These stories and the summaries of the corresponding articles found in our dataset are listed in Table \ref{tab:comparison_chatgpt}.\input{tables/comparison_chatgpt}

Second, we compared our list of stances to those in \citet{klostermann2022effect},\footnote{We also considered comparing our list of stances to those in \citet{bhagwat2020corporate} and \citet{conway2024consuming}, but the list of events used in these papers is either unavailable or de-identified, which prevents us from knowing which firm took which stances in their data.} which conducted a systematic keyword search on LexisNexis and Google News to manually identify 106 events of corporate political activism involving 92 companies between 2016 and 2018. Since their list is much smaller than ours, we expect our database to include all these events. An important caveat is that the list of companies and topics represented in \citet{klostermann2022effect} differs from ours. Out of the 33 events that refer to both a company and a topic in our sample, we find that 19 of them are included in our own database (58\%) while 14 are missing (42\%). A more thorough investigation of these missing stances revealed that they were all instances where multiple companies took a common stance (e.g., making a statement condoning Trump’s travel bans) and our database includes this stance for some companies but not all of them. 

This issue was already described in Appendix \ref{ap:factiva}: when an article mentions multiple companies, the automatic tagging of companies provided by Factiva may miss some of them. To quantify the magnitude of this issue, we picked a random sample of 200 articles from our final list of articles (i.e., those identified by our pipeline as describing a company taking a stance on one of our nine topics of interest), and read each of them. Out of those 200 articles, 12 referred to multiple companies (6\%). Factiva correctly tagged all companies mentioned in 6 of these 12 articles, but failed to tag some companies in the 6 remaining ones. In total, we miss 10 occurrences where a company of our sample took a stance common to other companies, indicating that our full database may miss about 5\% (10 out of 200) of relevant stances taken over our sample period.

Third, our final dataset may include articles that do not refer to a company of our sample taking a stance on one of our nine topics. This is all the more likely as we designed our first GPT prompt to minimize the number of false negatives while allowing for false positives (see Section \ref{ap:gpt_prompts}). Out of the 200 articles that we picked randomly and read carefully, we assess that only 7 of them (3.5\%) should not have been included in the final dataset. These articles either refer to the company facing a lawsuit related to the topic (rather than taking a stance on the topic) or refer to the company taking a stance on a different topic. 

Overall, these different tests indicate that our final dataset of company stances on societal issues misses about 9\% of the stances that were truly taken by our sample of companies. Conversely, less than 5\% of the articles included in the dataset are not about a company taking a stance on the topic of interest. This suggests that we slightly underestimate the total number of stances.

\subsection{Meta Content Library} \label{ap:meta}

We replicate our collection of corporate stances on an alternative source of information: Facebook posts. Using the Meta Content Library's page search tool, we identified relevant Facebook pages by manually searching for the names of the 836 companies in our main sample (see Appendix \ref{ap:sample_firms}). We restricted our search to a single Facebook page per company and included only verified pages. The search process followed these guidelines:

\begin{itemize}
    \item \textbf{Parent company page available}: When the parent company maintained an official Facebook presence, this page was selected.

    \item \textbf{No parent company page, but a closely matching subsidiary or product page}: In cases where the parent company did not have a Facebook page, we selected the page corresponding most closely to the company name (for example, for Textron Inc. we use the page of its subsidiary Textron Aviation, and for Live Nation Entertainment the page Live Nation Concerts). This approach was chosen to reflect how a typical user might search for a company on the platform. While we initially considered including all subsidiaries and brands, this proved infeasible at scale.

    \item \textbf{Multiple closely matching subsidiary or product pages}: If multiple pages closely matched the company name, the company was excluded to maintain the ``one page per company" rule.

    \item \textbf{No relevant page found}: Companies without a clearly identifiable or closely matching page were excluded.
\end{itemize}

Whenever possible, we prioritized U.S.-specific pages. If a U.S.\ page was unavailable, we selected a general or global company page.

A key challenge was that the Meta Content Library does not reliably distinguish between Facebook pages with identical names, particularly when they represent different geographic regions (for instance, the Meta Content Library returns several pages named "Mastercard" or "Netflix"). To resolve this ambiguity, we looked up each candidate page in the Meta Ad Library, a public database of advertising on Meta's platforms that reports, for each page, its unique page identifier and handle (e.g., @MastercardUS) as well as the countries from which it is managed. This enabled us to accurately select the U.S.-based page when available, or otherwise the primary global page. The verified page link was then used for data extraction within the Meta Content Library.

Overall, we found Facebook pages for 306 companies. From these pages, we then queried posts using topic-specific keyword groups, following the same search protocol as on Factiva (Appendix \ref{ap:factiva}):

\begin{itemize}
    \item \textbf{Abortion}:
    (reproductive health) \textbar{}
    (planned parenthood) \textbar{}
    (abortion) \textbar{}
    (roe v wade)

    \item \textbf{Climate}: climate change

    \item \textbf{Gender}:
    (paycheck fairness act) \textbar{}
    (gender discrimination) \textbar{}
    (gender equality) \textbar{}
    (gender inequality) \textbar{}
    (equal pay) \textbar{}
    (women's rights)

    \item \textbf{Guns}:
    (gun ownership) \textbar{}
    (gun control) \textbar{}
    (gun ban) \textbar{}
    (gun prevention) \textbar{}
    (gun violence) \textbar{}
    (firearms)

    \item \textbf{Immigration}:
    (border security) \textbar{}
    (illegal immigrants) \textbar{}
    (illegal immigration) \textbar{}
    (undocumented immigrants) \textbar{}
    (immigration reform) \textbar{}
    (immigration system) \textbar{}
    (DACA) \textbar{}
    (Deferred Action for Childhood Arrivals)

    \item \textbf{LGBTQ}:
    (LGBT) \textbar{}
    (sexual orientation) \textbar{}
    (equality act) \textbar{}
    (religious freedom) \textbar{}
    (religious liberty) \textbar{}
    (Defense of Marriage Act) \textbar{}
    (Respect for Marriage Act) \textbar{}
    (same sex marriage) \textbar{}
    (same-sex marriage) \textbar{}
    (marriage equality)

    \item \textbf{Minimum Wage}: minimum wage

    \item \textbf{Race}:
    (Black Lives Matter) \textbar{}
    (racial discrimination) \textbar{}
    (racial justice)

    \item \textbf{Voting}:
    (voting laws) \textbar{}
    (voter identification laws) \textbar{}
    (voter ID laws) \textbar{}
    (voting rights) \textbar{}
    (right to vote)
\end{itemize}

These queries returned posts containing at least one of the specified keyword groupings. Due to platform constraints, we could not require exact phrase matching. Instead, posts were included if they contained all words within a given term grouping, regardless of the order (e.g., a post containing both ``reproductive'' and ``health'' is included, even if these two words were not following each other in the text).

We restricted the dataset to posts published between January 1, 2008 and December 31, 2025. Only downloadable data was included. The Meta Content Library limits downloadable data to posts from pages with at least 15,000 likes or followers, and from profiles that are either verified or have at least 25,000 followers. Additionally, we did not enable the ``find text in images'' option, as this feature was only available for recent posts at the time of data collection. 

The Facebook dataset includes the date and textual content of each post for a total of 2,677 posts, across 223 companies of our sample.\footnote{We exclude the pages of three news organizations (The New York Times, Yahoo, and Fox News), whose posts report on our topics rather than take a stance on them, as well as posts not written in English. A post matching the keywords of several topics appears once for each topic; we classify it separately for each topic.} As for news articles, many of these posts do not correspond to corporate stances on our topics of interest. To identify stances, we implement the LLM classification described in Appendix \ref{ap:gpt_prompts}. More precisely, we run the following prompt on all posts:

\begin{quote}
\textit{I will input a Facebook post. Quietly determine whether the post describes an instance in which the company [company name] or its CEO publicly demonstrated, through statements or actions, support or opposition to the issue of [topic]. Respond `Yes' or `No', without further comments.}

\textit{Here is the post: [post text]}
\end{quote}

This classification identifies 2,088 posts as corporate stances, taken by 210 companies in our main sample between 2008 and 2024.

\subsection{Employee Demographics} \label{ap:revelio}

We construct firms' employee demographics by month  using the Revelio database, which covers all 836 S\&P 500 firms included in our sample except six that do not have a LinkedIn profile. Three of them are temporary companies that existed in between company merging. The three others are companies we cannot find data for: Harrah’s Entertainment, Columbia Pipeline Group, and Novell Inc.

\textbf{Location} We obtain the share of employees reported to work domestically and the share of international employees who work outside of the US from Revelio’s workforce dynamics data, which keeps track of each firm’s monthly headcount by country. From the individual-level data we also observe LinkedIn positions that report their work location, which we use to construct employee headcount by state for each firm and each month in our sample. 

\textbf{Ethnicity} For each individual user on LinkedIn, Revelio estimates her ethnicity using her first name, last name, and location (their prediction mobilizes US Census data). We classify an individual to be a certain race if Revelio’s predicted probability is greater than 95\%. Using this, we calculate the monthly share of Black employees for firms in our sample. 

\textbf{Gender} Similarly, Revelio predicts the gender of each LinkedIn user using her first name, training their prediction model with US Social Security data. We classify an individual to be male or female if Revelio’s predicted probability is greater than 95\%. We compute the share of female employees for each firm-month observation.

\subsection{Matching with Voter Records} \label{ap:matching_l2}

In the Revelio data, there are 12,742,465 LinkedIn profiles of individuals who worked in an S\&P 500 company in the U.S.\ during our sampling period. We use their most recent self-reported work location for matching. 495,016 individuals were missing a first name, last name or location on LinkedIn, and are dropped from the matching process. Our final matching sample consists of 12,247,449 individuals linked to their most recent job information. Our version of the L2 data contains voter snapshots from 2018 to 2025. There are 276,326,401 voters recorded in total. While L2 does not directly track voting outcomes, it does contain party registration records for voters in 33 states.\footnote{The 17 states where party registration is estimated are Alabama, Georgia, Hawaii, Illinois, Indiana, Michigan, Minnesota, Missouri, Montana, North Dakota, Ohio, South Carolina, Texas, Vermont, Virginia, Washington, and Wisconsin. We do not use any partisanship information for individuals matched to these states. However, we do not exclude them from the matching process, as we also use the voter file to obtain each matched individual’s date of birth.} Our partisanship data is restricted to these states only. As most people do not report their date of birth on LinkedIn, the age variable we use is estimated using each individual’s self-reported education or work history. We also map each LinkedIn user’s self-reported work location to their corresponding coordinates (latitude, longitude) to compute the distance from matched L2 residential addresses. 

We adopt 50 combinations of individual name, age, and work location to match LinkedIn and the voter file data. Our matching procedure follows \citet{chinoy2024political}, using a “zoomout” principle. We gradually relax restrictions on the distance between the LinkedIn work location and the voter’s residential address, starting at $\leq$ 5 miles, then 25 miles, then 50 miles, then 100 miles, and finally matching all records within the same state. We also first attempt to match birth years within a 3-year window, incrementally relaxing the difference while restricting it to be no greater than 10 years. We always enforce the age difference between L2 and LinkedIn to be less than 10 years if the LinkedIn age is observed.\footnote{If either the LinkedIn individual’s estimated age or the voter’s age is unavailable, we relax these age restrictions.} For names, we first match using first name, middle name and last name, then first name and last name, then common nickname and last name, and finally
using only the first four letters of first name and last name. 

Individuals who are matched to multiple L2 records in a year are dropped from the data. We also do not allow an L2 voter to be matched to multiple LinkedIn profiles. As our voter registration data only goes back to 2018 while our LinkedIn data ranges from 2008 to 2025, we backfill an individual’s party registration with the earliest year she is matched with. For example, if an individual worked in an S\&P 500 firm from 2014 to 2018 but only has voting records from 2018 to 2024, we would use her voter registration record from 2018 to determine her political affiliation.

Together, we are able to match 5,078,315 individuals, yielding a final matching rate of 41.5\%. This rate aligns with the matching
rate accomplished in \citet{chinoy2024political} (40.4\%).

We use similar matching procedures to determine the partisanship of executives and board members, obtained via Execucomp and ISS Board Members. There are 11,643 executives from firms in our sample. While Execucomp does not contain data on the executive’s work location, it does keep track of their gender and exact year of birth. Thus, we use combinations of executive name and age for matching instead, and restrict to matches of the same gender. In total, we are able to match 4,641 executives, yielding a matching rate of 39.9\%. Similarly, for board members, only gender and year of birth are available. We use the same matching procedure as the executives
above, and match 6,670 board members out of 17,550 in total, achieving a matching rate of 38\%.

\subsection{Peak Detection in Factiva and Google Trends} \label{ap:peak_detection}

To detect peaks in media attention on a given topic over our sample period, we use Factiva and collect the monthly number of articles published on each of our nine topics. We conduct the same search as described in Section \ref{ap:factiva} but without filtering on any company. This gives us, for each topic, a monthly time series of news article counts.

To detect peaks in public attention, we use the Google Trends API which provides, for each topic and at the monthly level, the relative number of Google searches related to that topic over our sample period, from 0 (the month with the lowest number of searches) to 100 (the month with the highest number of searches).\footnote{Since our nine societal topics are very popular, we do not need to conduct a keyword search as we are able to use directly the ``Google Topics" functionality of Google Trends.} 

We treat each source separately but use the same peak detection algorithm for both of them. We define as a peak a month or a short series of months in which the number of articles or Google searches is well above the underlying time trend. First, we predict the level in each month (either the number of articles or the relative number of Google searches) with a regression model where the included covariates are a cubic time trend and month-of-the-year fixed effects interacted with a linear time trend, all of these interacted with sub-period fixed effects where each sub-period is a third of the whole period (the sample period covers 210 months so the first sub-period includes the first 70 months, the second one includes the next 70 months, and the last one includes the last 70 months).
Second, we smooth the predicted values by applying a 5-month moving average: we replace each predicted value by the average of itself, the predicted values of the two preceding months, and the predicted values of the two following months. Third, we define as a peak candidate any month in which the difference between the actual value and the smoothed predicted value is larger than two standard deviations. A peak candidate can be a single month or a contiguous series of months above the two-standard-deviation threshold. When two peak candidates are separated by three months or less, they are considered the same peak. Finally, we keep a maximum of ten peaks per series, excluding those whose maximum height (i.e., the level of the month for a single-month peak and the level of the highest month for a multi-month peak) is not among the top ten maximum heights. This approach yields a total of 48 peaks for the Factiva series and 40 for the Google Trends series (see the distribution of peaks across topics in Table \ref{tab:dist_peaks}). On average, a peak lasts 1.7 months, with most peaks lasting only one month (65 out of a total of 88 peaks) while the longest peak lasts 7 months.

\begin{table}[htbp]
\centering
\caption{Number of Peaks by Topic and Source}
\label{tab:dist_peaks}
\begin{tabular}{lrr}
\toprule
\textbf{Topic} & \textbf{Factiva} & \textbf{Google} \\
\midrule
Abortion    & 4  & 3 \\
Climate     & 5  & 5 \\
Gender      & 5  & 1 \\
Guns        & 6  & 4 \\
Immigration & 5  & 6 \\
LGBTQ       & 8  & 6 \\
Minimum wage   & 5  & 8 \\
Race        & 2  & 2 \\
Voting      & 8  & 5 \\
\bottomrule
\end{tabular}
\end{table}

We provided two LLMs (ChatGPT and Gemini) with the list of peak dates (i.e., the month of the peak for single-month peaks and the first month of the peak for multi-month peaks) and asked them to identify, for each date, a prominent event that drew public attention on the topic of interest in the U.S.\ and which could correspond to that date. After carefully reading their answers, we find that the two LLMs identified the same event for all 88 peaks except 3. Even when they disagreed, at least one of them associated the peak date with a relevant event. Therefore, we are confident that our peak detection algorithm captures spikes in news and Google searches due to salient events on our topics of interest.

In our analysis (Section \ref{sec:goal}), we exclude peaks that are less than 6 months apart from another peak. The final list of peaks and their corresponding event according to Gemini and GPT are provided in Tables \ref{tab:peaks_factiva} and \ref{tab:peaks_google}.

\input{tables/peak_event_tables_factiva}

\input{tables/peak_event_tables_google}

\section{Additional Figures and Tables} \label{ap:figures}

\begin{figure}[H]
    \centering
    \caption{Quarterly Stances by Topic: Factiva vs.\ Facebook}
    \includegraphics[width=\linewidth]{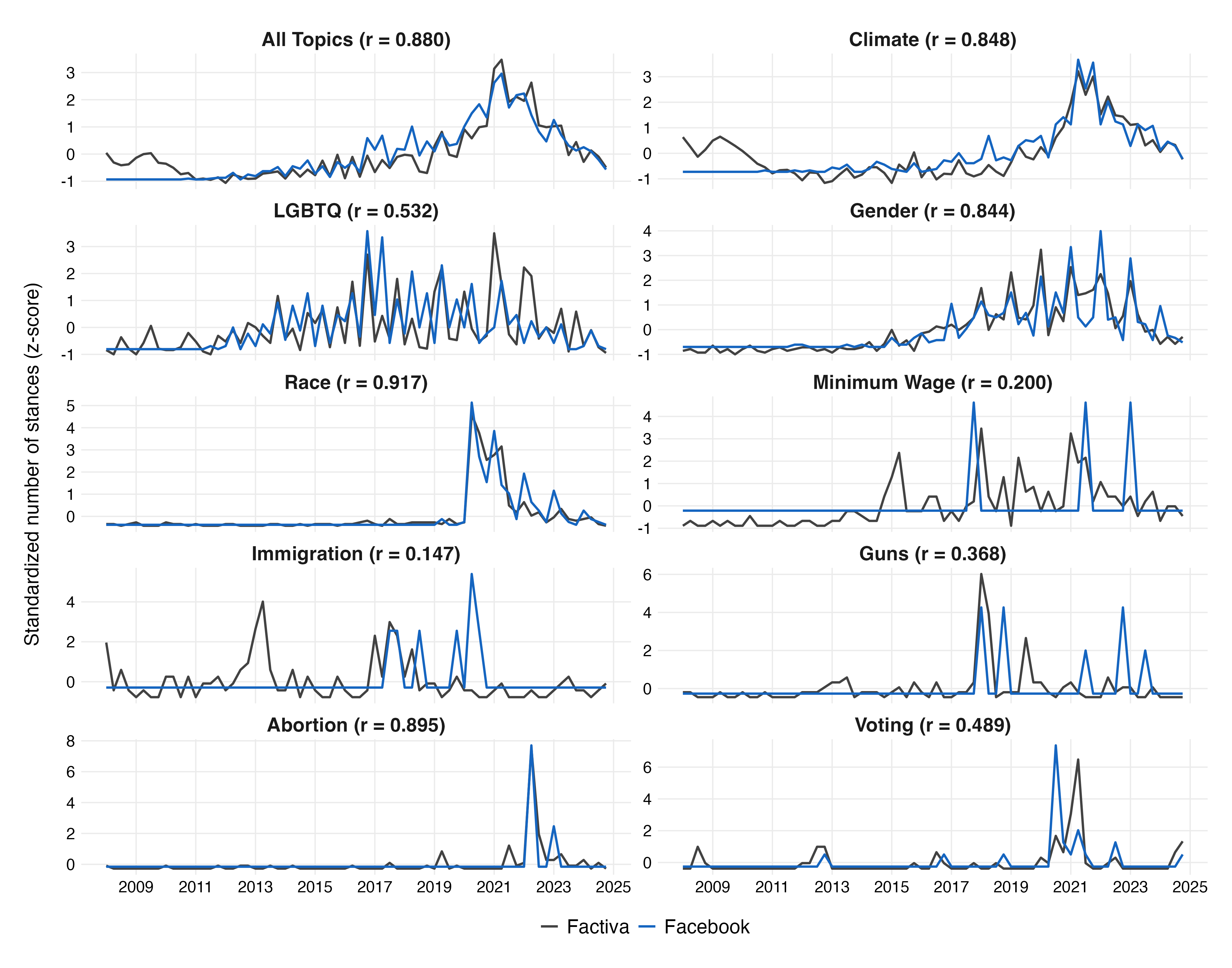}
    \label{fig:meta_val_timeline}
    \begin{minipage}{\linewidth}
    \footnotesize
        \textit{Notes:} Quarterly number of stances among news articles and press releases (for the 836 S\&P 500 companies in our sample) and of stances among Facebook posts (for the 223 companies with downloadable Facebook posts), 2008--2024, each standardized within topic. $r$ (in the panel titles) is the correlation between the two quarterly series. Minimum wage, abortion, guns, immigration and voting rights have fewer than 25 Facebook stance posts each.
    \end{minipage}

\end{figure}

\begin{figure}[H]
    \centering
    \caption{Company-Level Stances in Factiva and on Facebook}
    \includegraphics[width=0.7\linewidth]{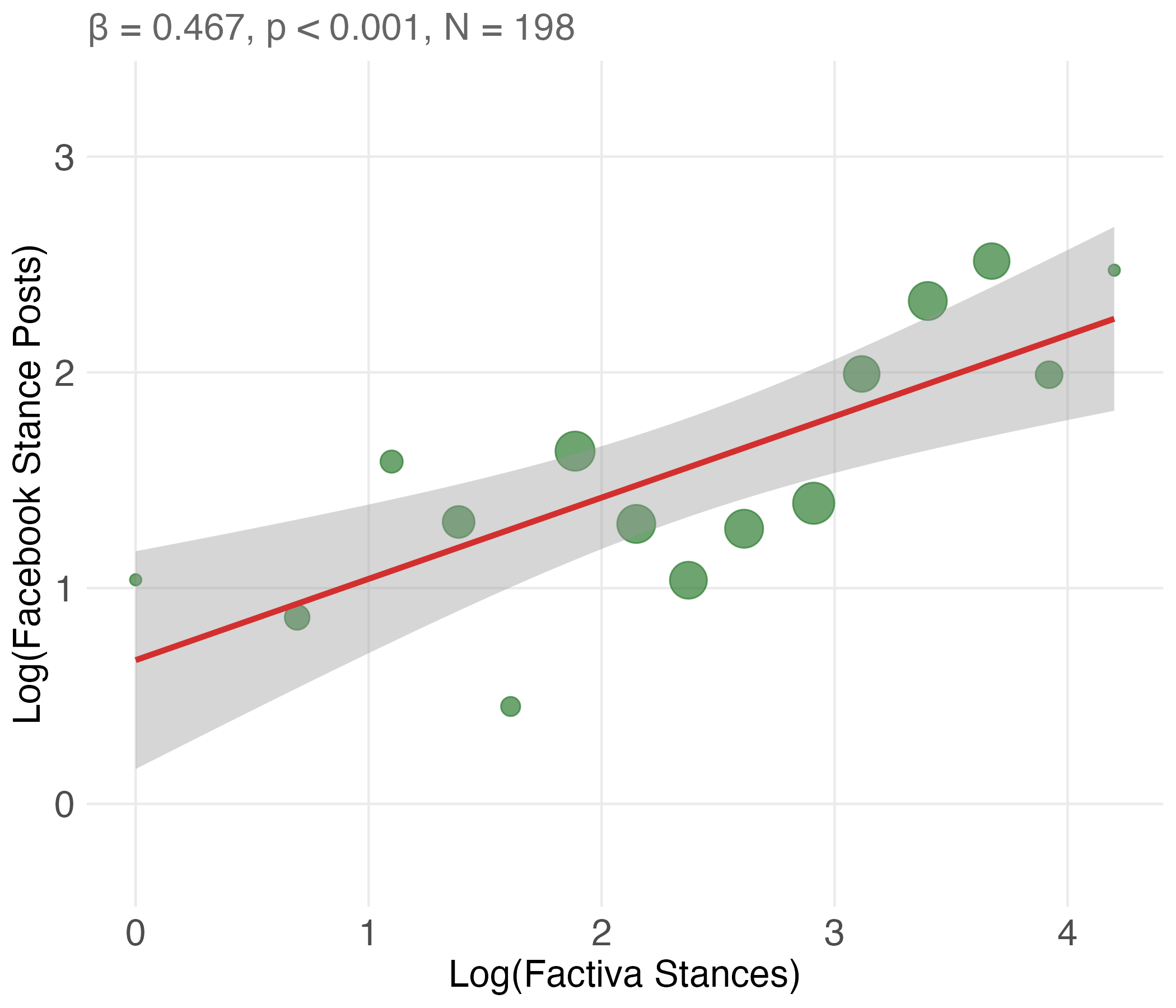}
    \label{fig:meta_val_loglog}
    \begin{minipage}{0.7\linewidth}
    \footnotesize
        \textit{Notes:} Binscatter showing the relationship between the number of stances taken on Facebook and the number of stances appearing in the news between 2008 and 2024, controlling for industry fixed effects and company headcount and revenue. We use only observations with strictly positive values for the two variables. Each point represents the mean of 20 equally-sized bins, with point size reflecting the number of observations per bin. The line is a linear fit with its 95\% confidence band. $\beta$ is estimated with the following equation:

\begin{equation*} \label{eq:facebook_loglog}
\ln(F_{i})=\lambda^I_i+\boldsymbol{\delta}'\!\mathit{Size}_{i}+\beta\,\ln(S_{i})+\epsilon_{i}
\end{equation*}

where $F_{i}$ is the average yearly number of stances taken by company $i$ on Facebook; $S_{i}$ is the average yearly number of stances in our news dataset; $\lambda^I_i$ is an industry fixed effect; $\mathit{Size}_{i}$ is a vector of controls including company headcount and revenue. Standard errors are clustered at the company level. Facebook posts by The New York Times, Yahoo and Fox News are excluded. $N = 198$ companies.
    \end{minipage}
\end{figure}

\begin{figure}[H]
    \centering
    \caption{Characteristics of Corporate Stance-Taking over Time}
    \includegraphics[width=0.75\linewidth]{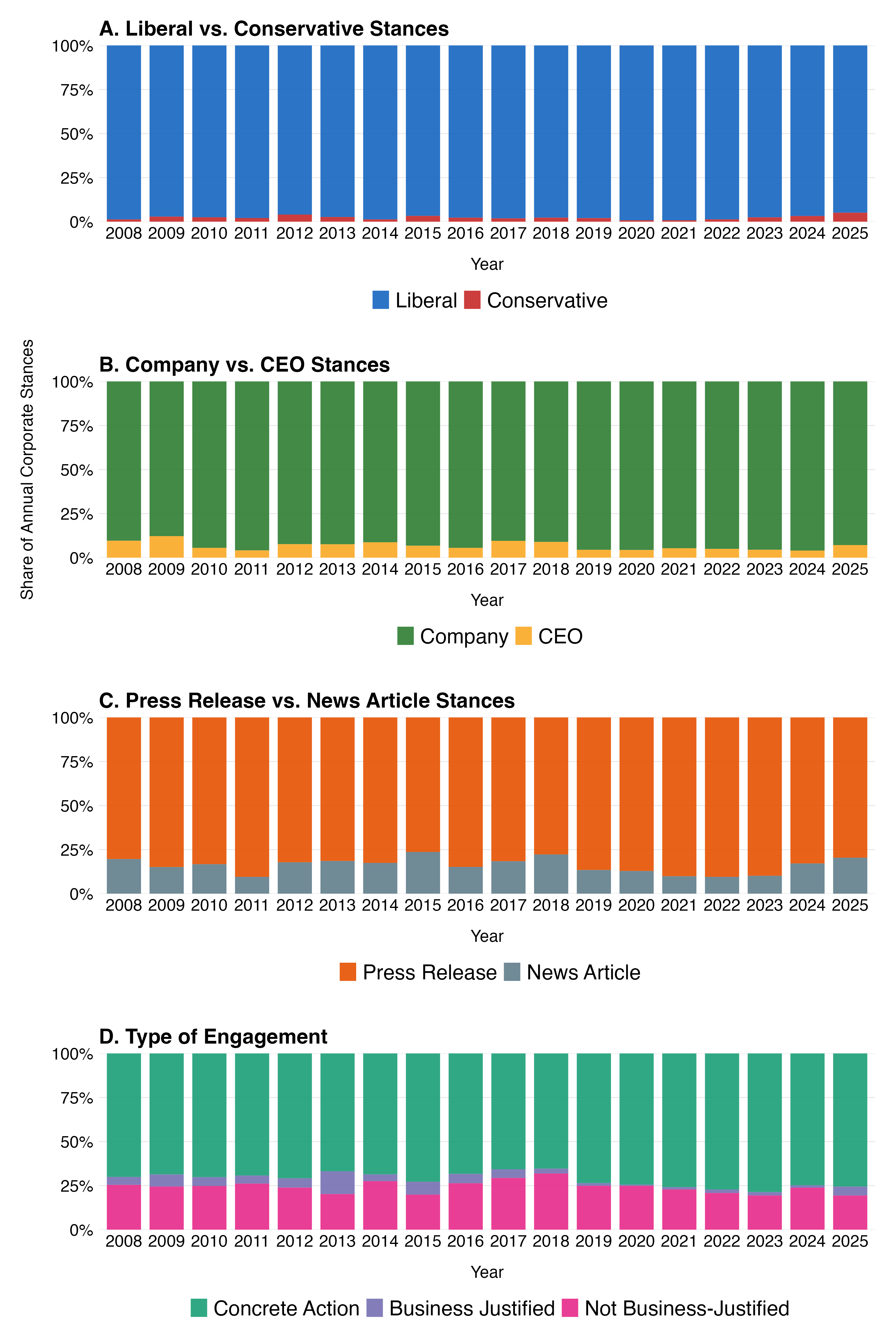}
    \label{fig:timeline_chars_prop}
    \begin{minipage}{0.75\linewidth}
    \footnotesize
        \textit{Notes:} Proportion of corporate stances taken by S\&P 500 companies, 2008--2025, colored by ideology (liberal vs.\ conservative), speaker (company vs.\ CEO), source (press release vs.\ news article) and type of engagement (concrete action vs.\ business-justified vs.\ not business-justified). 2025 includes data only up to June 1st.
    \end{minipage}
\end{figure}

\begin{figure}[H]
    \centering
    \caption{Corporate Stance-Taking by Industry Sector}
    \includegraphics[width=0.9\linewidth]{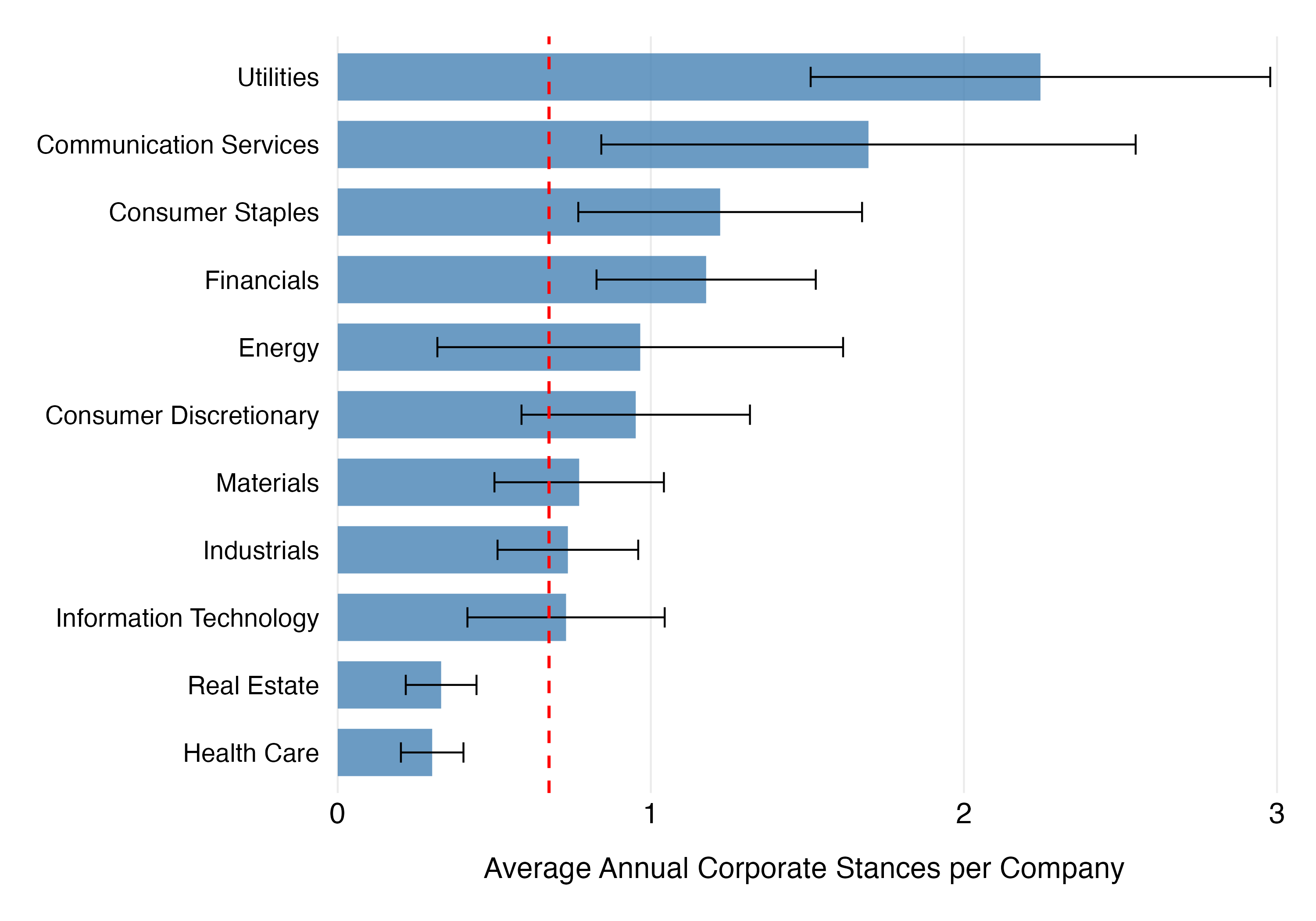}
    \label{fig:sector}
    \begin{minipage}{0.9\linewidth}
    \footnotesize
        \textit{Notes:} Average annual corporate stances per company for S\&P 500 companies, 2008--2025, by GICS sector. Error bars show 95\% confidence intervals. Red dashed line shows the overall cross-sector average.
    \end{minipage}
\end{figure}

\begin{figure}[H]
    \centering
    \caption{Top 10 Companies by Corporate Stance-Taking}
    \includegraphics[width=0.9\linewidth]{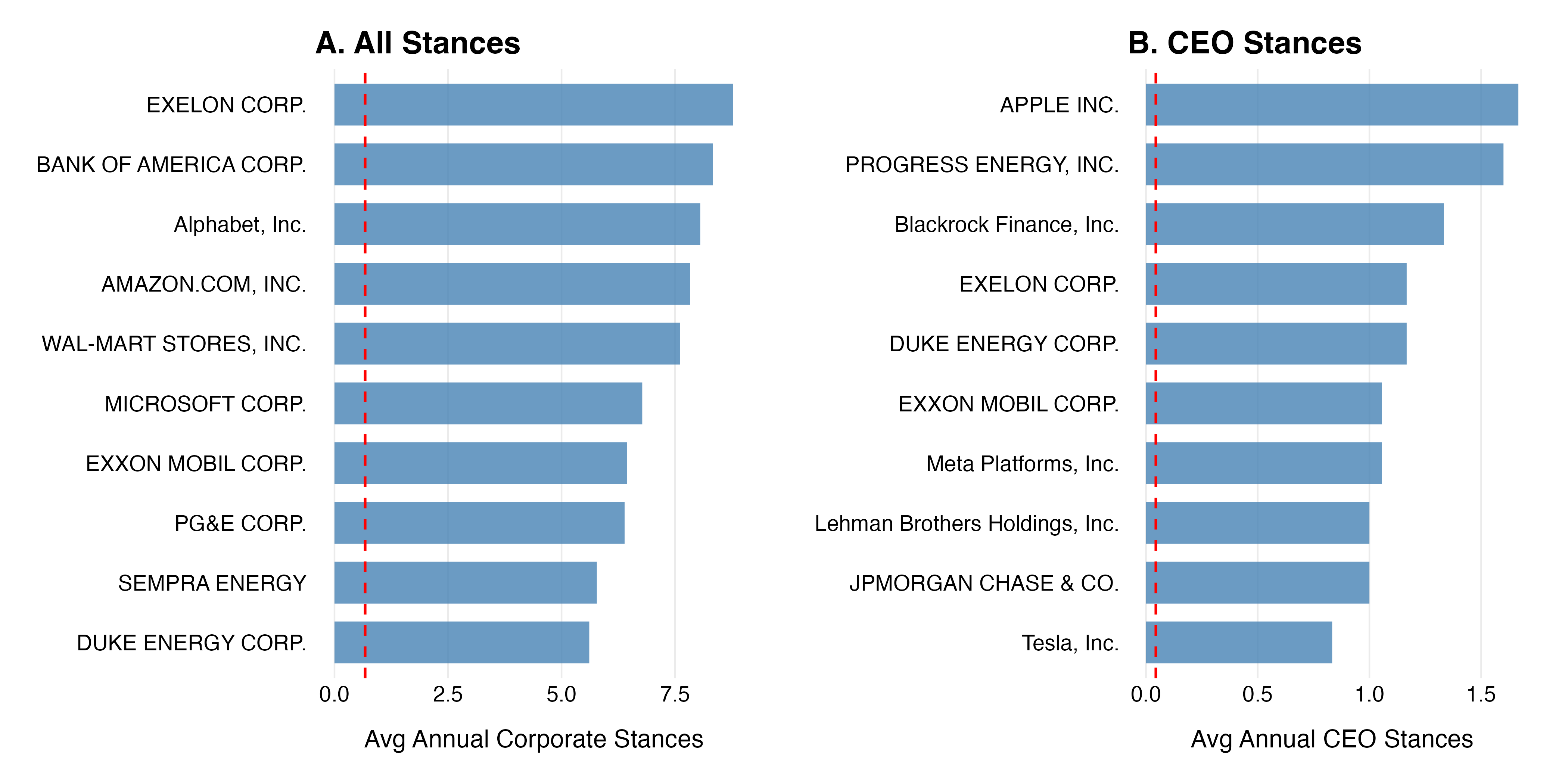}
    \label{fig:top10companies}
    \begin{minipage}{0.9\linewidth}
    \footnotesize
        \textit{Notes:} Top 10 S\&P 500 companies by average annual corporate stances, 2008--2025. Panel A ranks by total stances; Panel B ranks by stances attributed to the CEO. Red dashed lines show the overall average across companies.
    \end{minipage}
\end{figure}

\begin{figure}[H]
    \centering
    \caption{Stances, donations, and lobbying --- full sample with $\log(\cdot + 1)$.}
    \includegraphics[width=\linewidth]{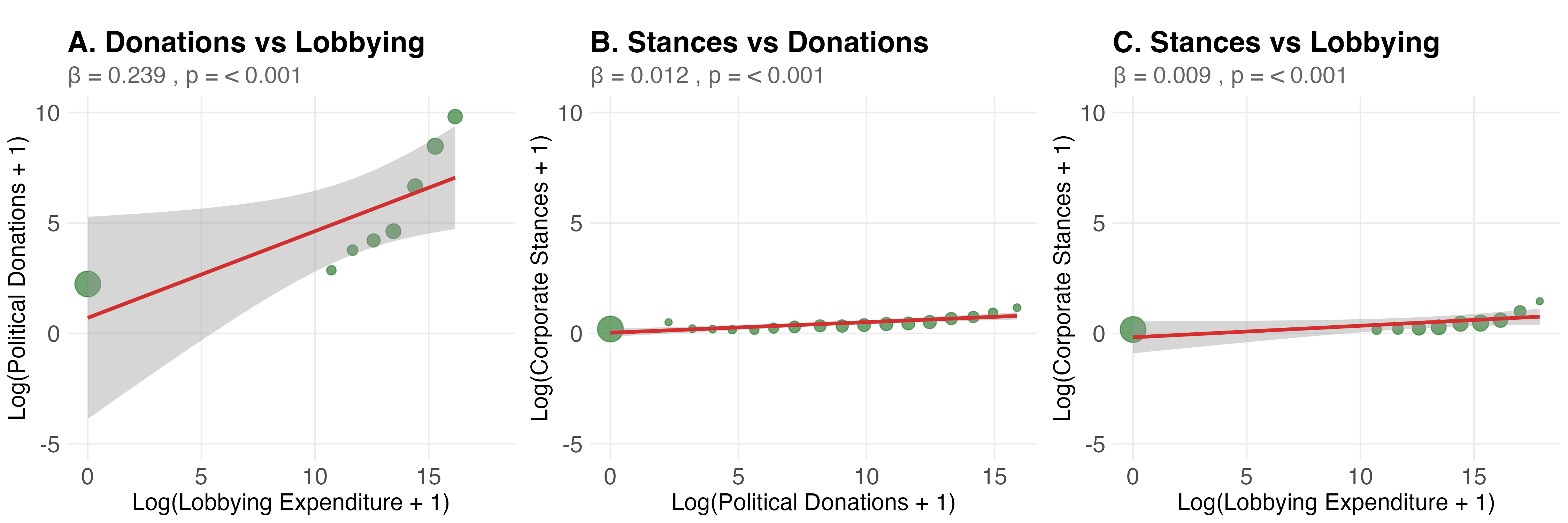}
    \label{fig:cpa-loglog-appendix}
    \begin{minipage}{\linewidth}
    \footnotesize
        \textit{Notes:} Same specification as Figure~\ref{fig:political_act} but estimated on the full sample of company-year observations (N = 13{,}748). To accommodate firm-years with zero stances, donations, or lobbying, both sides of each regression use the $\log(\cdot + 1)$ transformation. All other elements (controls, industry and year fixed effects, and firm-clustered standard errors) are identical to the main specification. The coefficients are 0.012 for stances on donations (p $<$ 0.001), 0.009 for stances on lobbying (p $<$ 0.001), and 0.239 for donations on lobbying (p $<$ 0.001).
    \end{minipage}
\end{figure}

\begin{figure}[H]
    \centering
    \caption{Effect of NGO Campaigns on Corporate Stance-Taking: Callaway and Sant'Anna Estimator}
    \begin{subfigure}[t]{0.49\linewidth}\centering
        \includegraphics[width=\linewidth]{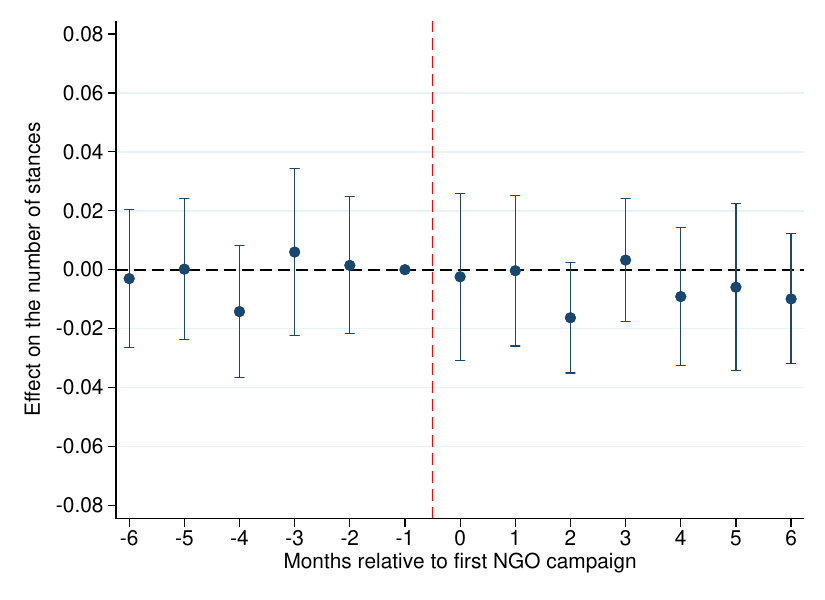}
        \caption{Any NGO campaign}
    \end{subfigure}\hfill
    \begin{subfigure}[t]{0.49\linewidth}\centering
        \includegraphics[width=\linewidth]{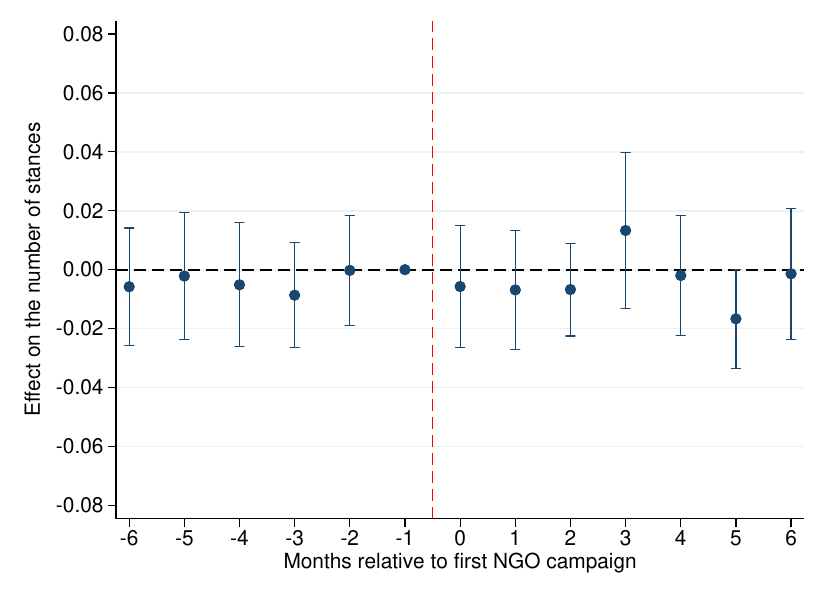}
        \caption{Environmental campaign}
    \end{subfigure}
    \label{fig:ngo_eventstudy_cs}
    \begin{minipage}{\linewidth}
    \footnotesize
        \textit{Notes:} We show point estimates and 95\% confidence intervals from the \citet{callaway_difference--differences_2021} estimator. We estimate the effect of the first NGO campaign (Panel a) or the first environmental campaign (Panel b) against a company on the monthly number of stances on all topics (Panel a) or the monthly number of liberal stances on climate change (Panel b), from six months before to six months after the campaign, relative to the month before the campaign. Companies never targeted by an NGO campaign serve as controls. Standard errors are clustered at the company level. $N = 490{,}893$ (Panel a) and $N = 483{,}197$ (Panel b) company-cohort-month observations.
    \end{minipage}
\end{figure}

\begin{figure}[H]
    \centering
    \caption{Effect of Topic-Specific peaks in News Articles and Google Searches with Alternative Fixed Effects}
    \begin{subfigure}[t]{0.49\linewidth}
        \centering
        \includegraphics[width=\linewidth]{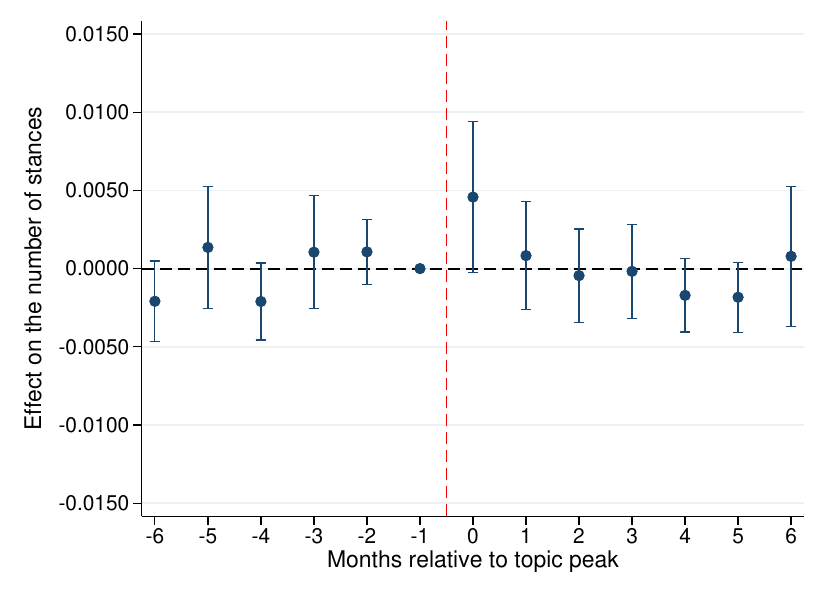}
        \caption{Factiva peaks}
    \end{subfigure}\hfill
    \begin{subfigure}[t]{0.49\linewidth}
        \centering
        \includegraphics[width=\linewidth]{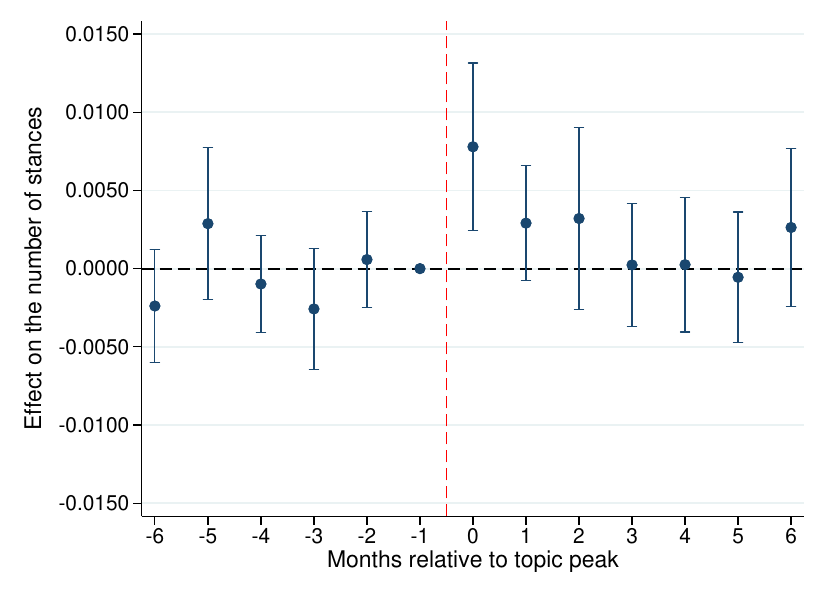}
        \caption{Google Trends peaks}
    \end{subfigure}
    \label{fig:peaks-iso6-specB}
    \begin{minipage}{0.97\linewidth}
    \footnotesize
        \textit{Notes:} We show point estimates and 95\% confidence intervals from specifications in the form of equation (\ref{eq:event_study_news}). We regress the monthly number of stances taken by a company on a given topic on dummy variables for the observation being recorded two to six months before a peak in news article (Panel a) or in Google searches (Panel b), as well as dummies for being recorded in the same month as a peak and one to six months after a peak. We also include dummies for being recorded seven months or more before or after a peak, respectively, and omit the dummy for being recorded one month before a peak. We control for company$\times$cohort, topic$\times$cohort, and cohort$\times$month fixed effects (rather than the company$\times$topic$\times$cohort and cohort$\times$month fixed effects used in Figure \ref{fig:peaks-iso6-specA}) as well as for company headcount and revenue. Standard errors are clustered two-way by peak and company. N = $13{,}006{,}580$ (Panel a) and $11{,}830{,}607$ (Panel b) company-month-topic-cohort observations.
    \end{minipage}
\end{figure}

\begin{figure}[H]
    \centering
    \caption{Stakeholders' Political Donations, Partisanship and Liberal Corporate Stance-Taking}
    \includegraphics[width=\linewidth]{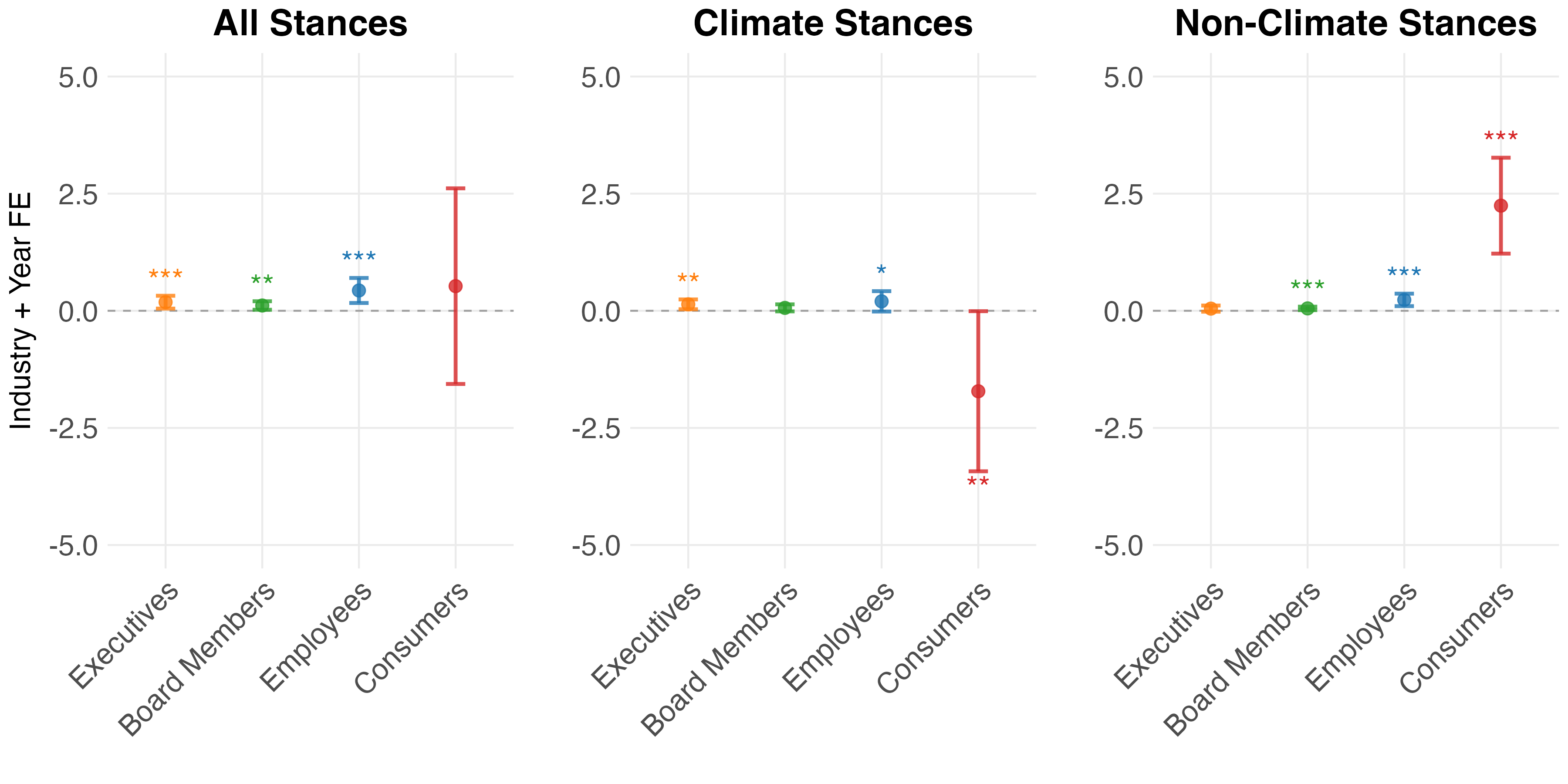}
    \label{fig:mv_stakeholders_donations}
    \begin{minipage}{\linewidth}
    \footnotesize
        \textit{Notes:} We show point estimates and 95\% confidence intervals from specifications in the form of equation \ref{eq:multivariate}. We regress the yearly number of liberal stances on the average CFscore of executives and board members who made campaign contributions (DIME), negatively coded so that higher values indicate more Democratic-leaning donations, the yearly share of Democratic employees, and the share of Democratic consumers (which does not vary over time). We control for industry and year fixed effects as well as company headcount and revenue. Standard errors are clustered at the company level. N = 13,748 company-year observations. Significance: $^{***}p<0.01$, $^{**}p<0.05$, $^{*}p<0.10$.
    \end{minipage}
\end{figure}

\begin{figure}[H]
    \centering
    \caption{Stakeholders Political Preferences and Liberal Corporate Stance-Taking}
    \includegraphics[width=1\linewidth]{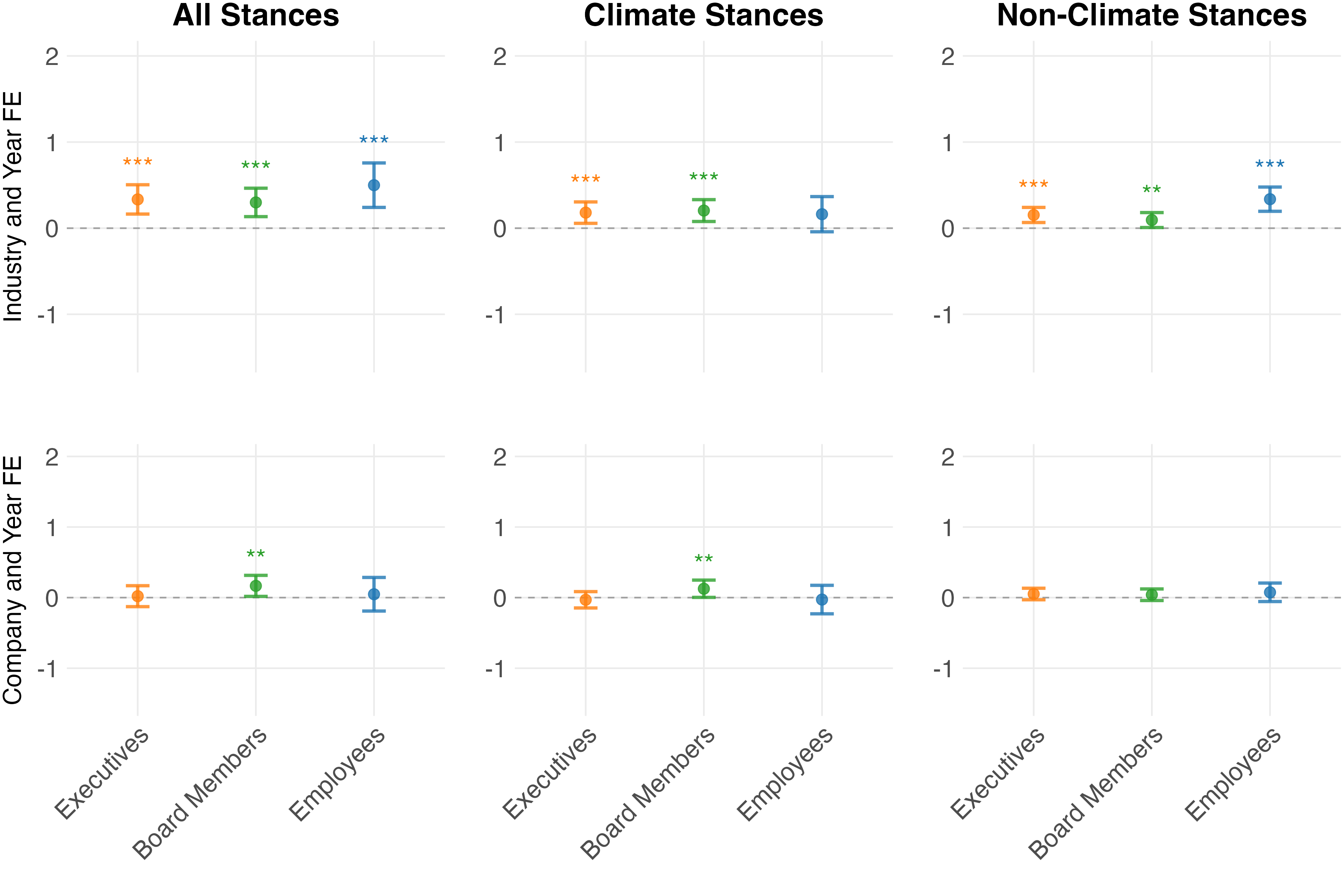}
    \label{fig:mv_stakeholders_politics}
    \begin{minipage}{\linewidth}
        \footnotesize
\textit{Notes:} We show point estimates and 95\% confidence intervals from specifications in the form of equation \ref{eq:multivariate}. We regress the yearly number of liberal stances on the yearly share of Democratic employees, executives, and board members. We control for company and year fixed effects as well as company headcount and revenue. Standard errors are clustered at the company level. N= 13,748 company-year observations. Significance: *** p < 0.01, ** p < 0.05, * p < 0.10.
    \end{minipage}
\end{figure}

    \begin{figure}[H]
    \centering
    \caption{Stakeholders Demographic Characteristics and Liberal Corporate Stance-Taking}
    \includegraphics[width=1\linewidth]{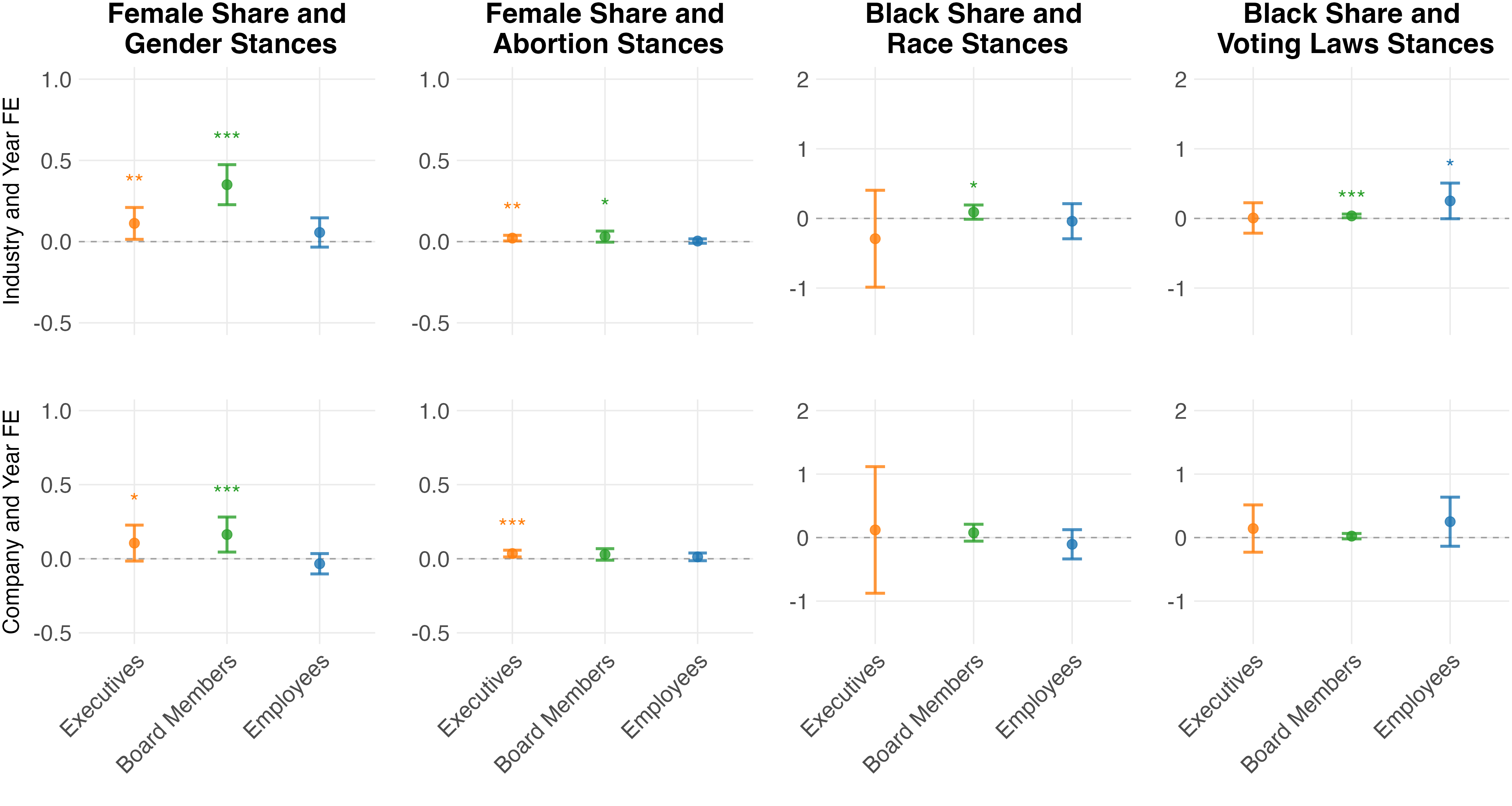}
    \label{fig:mv_stakeholders}
    \begin{minipage}{\linewidth}
        \footnotesize
        \textit{Notes:} We show point estimates and 95\% confidence intervals from specifications in the form of equation \ref{eq:multivariate}. We regress the yearly number of stances on gender issues or abortion on the yearly share of female employees, executives, and board members, and the yearly number of stances on race or voting laws on the corresponding shares of Black employees, executives, and board members. We control for company and year fixed effects as well as company headcount and revenue. Standard errors are clustered at the company level. N= 13,748 company-year observations. Significance: *** p < 0.01, ** p < 0.05, * p < 0.10.
    \end{minipage}
\end{figure}

\begin{figure}[H]
    \centering
    \caption{Gender Changes in Leadership and Stances on Gender and Abortion}
    \label{fig:es_arr_gender}
    \begin{subfigure}[t]{0.49\linewidth}\centering
        \includegraphics[width=\linewidth]{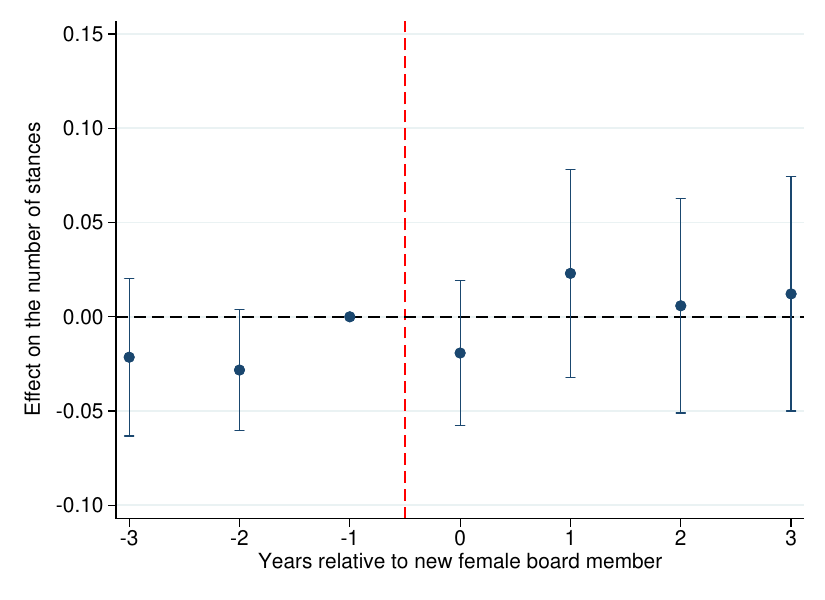}
        \caption{Board members: gender stances}
    \end{subfigure}\hfill
    \begin{subfigure}[t]{0.49\linewidth}\centering
        \includegraphics[width=\linewidth]{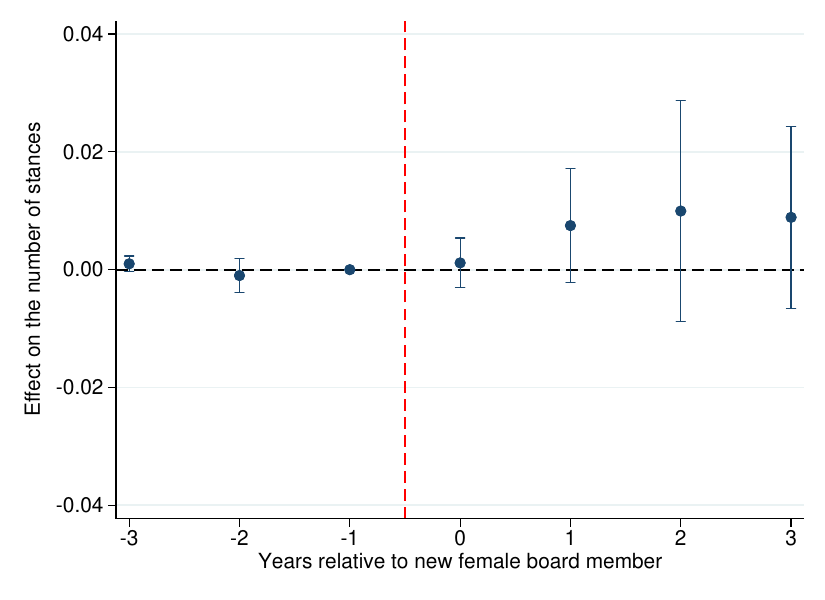}
        \caption{Board members: abortion stances}
    \end{subfigure}
    \par\medskip
    \begin{subfigure}[t]{0.49\linewidth}\centering
        \includegraphics[width=\linewidth]{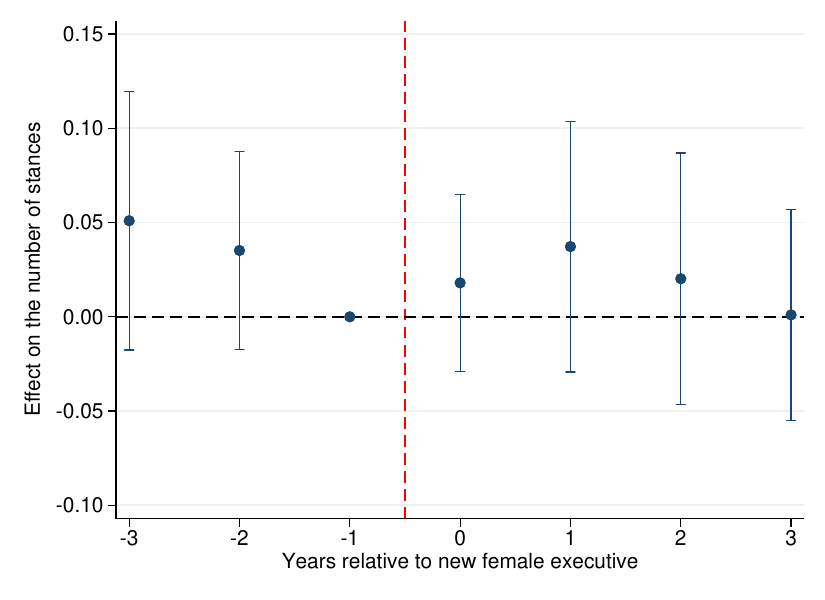}
        \caption{Executives: gender stances}
    \end{subfigure}\hfill
    \begin{subfigure}[t]{0.49\linewidth}\centering
        \includegraphics[width=\linewidth]{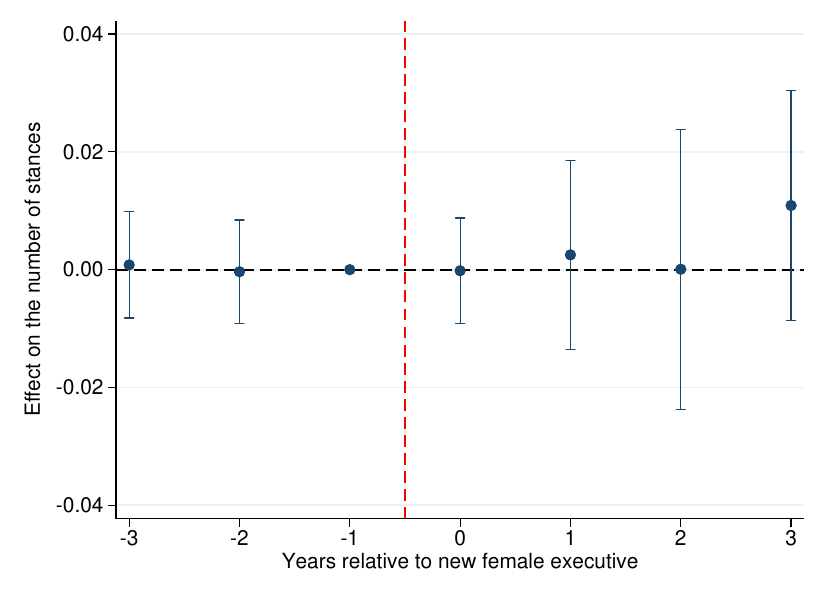}
        \caption{Executives: abortion stances}
    \end{subfigure}
    \begin{minipage}{\linewidth}
        \footnotesize
        \textit{Notes:} We show point estimates and 95\% confidence intervals from specifications in the form of equation~\eqref{eq:event_study_boardex}. The outcome is the yearly number of liberal stances on gender issues (Panels a and c) and abortion (Panels b and d). An event is the arrival of a new female board member (Panels a and b) or of a new female executive (Panels c and d) that increases the number of women on the board or in the executive team. The sample includes 363 events and 696 control companies (board members), and 297 events and 2,466 control companies (executives). $N = 17{,}803$ (board members) and $46{,}651$ (executives) company-year-cohort observations. Other notes as in Figure \ref{fig:es_part}.
    \end{minipage}
\end{figure}

\begin{figure}[H]
    \centering
    \caption{Racial Changes in Leadership and Stances on Race and Voting Laws}
    \label{fig:es_arr_race}
    \begin{subfigure}[t]{0.49\linewidth}\centering
        \includegraphics[width=\linewidth]{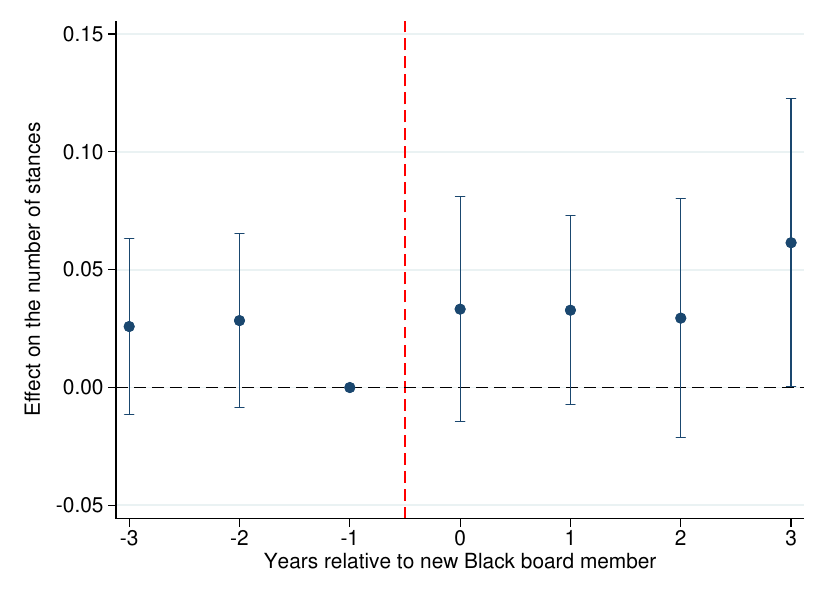}
        \caption{Stances on racial discrimination}
    \end{subfigure}\hfill
    \begin{subfigure}[t]{0.49\linewidth}\centering
        \includegraphics[width=\linewidth]{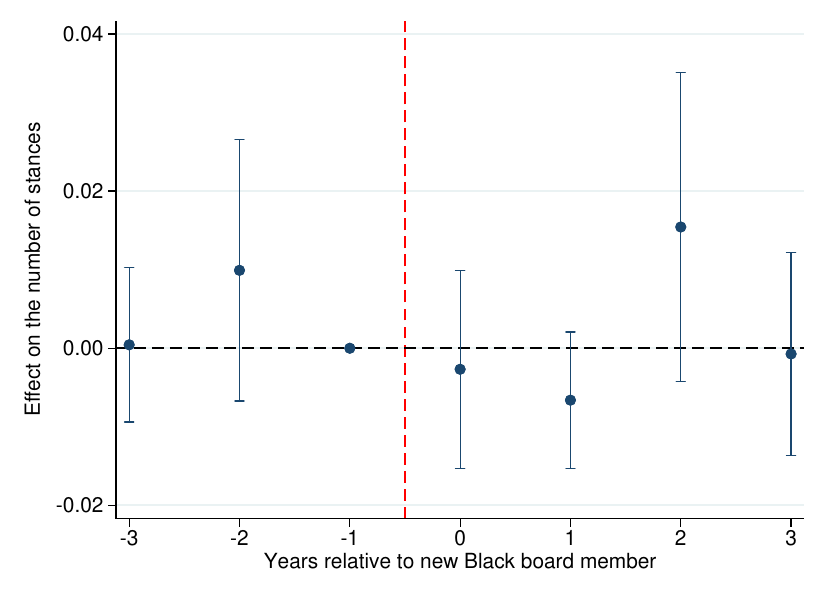}
        \caption{Stances on voting laws}
    \end{subfigure}
    \begin{minipage}{\linewidth}
        \footnotesize
        \textit{Notes:} We show point estimates and 95\% confidence intervals from specifications in the form of equation~\eqref{eq:event_study_boardex}. The outcome is the yearly number of liberal stances on racial discrimination (Panel a) and voting laws (Panel b). An event is the arrival of a new Black board member that increases the number of Black directors, in a year in which the ethnicity of all new and departing directors is known. The sample includes 231 events and 2,100 control companies. $N = 39{,}239$ company-year-cohort observations. Other notes as in Figure \ref{fig:es_part}.
    \end{minipage}
\end{figure}

\end{document}

%% file: preamble.tex
\usepackage{float}
\usepackage{amsmath}
\usepackage{amssymb}
\usepackage{graphicx}
\usepackage{setspace}
\usepackage[unicode=true,pdfusetitle,
 bookmarks=true,bookmarksnumbered=false,bookmarksopen=false,
 breaklinks=false,pdfborder={0 0 0},pdfborderstyle={},backref=false,colorlinks=false]
 {hyperref}
\makeatletter
\usepackage{adjustbox}
\usepackage{amsfonts}
\usepackage{amsmath,bbm}
\usepackage{textcomp}
\usepackage{appendix}
\usepackage{achicago}
\usepackage{bm}
\usepackage{bbm}
\usepackage{booktabs}
\usepackage{changepage}
\usepackage{dcolumn}
\usepackage[bottom]{footmisc}
\usepackage{fancyhdr}
\usepackage[letterpaper,top=1in,right=1in,left=1in,bottom=1in]{geometry}
\usepackage{graphicx}
\usepackage{hyperref}
\usepackage{pdflscape}
\usepackage{sectsty}
\usepackage{secdot}
\usepackage{tabularx}
\usepackage{threeparttablex}
\usepackage{threeparttable}
\usepackage{verbatim}
\usepackage{caption}
\usepackage{subcaption}
\usepackage{tikz}
\usetikzlibrary{shapes.geometric, arrows}
\usetikzlibrary{positioning}
\usetikzlibrary{calc,decorations.markings}
\usepackage{titling}
\newcommand*{\centerfloat}{\parindent \z@
  \leftskip \z@ \@plus 1fil \@minus \textwidth
  \rightskip\leftskip
  \parfillskip \z@skip}

\usepackage{siunitx}
\usepackage[T1]{fontenc}
\makeatother
\usepackage{threeparttable}
\usepackage{multirow}

\usepackage{bibentry}
\usepackage{achicago}
\usepackage{bm}
\usepackage{bbm}
\usepackage{natbib}
\usepackage{har2nat}

\usepackage{booktabs}
\usepackage{longtable}
\usepackage{pdflscape}
\usepackage{array}

%% file: tables/comparison_chatgpt.tex
\begin{landscape}
\scriptsize
\setlength{\tabcolsep}{4pt}
\renewcommand{\arraystretch}{1.15}

\begin{longtable}{p{1.2cm} p{1.0cm} p{2.2cm} p{5.3cm} p{2.0cm} p{7.0cm}}
\caption{Comparison with Stances provided by ChatGPT}\label{tab:comparison_chatgpt} \\

\toprule
\textbf{S\&P 500 Code} & \textbf{Year} & \textbf{Topic} & \textbf{Case described by GPT} & \textbf{In dataset} & \textbf{Summary of article in our dataset} \\
\midrule
\endfirsthead

\toprule
\textbf{S\&P 500 Code} & \textbf{Year} & \textbf{Topic} & \textbf{Case described by GPT} & \textbf{In dataset} & \textbf{Summary  of article in our dataset} \\
\midrule
\endhead

\midrule
\multicolumn{6}{r}{\emph{Continued on next page}} \\
\endfoot

\bottomrule
\endlastfoot

MSFT & 2022 & Abortion & Added travel coverage for abortion-related care after Dobbs decision. & Yes & Microsoft Corp announced it will cover travel expenses for U.S. employees seeking abortion and gender-affirming care, joining other companies in support of reproductive health access amid increasing state-level abortion restrictions. \\

JPM & 2022 & Abortion & Reimbursed employee travel for abortion access. & Yes & JPMorgan Chase \& Co. is considering extending abortion benefits for employees, including travel coverage, as companies respond to potential changes in U.S. abortion rights anticipated from a Supreme Court ruling. \\

AMZN & 2022 & Abortion & Reimbursed out-of-state abortion travel. & Yes & Amazon has announced it will reimburse U.S. employees up to \$4,000 in travel expenses for non-life threatening medical treatments, including abortion, as part of a broader effort to support access to reproductive health services amid tightening restrictions in various states. \\

META & 2022 & Abortion & Expanded benefits to cover abortion travel. & Yes & Following the U.S. Supreme Court's decision to overturn Roe v. Wade, companies like Meta Platforms, Inc. announced they would reimburse employees for travel expenses related to obtaining abortion services, while also limiting internal discussions on the topic. \\

NFLX & 2022 & Abortion & Updated health plan to include abortion travel. & No & \\

SBUX & 2022 & Abortion & Offered travel reimbursement for abortion services. & Yes & Starbucks has announced that it will reimburse U.S. employees for travel expenses related to abortions and gender-affirming procedures when such services are not available within 100 miles of their home. \\

TSLA & 2022 & Abortion & Covered travel expenses for medical care including abortion. & Yes & Tesla, Inc. will cover transportation costs for employees seeking out-of-state abortions, joining other companies in providing support in response to restrictive abortion laws in various states. \\

IBM & 2017 & Climate & Reaffirmed support for Paris Agreement post U.S. withdrawal. & Yes & Major U.S. companies, including IBM, announced their decision to remain on a presidential advisory council despite President Trump's withdrawal from the Paris climate accord, emphasizing the importance of engaging with the administration to influence climate policies. \\

AAPL & 2017 & Climate & Publicly backed Paris Agreement; 100\% renewable operations push. & Yes & Apple CEO Tim Cook declared on Twitter that the decision to withdraw from the Paris Agreement was incorrect and reaffirmed the company's commitment to combating climate change. \\

AMZN & 2019 & Climate & Launched The Climate Pledge to reach net-zero by 2040. & Yes & Amazon.com, Inc. announced a \$2 billion Climate Pledge Fund to invest in sustainable technologies, although the fund's small scale may limit its impact compared to the company's broader commitments to renewable energy and electric vehicles. \\

F & 2019 & Climate & Backed California's stricter emissions standards; invested in EVs. & Yes & Ford Motor Co. and other automakers reached a voluntary agreement with California to uphold stricter fuel efficiency standards, defying President Trump's efforts to roll back Obama-era regulations aimed at combating climate change. \\

MSFT & 2020 & Climate & Pledged to be carbon negative by 2030. & Yes & Microsoft Corp. is among nine leading companies launching the ``Transform to Net Zero'' initiative, aimed at accelerating the transition to a net zero global economy by providing research and actionable roadmaps for businesses to achieve net zero emissions by 2050. \\

BLK & 2020 & Climate & Declared climate risk investment risk; shifted portfolios. & Yes & BlackRock's commitment to prioritize sustainability in its investment strategy marks a significant shift in capital markets, as it aligns with the growing necessity to address climate risks and may influence broader investment management practices. \\

GM & 2021 & Climate & Targeted all-electric light-duty lineup by 2035. & Yes & General Motors announced its ambitious goal to eliminate emissions from most of its light-duty vehicles by 2035, aiming for carbon neutrality by 2040 in response to climate change concerns and industry shifts towards electric vehicles. \\

CRM & 2015 & Gender & Conducted pay-equity audits; adjusted compensation. & No &  \\

SBUX & 2018 & Gender & Announced U.S. pay equity across gender/race; shared methodology. & Yes & Starbucks Corp. has taken a leadership role in gender and racial pay equity by publicly disclosing its median pay gap data, contrasting with Microsoft, which is facing shareholder pressure to enhance transparency on the issue. \\

SBUX & 2013 & Guns & Asked customers not to bring firearms openly in stores. & Yes & Starbucks CEO Howard Schultz, in an open letter to customers, urged them to refrain from bringing guns into Starbucks stores, reflecting the company's neutral stance on gun control while emphasizing the aim to create a comfortable environment for all patrons. \\

C & 2018 & Guns & Set client restrictions, including age 21 and background checks, for firearm sales. & Yes & Citigroup Inc. announced new policies requiring its retail clients to enforce stricter gun sale regulations, including background checks and age limits, in response to a recent school shooting. \\

BAC & 2018 & Guns & Limited financing to makers of military-style firearms for civilians. & Yes & Bank of America Corp. announced plans to halt lending to manufacturers of military-style firearms for civilians, following the shooting at a Florida high school, marking a significant move in the ongoing national debate over gun control. \\

KR & 2018 & Guns & Requested no open carry; raised firearm purchase age to 21. & Yes & Kroger has announced it will raise the minimum age for purchasing firearms to 21 in response to the Parkland school shooting, joining other retailers in implementing stricter gun sale policies. \\

CRM & 2019 & Guns & Restricted use of commerce tools for certain gun sales. & Yes & Salesforce, Inc. has implemented a policy prohibiting the sale of military-style firearms through its software and has established a framework for ethical decision-making on social issues, reflecting its commitment to responsible corporate practices amid scrutiny over its business relationships. \\

WMT & 2019 & Guns & Ended certain ammo sales; asked customers not to open carry. & Yes & Walmart announced it will stop selling certain ammunition for handguns and short-barrel rifles and is requesting that customers refrain from openly carrying firearms in stores, following recent mass shootings, including one at a Walmart location. \\

AAPL & 2017 & Immigration & Publicly supported DACA; filed briefs and advocated for Dreamers. & Yes & Apple CEO Tim Cook expresses support for the company's immigrant employees, advocating for the preservation of the Deferred Action for Childhood Arrivals program as President Trump considers its elimination. \\

MSFT & 2017 & Immigration & Led tech opposition to travel ban; backed DACA employees. & Yes & Leaders from major tech companies, including Microsoft Corp., expressed strong opposition to President Trump's travel restrictions targeting citizens from seven Muslim-majority countries, highlighting concerns over its impact on their workforce and the importance of immigration to the industry. \\

CRM & 2015 & LGBTQ & Pressured Indiana over RFRA; curtailed events/investment. & Yes & Tech company leaders, including Salesforce CEO Marc Benioff, urge Indiana Governor Mike Pence not to sign the Religious Freedom Restoration Act, citing concerns it could lead to discrimination against LGBTQ+ individuals and negatively affect the state's economic appeal. \\

PYPL & 2016 & LGBTQ & Canceled Charlotte expansion over NC ``HB2'' bathroom law. & Yes & PayPal Holdings, Inc. canceled plans to open a global operations center in Charlotte, North Carolina, in response to the state's controversial law restricting transgender individuals' bathroom access, highlighting the company's commitment to LGBTQ+ rights. \\

DIS & 2022 & LGBTQ & Opposed Florida's ``Parental Rights in Education'' law; paused Florida political donations. & Yes & The article discusses Florida Governor Ron DeSantis's push to eliminate Walt Disney Co.'s special tax district in response to the company's opposition to the state's ``Don't Say Gay'' legislation, which restricts discussions of gender identity and sexual orientation in schools. \\

AMZN & 2018 & MinWage & Raised company minimum wage to \$15 and lobbied others to follow. & Yes & Amazon.com, Inc. has announced a raise in its minimum wage to \$15 for all U.S. employees, effective November 1, 2018, benefiting over 350,000 workers including full-time, part-time, temporary, and seasonal staff. \\

TGT & 2020 & MinWage & Raised companywide minimum wage to \$15; higher in costly markets. & Yes & Target Corp. announced it will permanently raise its starting minimum wage to \$15 an hour beginning July 5, 2020, while also offering additional benefits for employees during the pandemic. \\

NKE & 2018 & Race & Featured Colin Kaepernick in campaign supporting racial justice. & Yes & Nike's ad campaign featuring Colin Kaepernick as a symbol of racial justice and the Black Lives Matter movement has sparked both support and backlash, highlighting the company's willingness to engage with divisive social issues to connect with millennials while risking alienation of conservative customers. \\

NFLX & 2020 & Race & Committed \$100M to support Black-led banks and institutions. & Yes & Netflix, Inc. announced its commitment to invest \$100 million in financial institutions serving Black communities to enhance their lending capabilities, following nationwide protests for racial justice. \\

WMT & 2020 & Race & Announced \$100M Center for Racial Equity. & No &  \\

DAL & 2021 & Voting & Publicly criticized Georgia's voting law. & Yes & Delta Air Lines, along with other corporations, is facing backlash for its criticism of voting restrictions in Georgia, while similar legislative efforts in Texas have sparked opposition from American Airlines and raised concerns among voting rights advocates. \\

KO & 2021 & Voting & Spoke out against Georgia's voting law. & Yes & Civil rights groups are urging companies like Coca-Cola Co. and Delta Air Lines to take a stand against new voting restrictions in Georgia \\

AAL & 2021 & Voting & Opposed Texas voting-restriction proposals. & Yes & American Airlines strongly opposes a Texas voting restriction bill that aims to limit access to voting, emphasizing the importance of protecting voting rights for its employees and customers. \\

\end{longtable}
\end{landscape}

%% file: tables/peak_event_tables_factiva.tex
{
\begin{landscape}
\footnotesize
\setlength{\LTleft}{0pt}
\setlength{\LTright}{0pt}
\renewcommand{\arraystretch}{1.12}

\begin{longtable}{p{1.8cm}p{3cm}p{1.8cm}p{6.8cm}p{6.8cm}}
\caption{Peaks in News Articles from Factiva}\label{tab:peaks_factiva} \\
\toprule
Topic & Starting month & Duration (in months) & Event from Gemini & Event from GPT \\
\midrule
\endfirsthead
\toprule
Topic & Starting month & Duration (in months) & Event from Gemini & Event from GPT \\
\midrule
\endhead
\midrule
\multicolumn{5}{r}{\emph{Continued on next page}} \\
\midrule
\endfoot
\bottomrule
\endlastfoot
Abortion & September 2015 & 1 & Release of heavily edited videos by Center for Medical Progress alleging Planned Parenthood sold fetal tissue (summer/Sept 2015). & Release of undercover videos targeting Planned Parenthood, which reignited a national fight over abortion and federal funding. \\
Abortion & May 2019 & 1 & Several states (e.g., Alabama, Georgia, Missouri) pass near-total abortion bans or fetal heartbeat bills. & Wave of state abortion bans in 2019, highlighted by Alabama's near-total abortion ban and Georgia's ``heartbeat'' law. \\
Abortion & May 2022 & 7 & Leak of the Supreme Court draft opinion in \emph{Dobbs v. Jackson Women's Health Organization}, which would overturn \emph{Roe v. Wade}. & Leak of the Supreme Court's draft \emph{Dobbs} opinion signaling the likely end of \emph{Roe v. Wade}. \\
Abortion & November 2024 & 1 & 2024 Presidential Election featuring multiple state abortion ballot measures driving high turnout post-Dobbs. & 2024 election in which abortion rights ballot measures were on the ballot in 10 states and remained a central national issue. \\
Climate & December 2009 & 1 & COP15 UN Climate Change Conference in Copenhagen and the ``Climategate'' email leak controversy. & U.N. climate summit in Copenhagen (COP15), which drew major U.S. attention to international climate policy. \\
Climate & November 2016 & 1 & Donald Trump elected president, raising fears of US withdrawal from the Paris Agreement (which took effect Nov 4, 2016). & Paris climate agreement entered into force in November 2016, just as Donald Trump's election raised doubts about U.S. climate commitments. \\
Climate & June 2017 & 1 & Trump announces the US withdrawal from the Paris climate agreement. & Trump announced that the United States would withdraw from the Paris climate agreement. \\
Gender & March 2015 & 1 & National debate on the gender pay gap fueled by Patricia Arquette's Oscar speech (late Feb), and Ellen Pao's landmark gender discrimination trial in Silicon Valley. & Renewed national debate over campus sexual assault and gender inequality in early 2015, fueled by high-profile cases and activism. \\
Gender & January 2017 & 1 & The Women's March on Washington and worldwide protests the day after Trump's inauguration. & Women's Marches around Trump's inauguration made gender equality and women's rights a major national issue. \\
Gender & January 2018 & 1 & The 2018 Women's March and the height of the \#MeToo movement. & The Time's Up movement and the 2018 Women's March kept sexual harassment and gender inequality at the center of national attention. \\
Guns & January 2013 & 4 & President Obama unveils sweeping gun control proposals in response to the Sandy Hook shooting. & Obama unveiled major gun-control proposals after Sandy Hook, including universal background checks and assault-weapons restrictions. \\
Guns & October 2015 & 4 & The Umpqua Community College shooting in Oregon. & Umpqua Community College shooting in Oregon renewed national debate over mass shootings and gun laws. \\
Guns & June 2016 & 1 & The Pulse nightclub shooting in Orlando, Florida. & Pulse nightclub shooting in Orlando, one of the deadliest mass shootings in U.S. history. \\
Guns & February 2018 & 2 & The Stoneman Douglas High School shooting in Parkland, Florida. & Marjory Stoneman Douglas High School shooting in Parkland, Florida, and the student-led March for Our Lives movement that followed. \\
Guns & August 2019 & 1 & Back-to-back mass shootings in El Paso, Texas, and Dayton, Ohio. & Mass shootings in El Paso, Texas, and Dayton, Ohio, reignited national debate over gun violence in August 2019. \\
Guns & June 2022 & 1 & Passage of the Bipartisan Safer Communities Act following the Uvalde school shooting in May. & Robb Elementary School shooting in Uvalde, Texas, followed by passage of the Bipartisan Safer Communities Act. \\
Immigration & May 2010 & 3 & Massive protests and national debate over Arizona's strict anti-illegal immigration law, SB 1070 (signed April 2010). & Arizona's SB 1070 immigration law and the national legal and political backlash it sparked in spring 2010. \\
Immigration & November 2014 & 1 & President Obama announces executive actions on immigration, including the expansion of DACA and creation of DAPA. & Obama announced executive actions on immigration in November 2014, including deportation relief for millions of undocumented immigrants. \\
Immigration & September 2017 & 5 & The Trump administration announces the phase-out of the Deferred Action for Childhood Arrivals (DACA) program. & Trump administration announced the end of DACA in September 2017, triggering national protests and legal battles. \\
Immigration & January 2019 & 1 & The longest federal government shutdown in US history over funding for Trump's border wall. & Federal government shutdown over Trump's demand for border-wall funding dominated immigration politics in January 2019. \\
Immigration & February 2025 & 1 & Implementation and escalation of the Trump administration's border closures and mass deportation operations. & Trump administration's early-2025 immigration crackdown, including the Laken Riley Act and moves to end protections for some migrants, dominated attention. \\
LGBTQ & March 2013 & 4 & Supreme Court hears arguments on the Defense of Marriage Act (DOMA) and Prop 8; HRC's red equals sign goes viral. & Supreme Court arguments in the DOMA and Proposition 8 cases brought same-sex marriage to the center of national attention. \\
LGBTQ & June 2016 & 1 & The Pulse nightclub shooting in Orlando, Florida (an anti-LGBTQ hate crime). & Pulse nightclub shooting in Orlando focused national attention on violence against the LGBTQ community. \\
LGBTQ & June 2019 & 1 & WorldPride in NYC and the 50th anniversary of the Stonewall Riots. & Fiftieth anniversary of the Stonewall uprising and WorldPride 2019 in New York renewed nationwide attention to LGBTQ rights. \\
LGBTQ & June 2020 & 1 & Supreme Court rules in \emph{Bostock v. Clayton County} that the Civil Rights Act protects gay and transgender workers. & Supreme Court's \emph{Bostock} decision held that federal employment law protects gay and transgender workers from discrimination. \\
LGBTQ & June 2021 & 1 & Pride Month amid a record-breaking wave of anti-transgender bills passed in conservative state legislatures. & Pride Month in 2021 unfolded amid a nationwide wave of state bills targeting transgender youth and LGBTQ rights. \\
LGBTQ & June 2023 & 1 & Pride Month marked by intense conservative boycotts of brands (like Target and Bud Light) over LGBTQ+ marketing, and the \emph{303 Creative} SCOTUS ruling. & Supreme Court's \emph{303 Creative} decision in June 2023 reignited debate over LGBTQ rights and religious-liberty exemptions. \\
Minimum wage & January 2014 & 1 & President Obama signs an executive order raising the minimum wage for federal contractors to \$10.10. & Obama called for raising the federal minimum wage to \$10.10 in his 2014 State of the Union address. \\
Minimum wage & October 2014 & 2 & Lead up to midterm elections where multiple states had successful ballot measures to increase the minimum wage. & Campaigns over 2014 state minimum-wage ballot measures made wage policy a prominent issue ahead of the midterm elections. \\
Minimum wage & July 2015 & 1 & New York's wage board recommends a \$15 minimum wage for fast-food workers, a major victory for the Fight for \$15 movement. & New York's wage board recommended a \$15 minimum wage for fast-food workers, marking a major Fight for \$15 breakthrough. \\
Minimum wage & April 2016 & 1 & Governors of California and New York sign landmark legislation to gradually raise their statewide minimum wages to \$15 an hour. & California and New York adopted plans in spring 2016 to raise their minimum wages to \$15 an hour. \\
Minimum wage & February 2021 & 2 & The Senate Parliamentarian rules that a \$15 federal minimum wage increase cannot be included in the COVID-19 reconciliation bill. & Fight over including a \$15 federal minimum wage in the COVID relief bill peaked in February 2021. \\
Race & July 2016 & 1 & High-profile police shootings of Alton Sterling and Philando Castile, followed by the shooting of Dallas police officers. & Police killings of Alton Sterling and Philando Castile in July 2016 sparked nationwide protests over race and policing. \\
Race & June 2020 & 4 & Massive global protests against systemic racism and police brutality following the murder of George Floyd. & George Floyd's killing and the nationwide Black Lives Matter protests of June 2020 dominated public attention to race. \\
Voting & October 2008 & 1 & Lead up to the historic 2008 Presidential Election between Barack Obama and John McCain. & Final stretch of the 2008 presidential campaign, with intense national attention to turnout, registration, and voting rules. \\
Voting & June 2013 & 1 & Supreme Court strikes down a key provision of the Voting Rights Act of 1965 in \emph{Shelby County v. Holder}. & Supreme Court's \emph{Shelby County v. Holder} decision weakened the Voting Rights Act in June 2013. \\
Voting & March 2015 & 1 & 50th Anniversary of the Selma to Montgomery marches and the Voting Rights Act of 1965. & Fiftieth anniversary of the Selma marches renewed national attention to voting rights in March 2015. \\
Voting & November 2018 & 1 & The 2018 Midterm Elections, featuring high turnout and significant battles over voter suppression (e.g., in Georgia). & 2018 midterm elections produced exceptionally high turnout and intense attention to voting access and ballot counting. \\
Voting & January 2022 & 1 & Senate failure to pass the John Lewis Voting Rights Advancement Act and Freedom to Vote Act due to the filibuster. & Senate fight over federal voting-rights legislation peaked in January 2022 as Democrats failed to change filibuster rules. \\
Voting & July 2024 & 4 & President Biden drops out of the 2024 presidential race, profoundly altering the election landscape. & Biden's withdrawal from the 2024 presidential race and Kamala Harris's emergence as the Democratic nominee transformed the election campaign. \\
\end{longtable}
\end{landscape}
}

%% file: tables/peak_event_tables_google.tex
{
\begin{landscape}
\footnotesize
\setlength{\LTleft}{0pt}
\setlength{\LTright}{0pt}
\renewcommand{\arraystretch}{1.12}

\begin{longtable}{p{1.8cm}p{3cm}p{1.8cm}p{6.8cm}p{6.8cm}}\caption{Peaks in Google Searches from Google Trends}\label{tab:peaks_google} \\
\toprule
Topic & Starting month & Duration (in months) & Event from Gemini & Event from GPT \\
\midrule
\endfirsthead
\toprule
Topic & Starting month & Duration (in months) & Event from Gemini & Event from GPT \\
\midrule
\endhead
\midrule
\multicolumn{5}{r}{\emph{Continued on next page}} \\
\midrule
\endfoot
\bottomrule
\endlastfoot
Abortion & May 2019 & 1 & Several states (e.g., Alabama, Georgia, Missouri) pass near-total abortion bans or fetal heartbeat bills. & Wave of state abortion bans in 2019, highlighted by Alabama's near-total abortion ban and Georgia's ``heartbeat'' law. \\
Abortion & May 2022 & 3 & Leak of the Supreme Court draft opinion in \emph{Dobbs v. Jackson Women's Health Organization}, which would overturn \emph{Roe v. Wade}. & Leak of the Supreme Court's draft \emph{Dobbs} opinion signaling the likely end of \emph{Roe v. Wade}. \\
Abortion & November 2024 & 1 & 2024 Presidential Election featuring multiple state abortion ballot measures driving high turnout post-Dobbs. & 2024 election in which abortion rights ballot measures were on the ballot in 10 states and remained a central national issue. \\
Climate & January 2017 & 3 & Trump's inauguration and subsequent removal of climate change pages from the White House website, plus gag orders on EPA. & Donald Trump took office promising to roll back Obama-era climate rules and revive fossil-fuel production. \\
Climate & February 2019 & 1 & Rep. Alexandria Ocasio-Cortez and Sen. Ed Markey officially introduce the Green New Deal resolution. & Introduction of the Green New Deal resolution by Alexandria Ocasio-Cortez and Ed Markey in February 2019. \\
Climate & February 2021 & 1 & The US officially rejoins the Paris Agreement under President Biden; Texas suffers severe grid failure from a winter storm, sparking climate debates. & Severe Texas winter storm and power-grid failure intensified the national debate over climate resilience and extreme weather. \\
Climate & April 2023 & 1 & Biden administration proposes strict new EPA vehicle emissions limits and faces continued backlash over the March approval of the Willow Project. & Release of the IPCC synthesis report in March 2023 renewed U.S. attention to climate risks and the need for rapid emissions cuts. \\
Climate & April 2024 & 1 & Biden administration announces sweeping new EPA power plant rules and Earth Day climate actions. & Biden administration finalized major EPA climate rules in spring 2024, including tighter power-plant and vehicle-emissions standards. \\
Gender & September 2020 & 1 & Death of Supreme Court Justice and feminist icon Ruth Bader Ginsburg. & Death of Supreme Court Justice Ruth Bader Ginsburg in September 2020 focused national attention on women's rights and gender equality. \\
Guns & December 2012 & 2 & The Sandy Hook Elementary School shooting in Newtown, Connecticut. & Sandy Hook Elementary School shooting in Newtown, Connecticut, which killed 20 children and 6 adults. \\
Guns & November 2016 & 1 & 2016 US Presidential Election, in which gun control and Second Amendment rights were heavily debated. & 2016 presidential election, in which gun rights and gun control were major campaign issues. \\
Guns & February 2018 & 2 & The Stoneman Douglas High School shooting in Parkland, Florida. & Marjory Stoneman Douglas High School shooting in Parkland, Florida, and the student-led March for Our Lives movement that followed. \\
Guns & August 2019 & 5 & Back-to-back mass shootings in El Paso, Texas, and Dayton, Ohio. & Mass shootings in El Paso, Texas, and Dayton, Ohio, reignited national debate over gun violence in August 2019. \\
Immigration & May 2010 & 1 & Massive protests and national debate over Arizona's strict anti-illegal immigration law, SB 1070 (signed April 2010). & Arizona's SB 1070 immigration law and the national legal and political backlash it sparked in spring 2010. \\
Immigration & November 2014 & 1 & President Obama announces executive actions on immigration, including the expansion of DACA and creation of DAPA. & Obama announced executive actions on immigration in November 2014, including deportation relief for millions of undocumented immigrants. \\
Immigration & November 2016 & 4 & Election of Donald Trump, who campaigned heavily on building a border wall and restricting immigration. & Trump won the 2016 election after running a campaign centered on border security, a wall, and immigration restriction. \\
Immigration & June 2018 & 1 & Public outrage peaks over the Trump administration's ``zero tolerance'' family separation policy at the US-Mexico border. & Trump administration's ``zero tolerance'' policy and family separations at the U.S.-Mexico border drew intense national attention. \\
Immigration & May 2023 & 1 & Expiration of Title 42 border policy (May 11), leading to a surge in border crossings and massive media coverage. & End of the Title 42 border policy in May 2023 triggered intense attention to migration and border enforcement. \\
Immigration & January 2025 & 1 & Inauguration of Donald Trump for his second term, launching strict new border controls and mass deportation policies. & Trump returned to office in January 2025 and immediately issued sweeping immigration executive orders, including a border emergency declaration. \\
LGBTQ & June 2016 & 6 & The Pulse nightclub shooting in Orlando, Florida (an anti-LGBTQ hate crime). & Pulse nightclub shooting in Orlando focused national attention on violence against the LGBTQ community. \\
LGBTQ & June 2019 & 1 & WorldPride in NYC and the 50th anniversary of the Stonewall Riots. & Fiftieth anniversary of the Stonewall uprising and WorldPride 2019 in New York renewed nationwide attention to LGBTQ rights. \\
LGBTQ & June 2020 & 1 & Supreme Court rules in \emph{Bostock v. Clayton County} that the Civil Rights Act protects gay and transgender workers. & Supreme Court's \emph{Bostock} decision held that federal employment law protects gay and transgender workers from discrimination. \\
LGBTQ & June 2021 & 1 & Pride Month amid a record-breaking wave of anti-transgender bills passed in conservative state legislatures. & Pride Month in 2021 unfolded amid a nationwide wave of state bills targeting transgender youth and LGBTQ rights. \\
LGBTQ & June 2022 & 1 & Pride Month overshadowed by fears of \emph{Obergefell} being overturned after the \emph{Dobbs} ruling, and Florida's ``Don't Say Gay'' law debates. & Pride Month 2022 took place amid backlash to Florida's ``Don't Say Gay'' law and a broader wave of state-level anti-LGBTQ measures. \\
LGBTQ & June 2023 & 1 & Pride Month marked by intense conservative boycotts of brands (like Target and Bud Light) over LGBTQ+ marketing, and the \emph{303 Creative} SCOTUS ruling. & Supreme Court's \emph{303 Creative} decision in June 2023 reignited debate over LGBTQ rights and religious-liberty exemptions. \\
Minimum wage & January 2014 & 1 & President Obama signs an executive order raising the minimum wage for federal contractors to \$10.10. & Obama called for raising the federal minimum wage to \$10.10 in his 2014 State of the Union address. \\
Minimum wage & April 2016 & 1 & Governors of California and New York sign landmark legislation to gradually raise their statewide minimum wages to \$15 an hour. & California and New York adopted plans in spring 2016 to raise their minimum wages to \$15 an hour. \\
Minimum wage & January 2017 & 1 & Minimum wage increases take effect in 19 states at the start of the year. & Minimum-wage increases took effect in many states and cities on January 1, 2017, drawing broad national attention. \\
Minimum wage & January 2018 & 1 & Minimum wage increases take effect in 18 states at the start of the year. & Minimum-wage increases took effect across numerous states and localities on January 1, 2018. \\
Minimum wage & January 2021 & 2 & President Biden proposes raising the federal minimum wage to \$15 in his COVID-19 relief package; wage bumps take effect in multiple states. & Biden entered office backing a \$15 federal minimum wage as the issue moved into the center of the early 2021 policy agenda. \\
Minimum wage & January 2024 & 1 & Minimum wage increases take effect in 22 states, many reaching or surpassing the \$15 threshold. & Twenty-two states raised their minimum wages on January 1, 2024, bringing renewed national attention to wage floors. \\
Race & November 2016 & 1 & Election of Donald Trump, heavily polarizing the nation along racial lines. & Donald Trump's election in November 2016 intensified national debate over racism, white identity politics, and hate incidents. \\
Race & June 2020 & 1 & Massive global protests against systemic racism and police brutality following the murder of George Floyd. & George Floyd's killing and the nationwide Black Lives Matter protests of June 2020 dominated public attention to race. \\
Voting & October 2008 & 1 & Lead up to the historic 2008 Presidential Election between Barack Obama and John McCain. & Final stretch of the 2008 presidential campaign, with intense national attention to turnout, registration, and voting rules. \\
Voting & November 2012 & 1 & The 2012 Presidential Election between Barack Obama and Mitt Romney. & Barack Obama's reelection in the November 2012 presidential election focused national attention on voting and turnout. \\
Voting & November 2016 & 1 & The 2016 Presidential Election between Donald Trump and Hillary Clinton. & 2016 presidential election and Donald Trump's victory made voting access, turnout, and election rules central national issues. \\
Voting & October 2020 & 2 & Lead up to the 2020 Presidential Election, marked by an unprecedented surge in mail-in voting and voter fraud controversies. & Final phase of the 2020 presidential election, dominated by mail voting, early voting, and disputes over election rules. \\
Voting & November 2024 & 1 & The 2024 Presidential Election between Donald Trump and Kamala Harris. & 2024 presidential election, in which Donald Trump won a return to the White House. \\
\end{longtable}
\end{landscape}
}